\documentclass{amsart}
\usepackage{amssymb}

\usepackage{graphicx}
\usepackage{amscd}
\usepackage{amsmath}

\theoremstyle{plain}

\numberwithin{equation}{section}

\input{tcilatex}
\input tcilatex

\begin{document}
\title[Quantum Causality and Information]{Quantum Causality, Decoherence, Trajectories and Information}
\author{V P Belavkin}
\address{Mathematics Department, University of Nottingham, NG7 2RD, UK}
\email{vpb@maths.nott.ac.uk}
\thanks{}
\date{}
\dedicatory{In celebration of the 100th anniversary of the discovery of quanta}

\begin{abstract}
A history of the discovery of ``new'' quantum mechanics and the paradoxes of
its probabilistic interpretation are briefly reviewed from the modern point
of view of quantum probability and information. The modern quantum theory,
which has been developed during the last 20 years for treatment of quantum
open systems including quantum noise, decoherence, quantum diffusions and
spontaneous jumps occurring under continuous in time observation, is not yet
a part of the standard curriculum of quantum physics. It is argued that the
conventional formalism of quantum mechanics is insufficient for the
description of quantum events, such as spontaneous decays say, and the new
experimental phenomena related to individual quantum measurements, but they
all have received an adequate mathematical treatment in quantum stochastics
of open systems.

Moreover, the only reasonable probabilistic interpretation of quantum
mechanics put forward by Max Born was in fact in irreconcilable
contradiction with traditional mechanical reality and causality. This led to
numerous quantum paradoxes, some of them due to the great inventors of
quantum theory such as Einstein and Schr\"{o}dinger. They are reconsidered
in this paper from the point of view of quantum information.

The development of quantum measurement theory, initiated by von Neumann,
indicated a possibility for resolution of this interpretational crisis by
divorcing the algebra of the dynamical generators and the algebra of the
actual observables, or \textit{be}ables. It is shown that within this
approach quantum causality can be rehabilitated in the form of a
superselection rule for compatibility of the past beables with the potential
future. This rule, together with the self-compatibility of the measurements
insuring the consistency of the histories, is called the nondemolition, or
causality principle in modern quantum theory. The application of this rule
in the form of the dynamical commutation relations leads to the derivation
of the von Neumann projection postulate, and also to the more general
reductions, instantaneous, spontaneous, and even continuous in time. This
gives a quantum stochastic solution, in the form of the dynamical filtering
equations, of the notorious measurement problem which was tackled
unsuccessfully by many famous physicists starting with Schr\"{o}dinger and
Bohr.

It has been recently proved that the quantum stochastic model for the
continuous in time measurements is equivalent to a Dirac type boundary-value
problem for the secondary quantized input ''offer waves from future'' in one
extra dimension, and to a reduction of the algebra of the consistent
histories of past events to an Abelian subalgebra for the ``trajectories of
the output particles''. This supports the corpuscular-wave duality in the
form of the thesis that everything in the future are quantized waves,
everything in the past are trajectories of the recorded particles.
\end{abstract}

\maketitle
\urladdr{http://www.maths.nott.ac.uk/personal/vpb/}
\tableofcontents

\section{Introduction}

In 1918 Max Planck was awarded the Nobel Prize in Physics for his quantum
theory of blackbody radiation or as we would say now, quantum theory of
thermal noise based on the hypothesis of energy discontinuity. Invented in
1900, it inspired an unprecedented revolution in both physical science and
philosophy of the 20th century, with an unimaginable deep revision in our
way of thinking.

In 1905 Einstein, examining the photoelectric effect, proposed a quantum
theory of light, only later realizing that Planck's theory made implicit use
of this quantum light hypothesis. Einstein saw that the energy changes in a
quantum material oscillator occur in jumps which are multiples of $\omega $.
Einstein received Nobel prize in 1922 for his work on the photoelectric
effect.

In 1912 Niels Bohr worked in the Rutherford group in Manchester on his
theory of the electron in an atom. He was puzzled by the discrete spectra of
light which is emitted by atoms when they are subjected to an excitation. He
was influenced by the ideas of Planck and Einstein and addressed a certain
paradox in his work. How can energy be conserved when some energy changes
are continuous and some are discontinuous, i.e. change by quantum amounts?%
\textit{\ }Bohr conjectured that an atom could exist only in a discrete set
of stable energy states, the differences of which amount to the observed
energy quanta. Bohr returned to Copenhagen and published a revolutionary
paper on the hydrogen atom in the next year. He suggested his famous formula 
\begin{equation*}
E_{m}-E_{n}=\hbar \omega _{mn}
\end{equation*}
from which he derived the major laws which describe physically observed
spectral lines. This work earned Niels Bohr the 1922 Nobel Prize about $%
10^{5}$ Swedish Kroner.

Thus before the rise of quantum mechanics 75 years ago, quantum physics had
appeared first in the form of quantum stochastics, i. e. the statistics of
quantum thermal noise and rules for quantum spontaneous jumps which are
often called the ``old quantum theory'', but in fact the occurrence of the
discontinuous quantum jumps as individual events have never been explained
by the ``new'' time-continuous quantum mechanics.

Quantum theory is the greatest intellectual achievements of the past
century. Since the discovery of quanta by Max Planck exactly 100 years ago
on the basis of spectral analysis of quantum thermal noise, and the wave
nature of matter, it has produced numerous paradoxes and confusions even in
the greatest scientific minds such as those of Einstein, de Broglie,
Schr\"{o}dinger, and it still confuses many contemporary philosophers and
scientists who fail to accept the Aristotle's superadditivity law of Nature.
Rapid development of the beautiful and sophisticated mathematics for quantum
mechanics and the development of its interpretation by Bohr, Born,
Heisenberg, Dirac, von Neumann and many others who abandoned traditional
causality, were little help in resolving these paradoxes despite the
astonishing success in the statistical prediction of the individual quantum
phenomena. Both the implication and consequences of the quantum theory of
light and matter, as well as its profound mathematical, conceptual and
philosophical foundations are not yet understood completely by the majority
of quantum physicists.

Specialists in different narrow branches of mathematics and physics rarely
understand quantum theory as a common thread which runs through everything.
The creators of quantum mechanics, the theory invented for interpretation of
the dynamical laws of fundamental particles, were unable to find a
consistent interpretation of it since they were physicists with a classical
mathematical education. After inventing quantum mechanics they spent much of
their lives trying to tackle the Problem of Quantum Measurement, the ``only
remaining problem'' of quantum theory -- the problem of its consistent
interpretation. Modern quantum phenomenology deals with individual quantum
events such as quantum diffusions and jumps which simply do not exist in the
orthodox quantum mechanics. There is no place for quantum events neither in
the existing quantum field theories nor in the projects of quantum gravity.
Moreover, as we shall see, there can't be any, even ``hidden variable''
solution of the problem of a consistent statistical interpretation of \emph{%
quantum causality} in the orthodox quantum theory. But without such
interpretation even the unified quantum field theory and gravity would stay
`a thing in itself' as a quantum mechanics of the closed universe without a
possibility of any kind of quantum future prediction based on the results of
past observations.

In this paper we review the most obscure sides of quantum systems theory
which are related to quantum causality and its implications for the time
arrow, dynamical irreversibility, consistent histories and prediction of
future. These sides traditionally considered by mathematical physicists as
`ill-defined', are usually left to quantum philosophers and theoreticians
for vague speculations. Most theoretical physicists have a broad
mathematical education, but it tends to ignore two crucial aspects for the
solution of these problems -- information theory and statistical
conditioning. This is why most of theoretical physicists are not familiar
with the mathematical development for solving all these questions which has
been achieved in \emph{quantum probability, information and quantum
mathematical statistics} during last 25 years. Surely the ``professional
theoretical physicists ought to be able to do better'' (cited from J Bell, 
\cite{Bell87}, p. 173).

As we shall show, the solution to this most fundamental problem of quantum
theory can be found in the framework of quantum probability as the rigorous
conceptual basis for \emph{the algebraic quantum systems theory} which is a
part of a unified mathematics and physics rather than quantum mechanics, and
an adequate framework for the treatment of both classical and quantum
systems in a unified way. We shall prove that the problem of quantum
measurement \emph{has} a solution in this framework which admits the
existence of infinite systems and superselection rules, some of them are
already wellused in the algebraic quantum physics. This can be achieved only
by imposing a new superselection rule for quantum causality called \emph{the
nondemolition principle} \cite{Be94} which can be short-frased as \emph{`The
past is classical, and it is consistent with the quantum future'}. This
future-past superselection rule defines the time arrow which leads to a
certain restriction of the general relativity principle and the notion of
local reality, the profound philosophical implications of which should yet
be analyzed. We may conjecture that the difficulties of quantum gravity are
fundamentally related to the orthodox point of view that it should be a
unification, or consistent with the general relativity and the quantum
theory, without taking into account the implications imposed by the
causality superselection rule on these theories. In any unified quantum
theory with a consistent interpretation addressing modern physics there
should be a place for quantum events and Bell's \emph{be}ables as elements
of classical reality and quantum causality. Without this it will never be
tested experimentally: any experiment deals with the statistics of random
events as the only elements of reality, and any reasonable physical theory
should admit the predictions based on the statistical causality.

The new infinity and superselection arise with the unitary dilation
constructed in this paper for the dynamical derivation of the projection or
any other reduction postulate which is simply interpreted as the Bayesian
statistical inference for the posterior states based on the prior
information in the form of the classical measurement data and the initial
quantum state. Such quantum statistical inference called quantum filtering
has no conceptual difficulty for the interpretation: it is understood as the
prior-posterior transition from \emph{quantum possibilities} to \emph{%
classical actualities} exactly in the same way as it is understood in
classical statistics where there is no problem of measurement. Indeed, the
collapse of the prior (mixed) state to the posterior (pure) state after the
measurement in classical probability is only questioned by those who don't
understand that this is purely in the nature of information: it is simply
the result of inference due to the gaining information about the existing
but \emph{a priori} unknown pure state. The only distinction of the
classical theory from quantum is that the prior mixed states cannot be
dynamically achieved from pure initial states without a procedure of either
statistical or chaotic mixing. In quantum theory, however, the mixed, or
decoherent states can be dynamically induced on a subsystem from the initial
pure disentangled states of a composed system simply by a unitary
transformation. And the quantum statistical inference (filtering) can result
effectively in the ``collapse''\ of a pure initial state to the posterior
state corresponding to the result of the measurement (This never happens in
classical statistics: a pure prior state induced dynamically by a pure
initial state doesn't collapse as it coincide with the posterior pure state
as the predetermined result of the inference with no gain of information
about the \emph{a priori} known pure state.)

The main aim of this paper is to give a comprehensive review of the recent
progress in this modern quantum theory now known also as \emph{quantum
stochastics} from the historical perspective of the discovery the
deterministic quantum evolutions by Heisenberg and Schr\"{o}dinger to the
stochastic evolutions of quantum jumps and quantum diffusion in quantum
noise. We will argue that this is the direction in which quantum theory
would have developed by the founders if the mathematics of quantum
stochastics had been discovered by that time. After resolving the typical
paradox of \ the Schr\"{o}dinger's cat by solving as suggested the related
instantaneous quantum measurement problem we move to the dynamical reduction
problems for the sequential and time-continuous measurements.

In order to appreciate the mathematical framework and rigorous formulations
for resolving these fundamental problems of quantum theory it is quite
instructive to start with the typical practical problems of the dynamical
systems theory, such as quantum sequential measurements, quantum statistical
prediction and quantum feedback control in real time. These problems were
indeed set up and analyzed even in the continuous time in the framework of
quantum theory first for the linear quantum dynamical systems (quantum open
oscillators) over 20 years ago in \cite{Be79, Be80} pioneering the new
quantum stochastic approach, and they have been now developed in full
generality within this modern rigorous approach to quantum open systems.

We shall concentrate on the modern quantum stochastic approach to \emph{%
quantum consistent histories}, \emph{quantum trajectories} and \emph{quantum
diffusions} originated in \cite{Be80, Be85, BaLu85, Be88, Dio88, Be89a,
Gis89, Be90b, GPR90, Be90c}. In its linear renormalized version this
approach gives the output statistics of \emph{quantum continuous measurements%
} as a result of the solution of a stochastic differential equation. This
allows the direct application of quantum conditioning and filtering methods
to tackle the dynamical problem of quantum individual dynamics under
continual (trajectorial) measurement. Here we refer mainly to the pioneering
and original papers on quantum diffusions in which the relevant quantum
structures as mathematical notions and methods were first invented. Most of
these results were rigorously proved and published in the mathematical
physics literature and is not well known for the larger physical audience.

We shall give a brief account of the relevant mathematics which plays the
same role for quantum stochastics as did the classical differential calculus
for Newtonian dynamics, and concentrate on its application to the dynamical
solution of quantum measurement problems \cite{Wig63, Dav76, Be78, BLP82,
Hol82, Be83, Bar83}, rather than give the full account of all related
theoretical papers which use more traditional ''down to earth'' methods.
Among these we would like to mention the papers on quantum decoherence \cite
{Zeh70, UnZu89}, dynamical state reduction program \cite{Per76, Gis83},
consistent histories and evolutions \cite{GeHa90, Haa95}, spontaneous
localization and events \cite{GRW86, BlJa95}, restricted and unsharp path
integrals \cite{Men93, AKS97} and their numerous applications to quantum
countings, jumps and trajectories in quantum optics and atomic physics \cite
{MiWa84, WCM85, Car86, ZMW87, Bar87, HMW89, Ued90, MiGa92}. Most of these
papers develop a phenomenological approach which is based on a
non-Hamiltonian ``instrumental'' linear master equation giving the
statistics of quantum measurements, but is not well adapted for the
description of individual and conditional behavior under the continuous
measurements. Pearl and Gisin took an opposite, nonlinear, initially even
deterministic approach for the individual evolutions, without considering
the statistics of measurements \cite{Per76, Gis83, Gis84}.

During the 90's many ''new'' quantum theories appeared in the theoretical
and applied physics literature, in particular, the quantum state diffusion
theory \cite{GiPe92, GiPe93}, where the nonlinear quantum filtrering
stochastic equations for diffusive measurements have been used without even
a reference to the continuous measurements, and quantum trajectories in
quantum optics \cite{Car93, WiMi93, GoGr93, WiMi94, GoGr94, Car94}, where
the stochastic solutions to quantum jump equations have been constructed
even without a reference to quantum stochastic filtering equations. However
the transition from nonlinear to linear stochastic equations and the quantum
stochastic unitary models for the underlying Hamiltonian microscopic
evolutions remain unexplained in these papers. Moreover, most of these
papers claim primarity or universality of the stochastic evolution but treat
very particular phenomenological models which are based only on the
counting, or sometimes diffusive (homodyne and heterodyne) models of quantum
noise and output process, and reinvent many notions such as quantum
conditioning and adaptedness with respect to the individual trajectories,
without references to the general quantum stochastic measurement and
filtering (conditioning) theory. An exception occurred only in \cite{GoGr94,
GGH95}, where our quantum stochastic filtering theory which had been
developed for these purposes in the 80's, was well understood both at a
macroscopic and microscopic level. This explains why a systematic review of
this kind is needed.

In order to appreciate the quantum drama which has been developing through
the whole century, it seems useful to give a account of the discovery of
quantum mechanics and its probabilistic interpretation at the 20th of the
past century. This is briefly done in the first sections Chapter 1. More
about the discovery of the ``old'' and ``new'' quantum mechanics starting
from the Plank's quanta \cite{Plnkab} and thei relation to the modern theory
of quantum noise and applications to quantum bits one can find in \cite
{Be01b}. Readers who are not interested in the historical perspective of
this subject and paradoxes of its interpretation are advised to start with
the Chapter 2 dealing with the famous problem of quantum measurement. The
specialists who are familiar with this might still find intersting a review
on quantum stochastics, causality, consistent trajectories, continual
measurements, quantum jumps and diffusions, and will find the origin and
explanation of these modern quantum theories in the last sections.

\section{Quantum Mechanics, Probabilities and Paradoxes}

\medskip\medskip

\begin{quote}
\textit{If anyone says he can think about quantum problems without getting
giddy, that only shows he has not understood the first thing about them} --
Max Planck.
\end{quote}

\subsection{At the Origin of Quantum Mechanics}

\medskip\medskip

\begin{quote}
\textit{The whole is more than the sum of its parts }-- Aristotle.
\end{quote}

This is the famous superadditivity law from Aristotle's \textit{Metaphysics}
which studies `the most general or abstract features of reality and the
principles that have universal validity'. Certainly in this broad definition
quantum physics is the most essential part of metaphysics.

Quantum theory is a mathematical theory which studies the most fundamental
features of reality in a unified form of waves and matter, it raises and
solves the most fundamental riddles of Nature by developing and utilizing
mathematical concepts and methods of all branches of modern mathematics,
including probability and statistics. Indeed, it began with the discovery of
new laws for `quantum' numbers, the natural objects \ which are the
foundation of pure mathematics. (`God made the integers; the rest is man's
work' -- Kronecker). Next it invented new applied mathematical methods for
solving quantum mechanical matrix and partial differential equations. Then
it married probability with algebra to obtain unified treatment of waves and
particles in nature, giving birth to quantum probability and creating new
branches of mathematics such as quantum logics, quantum topologies, quantum
geometries, quantum groups. It inspired the recent creation of quantum
analysis and quantum calculus, as well as quantum statistics and quantum
stochastics.

\subsubsection{The discovery of matrix mechanics}

In 1925 a young German theoretical physicist, Heisenberg, gave a preliminary
account of a new and highly original approach to the mechanics of the atom 
\cite{Heis25}. He was influenced by Niels Bohr and proposed to substitute
for the position coordinate of an electron in the atom complex arrays 
\begin{equation*}
q_{mn}\left( t\right) =q_{mn}e^{i\omega _{mn}t}
\end{equation*}
oscillating with Bohr's frequencies $\omega _{mn}=\hbar ^{-1}\left(
E_{m}-E_{n}\right) $. He thought that they would account for the random
jumps $E_{m}\mapsto E_{n}$ in the atom corresponding to the spontaneous
emission of the energy quanta $\hbar \omega _{mn}$. His Professor, Max Born,
was a mathematician who immediately recognized an infinite matrix algebra in
Heisenberg's multiplication rule for the tables $\mathrm{Q}\left( t\right) =%
\left[ q_{mn}\left( t\right) \right] $. The classical momentum was also
replaced by a similar matrix, 
\begin{equation*}
\mathrm{P}\left( t\right) =\left[ p_{mn}e^{i\omega _{mn}t}\right] ,
\end{equation*}
as the Planck's electro- magnetic quanta were thought being induced by
oscillator equations 
\begin{equation*}
\frac{\mathrm{d}}{\mathrm{d}t}q_{mn}\left( t\right) =i\omega
_{mn}q_{mn}\left( t\right) ,\;\frac{\mathrm{d}}{\mathrm{d}t}p_{mn}\left(
t\right) =i\omega _{mn}p_{mn}\left( t\right) .
\end{equation*}
These equations written in terms of the matrix algebra as 
\begin{equation}
\frac{\mathrm{d}}{\mathrm{d}t}\mathrm{Q}\left( t\right) =\frac{i}{\hbar }%
\left[ \mathrm{H},\mathrm{Q}\left( t\right) \right] ,\;\frac{\mathrm{d}}{%
\mathrm{d}t}\mathrm{P}\left( t\right) =\frac{i}{\hbar }\left[ \mathrm{H},%
\mathrm{P}\left( t\right) \right] ,  \label{1.1}
\end{equation}
now are known as the Heisenberg equations, where $\mathrm{H}$ called the
Hamiltonian, is the diagonal matrix $\mathrm{E}=\left[ E_{n}\delta _{mn}%
\right] $, and $\left[ \mathrm{H},\mathrm{B}\right] $ denotes the matrix
commutator $\mathrm{HB}-\mathrm{BH}$. In order to achieve the correspondence
with the classical mechanics, their young colleague Jordan suggested to
postulate the canonical (Heisenberg) commutation relations 
\begin{equation}
\left[ \mathrm{Q}\left( t\right) ,\mathrm{P}\left( t\right) \right] =i\hbar 
\mathrm{I},  \label{1.2}
\end{equation}
where $\mathrm{I}$ is the unit matrix $\left[ \delta _{mn}\right] $, \cite
{BHJ26}. This made the equations (\ref{1.1}) formally equivalent to the
Hamiltonian equations 
\begin{equation*}
\frac{\mathrm{d}}{\mathrm{d}t}\mathrm{Q}\left( t\right) =H_{p}\left( \mathrm{%
Q}\left( t,\mathrm{P}\left( t\right) \right) \right) ,\;\frac{\mathrm{d}}{%
\mathrm{d}t}\mathrm{P}\left( t\right) =-H_{q}\left( \mathrm{Q}\left(
t\right) ,\mathrm{P}\left( t\right) \right) ,
\end{equation*}
but the derivatives $H_{p},H_{q}$ of the Hamiltonian function $H\left(
q,p\right) $ were now replaced by the appropriate matrix-algebra functions
of the noncommuting $\mathrm{P}$ and $\mathrm{Q}$ such that $H\left( \mathrm{%
Q},\mathrm{P}\right) =\mathrm{H}$. For the non-relativistic electron in a
potential field $\phi $ this yielded the Newton equation 
\begin{equation}
m\frac{\mathrm{d}^{2}}{\mathrm{d}t^{2}}\mathrm{Q}\left( t\right) =-\nabla
\phi \left( \mathrm{Q}\left( t\right) \right) ,  \label{1.3}
\end{equation}
but with the non-commuting initial conditions $\mathrm{Q}\left( 0\right) $
and $\frac{\mathrm{d}}{\mathrm{d}t}\mathrm{Q}\left( 0\right) =\frac{1}{m}%
\mathrm{P}\left( 0\right) $, where the potential force is replaced by the
corresponding matrix function of $\mathrm{Q}=\left[ q_{mn}\right] $.

Thus the new, quantum mechanics was first invented in the form of \emph{%
matrix mechanics}, emphasizing the possibilities of quantum transitions, or
jumps between the stable energy states $E_{n}$ of an electron. However there
was no mechanism suggested to explain the actualities of these spontaneous
transitions in the continuous time. In 1932 Heisenberg was awarded the Nobel
Prize for his pioneering work on the mathematical formulation of the new
physics.

Conceptually, the new atomic theory was based on the positivism of Mach as
it operated not with real space-time but with only observable quantities
like atomic transitions. However many leading physicists were greatly
troubled by the prospect of loosing reality and deterministic causality in
the emerging quantum physics. Einstein, in particular, worried about the
element of `chance' which had entered physics. In fact, this worries came
rather late since Rutherford had introduced a spontaneous effect when
discussing radio-active decay in 1900.

\subsubsection{The discovery of wave mechanics}

In 1923 de Broglie, inspired by the works of Einstein and Planck, extended
the wave-corpuscular duality also to material particles. He used the
Hamilton-Jacobi theory which had been applied both to particles and waves.
In 1928 de Broglie received the Nobel Prize for this work.

In 1925, Schr\"{o}dinger gave a seminar on de Broglie's material waves, and
a member of the audience suggested that there should be a wave equation.
Within a few weeks Schr\"{o}dinger found his celebrated wave equation, first
in a relativistic, and then in the non-relativistic form \cite{Schr26}.
Instead of seeking the classical solutions to the Hamilton-Jacobi equation 
\begin{equation*}
H\left( q,\frac{\hbar }{i}\frac{\partial }{\partial q}\ln \psi \right) =E
\end{equation*}
he suggested finding those wave functions $\psi \left( q\right) $ which
satisfy the linear equation 
\begin{equation}
H\left( q,\frac{\hbar }{i}\frac{\partial }{\partial q}\right) \psi =E\psi
\label{1.4}
\end{equation}
(It coincides with the former equation only if the Hamiltonian function $%
H\left( q,p\right) $ is linear with respect to $p$ but not for the
non-relativistic $H\left( q,p\right) =\frac{1}{2m}p^{2}+\phi \left( q\right) 
$).

Schr\"{o}dinger published his revolutionary \textit{wave mechanics} in a
series of six papers \cite{Schr26c} in 1926 during a short period of
sustained creative activity that is without parallel in the history of
science. Like Shakespeare, whose sonnets were inspired by a dark lady,
Schr\"{o}dinger was inspired by a mysterious lady of Arosa where he took ski
holidays during the Christmas 1925 but `had been distracted by a few
calculations'. This was the second formulation of the new quantum theory,
which he successfully applied to the Hydrogen atom, oscillator and other
quantum mechanical systems, solving the corresponding Sturm-Liouville
boundary-value problems of mathematical physics. The mathematical
equivalence between the two formulations of quantum mechanics was understood
by Schr\"{o}dinger in the fourth paper where he suggested the non-stationary
wave equation written in terms of the Hamiltonian operator $\mathrm{H}%
=H\left( q,\frac{\hbar }{i}\frac{\partial }{\partial q}\right) $ for the
complex\ time-dependent wave-function $\psi \left( t,q\right) $ simply as 
\begin{equation}
i\hbar \frac{\partial }{\partial t}\psi \left( t\right) =\mathrm{H}\psi
\left( t\right) ,  \label{1.5}
\end{equation}
and he also introduced operators associated with each dynamical variable.

Unlike Heisenberg and Born's matrix mechanics, the general reaction towards
wave mechanics was immediately enthusiastic. Plank described
Schr\"{o}dinger's wave mechanics as `epoch-making work'. Einstein wrote:
`the idea of your work springs from true genius...'. Next year
Schr\"{o}dinger was nominated for the Nobel Prize, but he failed to receive
it in this and five further consecutive years of his nominations by most
distinguished physicists of the world, the reason behind his rejection being
`the highly mathematical character of his work'. Only in 1933 did he receive
his prize, this time jointly with Dirac, and this was the first, and perhaps
the last, time when the Nobel Prize for physics was given to true
mathematical physicists.

Following de Broglie, Schr\"{o}dinger initially thought that the wave
function corresponds to a physical vibration process in a real continuous
space-time because it was not stochastic, but he was puzzled by the failure
to explain the blackbody radiation and photoelectric effect from this wave
point of view. In fact the wave interpretation applied to light quanta leads
back to classical electrodynamics, as his relativistic wave equation for a
single photon coincides mathematically with the classical wave equation.
However after realizing that the time-dependent $\psi $ in (\ref{1.5}) must
be a complex function, he admitted in his fourth 1926 paper \cite{Schr26c}
that the wave function $\psi $ cannot be given a direct interpretation, and
described the wave density $\left| \psi \right| ^{2}=\psi \bar{\psi}$ as a
sort of weight function for superposition of point-mechanical configurations.

Although Schr\"{o}dinger was a champion of the idea that the most
fundamental laws of the microscopic world are absolutely random even before
he discovered wave mechanics, he failed to see the probabilistic nature of $%
\psi \bar{\psi}$. Indeed, his equation was not stochastic, and it didn't
account for the individual random jumps $E_m\rightarrow E_n$ of the
Bohr-Heisenberg theory but rather opposite, it did prescribe the
preservation of the eigenvalues $E=E_n$.

For the rest of his life Schr\"{o}dinger was trying to find apparently
without a success a more fundamental equation which would be responsible for
the energy transitions in the process of measurement of the quanta $\hbar
\omega _{mn}$. As we shall see, he was right assuming the existence of such
equation.

\subsubsection{Interpretations of quantum mechanics}

The creators of the rival matrix quantum mechanics were forced to accept the
simplicity and beauty of Schr\"{o}dinger's approach. In 1926 Max Born put
forward the statistical interpretation of the wave function by introducing
the statistical means 
\begin{equation*}
\left\langle \mathrm{X}\right\rangle =\int \bar{\psi}\left( x\right) x\psi
\left( x\right) \mathrm{d}x
\end{equation*}
for Hermitian dynamical variables $\mathrm{X}$ in the physical state,
described in the eigen-representation of $\mathrm{X}$ by a complex function $%
\psi \left( x\right) $ normalized as $\left\langle \mathrm{I}\right\rangle
=1 $. Thus he identified the quantum states with one-dimensional subspaces
of a Hilbert space $\mathcal{H}$ corresponding to the normalized $\psi $
defined up to a complex factor $\mathrm{e}^{i\theta }$. This was developed
in Copenhagen and gradually was accepted by almost all physicists as the
``Copenhagen interpretation''. Born by education was a mathematician, but he
received the Nobel Prize in physics for his statistical studies of wave
functions later in 1953 as a Professor of Natural Philosophy at Edinburgh.
Bohr, Born and Heisenberg considered electrons and quanta as unpredictable
particles which cannot be visualized in the real space and time.

The most outspoken opponent of a/the probabilistic interpretation was
Einstein. Albert Einstein admired the new development of quantum theory but
was suspicious, rejecting its acausality and probabilistic interpretation.
It was against his scientific instinct to accept statistical interpretation
of quantum mechanics as a complete description of physical reality. There
are famous sayings of his on that account:

\begin{quote}
\textit{`God is subtle but he is not malicious', `God doesn't play dice'}
\end{quote}

During these debates on the probabilistic interpretation of quantum
mechanics of Einstein between Niels Bohr, Schr\"{o}dinger often sided with
his friend Einstein, and this may explain why he was distancing himself from
the statistical interpretation of his wave function. Bohr invited
Schr\"{o}dinger to Copenhagen and tried to convince him of the
particle-probabilistic interpretation of quantum mechanics. The discussion
between them went on day and night, without reaching any agreement. The
conversation, however deeply affected both men. Schr\"{o}dinger recognized
the necessity of admitting both wave and particles, but he never devised a
comprehensive interpretation rival to Copenhagen orthodoxy. \ Bohr ventured
more deeply into philosophical waters and emerged with his concept of
complementarity:

\begin{quote}
\textit{Evidence obtained under different experimental conditions cannot be
comprehended within a single picture, but must be regarded as complementary
in the sense that only the totality of the phenomena exhausts the possible
information about the objects.}
\end{quote}

In his later papers Schr\"{o}dinger accepted the probabilistic
interpretation of $\psi \bar{\psi}$, but he did not consider these
probabilities classically, instead he interpreted them as the strength of
our belief or anticipation of an experimental result. In this sense the
probabilities are closer to propensities than to the frequencies of the
statistical interpretation of Born and Heisenberg. Schr\"{o}dinger had never
accepted the subjective positivism of Bohr and Heisenberg, and his
philosophy is closer to that called representational realism. He was content
to remain a critical unbeliever.

There have been many other attempts to retain the deterministic realism in
the quantum world, the most extravagant among these being the ensemble-world
interpretations of Bohm \cite{Boh52} and Everett \cite{Eve57}. The first
interpretational theory, known as the pilot-wave theory, is based on the
conventional Schr\"{o}dinger equation which is used to define the flow of a
classical fluid in the configuration space. The predictions of this
classical macroscopic theory coincide with the statistical predictions of
the orthodox quantum theory only for the ensembles of coordinate-like
observables with the initial probability distribution over the many worlds
given by the initial pilot wave. Other observables like momenta which are
precisely determined at each point by the velocity of this fluid, have no
uncertainty under the fixed coordinates. This is inconsistent with the
prediction of quantum theory for individual systems, and there is no way to
incorporate the stochastic dynamics of sequentially monitored individual
quantum particles in the single world into this fluid dynamics. Certainly
this is a variation of the de Broglie-Schr\"{o}dinger old interpretation,
and it doesn't respect the Bell's first principle for the interpretational
theories ``that it should be possible to formulate them for small systems'' 
\cite{Bell87}, p. 126.

The Everett's many-world interpretation also assumes that the classical
configurations at each time are distributed in the comparison class of
possible worlds worth probability density $\psi \bar{\psi}$. However no
continuity between present and past configurations is assumed, and all
possible outcomes of their measurement are realized every time, each in a
different edition of the continuously multiplying universe. The observer in
a given brunch of the universe is aware only of what is going on in that
particular branch, and this results in the reduction of the wave-function.
This would be macroscopically equivalent to the pilot-wave theory if the de
Broglie-Schr\"{o}dinger-Bohm fluid dynamics could be obtained as the average
of wave equations over all brunches. An experienced statistician would
immediately recognize in this many-world interpretation an ensemble model
for a continuously branching stochastic process, and would apply the
well-developed stochastic analysis and differential calculus to analyze this
dynamical model. However no stochastic equation for a continuously monitored
branch was suggested in this theory, although there should be many if at
all, corresponding to many possible choices of classical configurations
(e.g. positions or momenta) for a single many-world Schr\"{o}dinger equation.

Living simultaneously in many worlds would have perhaps certain advatages,
but from the philosophical and practical point of view, however, to have an
infinite number (continuum product of continua?) of real worlds at the same
time without their communication seems not better than to have none. As Bell
wrote in \cite{Bell87}, p. 134: ``to have multiple universes, to realize all
possible configurations of particles, would have seemed grotesque''. Even if
such a weighted many-world dynamical theory had been developed to a
satisfactory level, it would have been reformulated in terms of
well-established mathematical language as a stochastic evolutionary theory
in the single world with the usual statistical interpretation. In fact, the
stochastic theory of continuously observed quantum systems has been already
derived, not just developed, in full generality and rigor in quantum
stochastics, and it will be presented in the last sections. But first we
shall demonstrate the underlying ideas on the elementary single-transition
level.

\subsection{Uncertaities and Quantum Probabilities}

\medskip\medskip

\begin{quote}
\textit{In mathematics you don't understand things. You just get used to them%
} - John von Neumann.
\end{quote}

In 1932 von Neumann put quantum theory on firm theoretical basis by setting
the mathematical foundation for new, quantum, probability theory, the
quantitative theory for counting non commuting events of quantum logics.
This noncommutative probability theory is based on essentially more general
axioms than the classical (Kolmogorovian) probability of commuting events,
which form common sense Boolean logic, first formalized by Aristotle. The
main idea of this theory is based upon the empirical fact that the maximal
number of alternatives in each experiment over a quantum system is smaller
than the quantum probability dimensionality of the system, i.e. the
dimensionality of the linear space of all propensities, or empirical
frequencies defining the quantum state. Unlike in the classical world were
the maximal number of alternatives always coincides with the dimensionality
of the probability space. Actually the quantum dimensionality squares the
classical one, is the sum of such squares for the hybrid systems, and this
defines every quantum probability space as the space of density matrices
rather than space of density functions.

Quantum probability has been under extensive development during the last 30
years since the introduction of algebraic and operational approaches for
treatment of noncommutative probabilities, and currently serves as the
mathematical basis for quantum stochastics and information theory.
Unfortunately its recent development was more in parallel with classical
probability theory rather than with physics, and many mathematical
technicalities of quantum calculi prevented its acceptance in physics.

In the next section we shall demonstrate the main ideas of quantum
probability arising from the application of classical probability theory to
quantum phenomena on the simple quantum systems. The most recent
mathematical development of these models and methods leads to a profound
stochastic theory of quantum open systems with many applications including
quantum information, quantum measurement, quantum filtering and prediction
and quantum feedback control, some of them are presented in the last
sections.

\subsubsection{Heisenberg uncertainty relations}

In 1927 Heisenberg derived \cite{Heis27} his famous uncertainty relations 
\begin{equation}
\Delta \mathrm{Q}\Delta \mathrm{P}\geq \hbar /2,\quad \Delta \mathrm{T}%
\Delta \mathrm{E}\geq \hbar /2  \label{2.1}
\end{equation}
which gave mathematical support to the revolutionary complementary principle
of Bohr. As Dirac stated:

\begin{quote}
\textit{Now when Heisenberg noticed that, he was really scared.}
\end{quote}

The first relation can be easily understood in the Schr\"{o}dinger
representations $\mathrm{Q}=x$, $\mathrm{P}=\frac{\hbar }{i}\frac{\partial }{%
\partial x}$ in terms of the standard deviations 
\begin{equation}
\Delta \mathrm{Q}=\left\langle \widetilde{\mathrm{Q}}^{2}\right\rangle
^{1/2},\quad \Delta \mathrm{P}=\left\langle \widetilde{\mathrm{P}}%
^{2}\right\rangle ^{1/2},  \label{2.2}
\end{equation}
where $\widetilde{\mathrm{Q}}=\mathrm{Q}-\left\langle \mathrm{Q}%
\right\rangle \mathrm{I}$ and $\widetilde{\mathrm{P}}=\mathrm{P}%
-\left\langle \mathrm{P}\right\rangle \mathrm{I}$ have the same commutator $%
\left[ \widetilde{\mathrm{Q}},\widetilde{\mathrm{P}}\right] =\frac{\hbar }{i}%
\mathrm{I}$ as $\mathrm{Q}$ and $\mathrm{P}$. To this end one can use the
Schwarz inequality $\left\langle \widetilde{\mathrm{Q}}^{2}\right\rangle
\left\langle \widetilde{\mathrm{P}}^{2}\right\rangle \geq \left|
\left\langle \widetilde{\mathrm{Q}}\widetilde{\mathrm{P}}\right\rangle
\right| ^{2}$ and that 
\begin{equation*}
\left| \left\langle \widetilde{\mathrm{Q}}\widetilde{\mathrm{P}}%
\right\rangle \right| \geq \left| \func{Im}\left\langle \widetilde{\mathrm{Q}%
}\widetilde{\mathrm{P}}\right\rangle \right| =\frac{1}{2}\left| \left\langle %
\left[ \widetilde{\mathrm{Q}},\widetilde{\mathrm{P}}\right] \right\rangle
\right| =\frac{\hbar }{2}.
\end{equation*}
.

The second uncertainty relation, which was first stated by analogy of $t$
with $x$ and of $\mathrm{E}$ with $\mathrm{P}$, cannot be proved in the same
way as the time operator $\mathrm{T}$ does not exist in the
Schr\"{o}dinger's Hilbert space $\mathcal{H}$ of wave functions $\psi \left(
x\right) $. So, it is usually interpreted in terms of the time $\Delta 
\mathrm{T}$ required for a quantum nonstationary state with spread in energy 
$\Delta \mathrm{E}$ to evolve to an orthogonal and hence distiguishable
state \cite{AhBo61}. However it can also be proved \cite{Be76, Hol80} in
terms of the standard deviation $\Delta \mathrm{T}$ of the \ optimal
statistical estimate for the time $t$ of the wave packet in the
Schr\"{o}dinger's picture $\psi \left( t-s\right) $ with respect to an
unknown initial $t_{0}=s$, with the energy spread $\Delta \mathrm{E}$. The
similar problem for the shift parameter $q$ in the wave packet $\psi \left(
x-q\right) $ defines the optimal estimate as the measurement of the
coordinate operator $\mathrm{Q}=x$. Although the optimal estimation of $t$
cannot be treated as the usual quantum mechanical measurement of a
self-adjoint operator in $\mathcal{H}$, the optimal estimation can been
realized \cite{BePer98} by the measurement of the self-adjoint operator $%
\mathrm{T}=s$ in an extended (doubled) Hilbert state space $\mathbb{H=}%
\mathcal{H}\oplus \mathcal{H}$ of all the functions 
\begin{equation*}
\Psi \left( t,s\right) =\psi \left( t-s\right) \oplus \overline{\varphi
\left( t-s\right) }=\Psi \left( 0,s-t\right) ,\quad \psi ,\varphi \in 
\mathcal{H}.
\end{equation*}
Note that such extension is simply a new reducible representation (time
representation) of the quantum system in which the Hamiltonian $\mathrm{H}$
is the momentum operator $\mathrm{E}=\frac{\hbar }{i}\frac{\partial }{%
\partial s}$ along the time coordinate $s$ if the initial states $\Psi
\left( 0\right) $ are restricted to the embedded subspace $\mathcal{H}$ by
the initial data constraint $\Psi \left( 0,s\right) =\psi \left( -s\right) $
($\varphi =0$). This subspace is not invariant under the (measurement of) $%
\mathrm{T}$, and after this measurement it should be projected back onto $%
\mathcal{H}$ .

Einstein launched an attack on the uncertainty relation at the Solvay
Congress in 1927, and then again in 1930, by proposing cleverly devised
thought experiments which would violate this relation. Most of these
imaginary experiments were designed to show that interaction between the
microphysical object and the measuring instrument is not so inscrutable as
Heisenberg and Bohr maintained. He suggested, for example, a box filled with
radiation with a clock described by the pointer coordinate $x$. The clock is
designed to open a shutter and allow one photon to escape at the time $%
\mathrm{T}$. He argued that the time of escape and the photon energy $%
\mathrm{E}$ can both be measured with arbitrary accuracy by measuring the
pointer coordinate and by weighing the box before and after the escape as
the difference of the weights $y$.

After proposing this argument Einstein is reported to have spent a happy
evening, and Niels Bohr an unhappy one. However after a sleepless night Bohr
showed next morning that Einstein was wrong. Mathematically Bohr's
explanation of the Einstein experiment can be expressed as the usual
measurement of two compatible variables $x$ and $y$ of the of the total
system under the question by the following simple`signal plus noise' formula 
\begin{equation}
X=s+Q,\quad Y=\frac{\hbar }{i}\frac{\partial }{\partial s}-P,  \label{2.3}
\end{equation}
where $Q$ and $P$ are the position and momentum operators of the
compensation weight under the box. Here the measuring quantity $X$, the
pointer coordinate of the clock, realizes an unsharp measurement of the
self-adjoint time operator $\mathrm{T}=s$ representing the time in the
extended Hilbert space $\mathbb{H}$, and the observable $Y$ realizes the
indirect measurement of photon energy $\mathrm{E}=i\hbar \frac{\partial }{%
\partial s}$ in $\mathbb{H}$. Due to the initial independence of the weight,
the commuting observables $X$ and $Y$ in the Einstein experiment will have
even greater uncertainty 
\begin{equation}
\Delta X\Delta Y=\Delta \mathrm{T}\Delta \mathrm{E}+\Delta Q\Delta P\geq
\hbar  \label{2.4}
\end{equation}
than that predicted by Heisenberg uncertainty $\Delta \mathrm{T}\Delta 
\mathrm{E}\geq \hbar /2$ as it is the sum with $\Delta Q\Delta P\geq \hbar
/2 $. This uncertainty remains obviously valid if the states $\psi \in 
\mathbb{H}$ of the `extended photon' are restricted to only physical photon
states $\psi \in \mathcal{H}$ corresponding to the positive spectrum of $%
\mathrm{E}$.

\subsubsection{Nonexistence of hidden variables}

Einstein hoped that eventually it would be possible to explain the
uncertainty relations by expressing quantum mechanical observables as
functions of some hidden variables $\lambda $ in deterministic physical
states such that the statistical aspect will arise as in classical
statistical mechanics by averaging these observables over $\lambda $.

Von Neumann's monumental book \cite{Neum32} on the mathematical foundations
of quantum theory was therefore a timely contribution, clarifying, as it
did, this point. \ Inspired by Lev Landau, he introduced, for the unique
characterization of the statistics of a quantum ensemble, the statistical
density operator $\rho $ which eventually, under the name normal, or regular
state, became a major tool in quantum statistics. He considered the linear
space $\frak{L}$ of all bounded Hermitian operators $\mathrm{L}=\mathrm{L}%
^{\dagger }$ as potential observables in a quantum system described by a
Hilbert space $\mathbb{H}$ of all normalizable wave functions $\psi $.
Although von Neumann considered any complete inner product complex linear
space as the Hilbert space, it is sufficient to reproduce his analysis for a
finite-dimensional $\mathbb{H}$. He defined the expectation $\left\langle 
\mathrm{L}\right\rangle $ of each $\mathrm{L}\in \frak{L}$ in a state $\rho $
by the linear functional $\mathrm{L}\mapsto $ $\left\langle \mathrm{L}%
\right\rangle $ of the regular form $\left\langle \mathrm{L}\right\rangle =%
\mathrm{TrL}\rho $, where $\mathrm{Tr}$ denotes the linear operation of
trace applied to the product of all operators on the right. He noted that in
order to have positive probabilities for the potential quantum mechanical
events $E$ as the expectations $\left\langle E\right\rangle $ of yes-no
observables described by the Hermitian projectors $\mathrm{E}\in \frak{L}$
(i.e. with $\left\{ 0,1\right\} $ spectrum), and probability one for the
identity event $I=1$ described by the identity operator $\mathrm{I}$ , 
\begin{equation}
\Pr \left\{ E=1\right\} =\mathrm{Tr}E\rho \geq 0,\quad \Pr \left\{
I=1\right\} =\mathrm{Tr}\rho =1,  \label{2.5}
\end{equation}
the statistical operator $\rho $ must be positive-definite and have trace
one. Then he proved that any linear (and even any additive) physically
continuous\emph{\ }functional $\mathrm{L}\mapsto \left\langle \mathrm{L}%
\right\rangle $ is regular, i.e. has such trace form. He applied this
technique to the analysis of the completeness problem of quantum theory,
i.e. whether it constitutes a logically closed theory or whether it could be
reformulated as an entirely deterministic theory through the introduction of
hidden parameters (additional variables which, unlike ordinary observables,
are inaccessible to measurements). He came to the conclusion that

\begin{quote}
\textit{the present system of quantum mechanics would have to be objectively
false, in order that another description of the elementary process than the
statistical one may be possible}

(quoted on page 325 in \cite{Neum32})
\end{quote}

To prove this theorem, von Neumann showed that there is no such state which
would be dispersion-free simultaneously for all possible quantum events $%
E\in \frak{L}$ described by all Hermitian projectors $E^{2}=E$. For each
such state, he argued, 
\begin{equation}
\left\langle E^{2}\right\rangle =\left\langle E\right\rangle =\left\langle
E\right\rangle ^{2}  \label{2.6}
\end{equation}
for all such $E$ would imply that $\rho =\mathrm{O}$ ($\mathrm{O}$ denotes
the zero operator) which cannot be statistical operator as $\mathrm{TrO}%
=0\neq 1$. Thus no state can be considered as a mixture of dispersion-free
states, each of them associated with a definite value of hidden parameters.
There are simply no such states, and thus, no hidden parameters. In
particular this implies that the statistical nature of pure states, which
are described by one-dimensional projectors $\rho =P_{\psi }$ corresponding
to wave functions $\psi $, cannot be removed by supposing them to be a
mixture of dispersion-free substates.

It is widely believed that in 1966 John Bell showed that von Neuman's proof
was in error, or at least his analysis left the real question untouched \cite
{Bell66}. To discredit the von Neumann's proof he constructed an example of
dispersion-free states parametrized for each quantum state $\rho $ by a real
parameter $\lambda $ for a simplest quantum system corresponding to the two
dimensional $\mathbb{H=}\frak{h}$ (we shall use the little $\frak{h}\simeq 
\mathbb{C}^{2}$ for this simplest state space and Pauli matrix calculus in
notation of the Appendix 1). He succeeded to do this by weakening the
assumption of the additivity for such states, requiring it only for the
commuting observables in $\frak{L}$, and by abandoning the linearity of the
constructed expectations in $\rho $. There is no reason, he argued, to keep
the linearity in $\rho $ for the observable eigenvalues determined by $%
\lambda $ and $\rho $, and to demand the additivity for non-commuting
observables as they are not simultaneously measurable: The measured
eigenvalues of a sum of noncommuting observables are not the sums of the
eigenvalues of this observables measured separately. For each spin-operator $%
\mathrm{L}=\sigma \left( \mathbf{l}\right) $ given by a 3-vector $\mathbf{l}%
\in \mathbb{R}^{3}$ as in the Appendix 1 Bell found a discontinuous family $%
s_{\lambda }\left( \mathbf{l}\right) $ of dispersion-free values $\pm l$, $%
l=\left| \mathbf{l}\right| $, parameterized by $\left| \lambda \right| \leq
1/2$, which reproduce the expectation $\left\langle \sigma \left( \mathbf{l}%
\right) \right\rangle =\mathbf{l\cdot r}$ in the pure quantum state
described by a unit polarization vector $\mathbf{r}$ in $\mathbb{R}^{3}$
when uniformly averaged over the $\lambda $.

Although the Bell's analysis of the von Neumann theorem is mathematically
incomplete as it ignores physical continuity which was assumed by von
Neumann in his definition of physical states, this is not the main reason
for failure of the Bell's. The reason for failure of the Bell's and others
hidden variable arguments is given in the Appendix 1 where is shown that all
dispersion free states even if they existed, not just the one constructed by
Bell for the exceptional case $\dim \frak{h}=2$, cannot be extended to the
quantum composed systems. \emph{All hidden variable theories are
incompatible with quantum composition principle which multiples the
dimensionally of the Hilbert space by the dimensionality of the state-vector
space of the additional quantum system.} In higher dimensions of $\mathbb{H}$
all such irregular states are ruled out by Gleason's theorem \cite{Glea57}
who proved that there is no even one additive zero-one value probability
function if $\dim \mathbb{H}>2$. In order that a hidden variable description
of the elementary quantum process may be possible, the present postulates of
quantum mechanics such as the composition principle would have to be
objectively false.

\subsubsection{Complementarity and common sense}

In view of the decisive importance of this analysis for the foundations of
quantum theory, Birkhoff and von Neumann \cite{BiNe36} setup a system of
formal axioms for the calculus of logico-theoretical propositions concerning
results of possible measurements in a quantum system described by a Hilbert
space $\mathbb{H}$. They started by formalizing the calculus of \emph{%
quantum propositions} corresponding to the potential idealized events $E$
described by orthoprojectors in $\mathbb{H}$, the projective operators $%
E=E^{2}$ which are orthogonal to their complements $E^{\bot }=\mathrm{I}-E$
in the sense $E^{\dagger }E^{\bot }=\mathrm{O}$, where $\mathrm{O}$ denotes
the multiplication by $0.$ The set $\mathcal{P}\left( \mathbb{H}\right) $ of
all othoprojectors, equivalently defined by 
\begin{equation*}
E^{\dagger }=E^{\dagger }E=E,
\end{equation*}
is the set of all Hermitian projectors $E\in \frak{L}$ as the only
observables with two eigenvalues $\left\{ 1,0\right\} $ (``yes'' and
``no''). Such calculus coincides with the calculus of linear subspaces $%
\frak{e}\subseteq \mathbb{H}$ including $0$-dimensional subspace $\mathbb{O}$%
, in the same sense as the common sense propositional calculus of classical
events coincides with the calculus in a Boolean algebra of subsets $%
E\subseteq \Omega $ including empty subset $\varnothing $. The subspaces $%
\frak{e}$ are defined by the condition $\frak{e}^{\bot \bot }=\frak{e}$,
where $\frak{e}^{\bot }$ denotes the orthogonal complement $\left\{ \phi \in 
\mathbb{H}:\left\langle \phi |\psi \right\rangle =0,\psi \in \frak{e}%
\right\} $ of $\frak{e}$, and they uniquely define the propositions $E$ as
the orthoprojectors $P\left( \frak{e}\right) $ onto the ranges 
\begin{equation}
\frak{e}=\mathrm{range}E:=E\mathbb{H}  \label{2.8}
\end{equation}
of $E\in \mathcal{P}\left( \mathbb{H}\right) $.\ In this calculus the
logical ordering $E\leq F$ implemented by the algebraic relation $EF=E$
coincides with 
\begin{equation*}
\mathrm{range}E\subseteq \mathrm{range}F,
\end{equation*}
the conjunction $E\wedge F$ corresponds to the intersection, 
\begin{equation*}
\mathrm{range}\left( E\wedge F\right) =\mathrm{range}E\cap \mathrm{range}F,
\end{equation*}
however the disjunction $E\vee F$ is represented by the linear sum $\frak{e}+%
\frak{f}$ of the corresponding subspaces but not their union 
\begin{equation*}
\mathrm{range}E\cup \mathrm{range}F\subseteq \mathrm{range}\left( E\vee
F\right) ,
\end{equation*}
and the smallest orthoprojector $\mathrm{O}$ corresponds to zero-dimensional
subspace $\mathbb{O}=\left\{ 0\right\} $ but not the empty subset $%
\varnothing $ (which is not linear subspace). Note that although $\mathrm{%
range}\left( E+F\right) =\frak{e}+\frak{f}$ $\ $for any $E,F\in \mathcal{P}%
\left( \mathbb{H}\right) $, the operator $E+F$ is not the orthoprojector $%
E\vee F$ corresponding to $\frak{e}+\frak{f}$ unless $EF=\mathrm{O}$. This
implies that the distributive law characteristic for propositional calculus
of classical logics no longer holds. However it still holds for the
compatible propositions described by commutative orthoprojectors due to the
orthomodularity property 
\begin{equation}
E\leq I-F\leq G\Longrightarrow \left( E\vee F\right) \wedge G=E\vee \left(
F\wedge G\right) .  \label{2.9}
\end{equation}
\ 

Actually as we shall see, a propositions $E$ can become an \emph{event} if
and only if it may serve as a condition for any other proposition $F$. In
terms of the natural order of the propositions this can be written as 
\begin{equation}
E^{\perp }\wedge F=\mathrm{I}-E\wedge F\text{ \quad }\forall F,  \label{2.7}
\end{equation}
where $E^{\perp }=\mathrm{I}-E$, and it is equivalent to the compatibility $%
EF=FE$ of the event-orthoprojector $E$ with any other orthoprojector $F$ of
the system.

For each regular state corresponding to a density operator $\rho $, one can
obtain the probability function $\left\langle E\right\rangle =\mathrm{Tr}%
E\rho $ on $\mathcal{P}\left( \mathbb{H}\right) $ called the \emph{quantum
probability measure}. It can also be defined as a function 
\begin{equation}
\mathsf{P}\left( \frak{e}\right) =\Pr \left\{ P\left( \frak{e}\right)
=1\right\} =\left\langle P\left( \frak{e}\right) \right\rangle  \label{2.10}
\end{equation}
on the set $\mathcal{E}$ of all subspaces $\frak{e}$ of $\mathbb{H}.$ It is
obviously positive, $\mathsf{P}\left( \frak{e}\right) >0$, with $\mathsf{P}%
\left( \mathbb{O}\right) =0$, normalized, $\mathsf{P}\left( \mathbb{H}%
\right) =1$, and additive but only for orthogonal $\frak{e}$ and $\frak{f}$: 
\begin{equation*}
\frak{e}\perp \frak{f}\Rightarrow \mathsf{P}\left( \frak{e}+\frak{f}\right) =%
\mathsf{P}\left( \frak{e}\right) +\mathsf{P}\left( \frak{f}\right) .
\end{equation*}
These properties are usually taken as definition of a probabilistic state on
the quantum logic $\mathcal{E}$. Note that not any ortho-additive function $%
\mathsf{P}$ is \emph{a priori} regular, i.e. induced by a density operator $%
\rho $ on $\mathcal{E}$ as $\mathsf{P}\left( \frak{e}\right) =\mathrm{Tr}%
P\left( \frak{e}\right) \rho $. However Gleason proved \cite{Glea57} that
every ortho-additive normalized function $\mathcal{E}\rightarrow \left[ 0,1%
\right] $ \emph{is} regular in this sense if $2<\dim \mathbb{H}\mathbf{%
<\infty }$ (he proved this also for the case $\dim \mathbb{H}=\infty $ under
the natural assumption of countable ortho-additivity, and it is also true
for $\dim \mathbb{H}=2$ under the quantum composition assumption, see the
Appendix 1). Any statistical mixture of such (regular) probability functions
is obviously a (regular) probability function, and the extreme functions of
this convex set correspond to the pure (regular) states $\rho =\psi \psi
^{\dagger }$.

Two propositions $E,F$ are called \emph{complementary} if $E\vee F=\mathrm{I}
$, orthocomplementary if $E+F=\mathrm{I}$, \emph{incompatible} or \emph{%
disjunctive} if $E\wedge F=\mathrm{O}$, and \emph{contradictory} or \emph{%
orthogonal} if $EF=\mathrm{O}$. As in the classical, common sense case,
logic contradictory propositions are incompatible. However \emph{%
incompatible propositions are not necessary contradictory }as can be easily
seen for any two nonorthogonal but not coinciding one-dimensional subspaces.
In particular, in quantum logics there exist complementary incompatible
pairs $E,F$, $E\vee F=\mathrm{I}$, $E\wedge F=\mathrm{O}$ which are not
ortho-complementary in the sense $E+F\neq \mathrm{I}$, i.e. $EF\neq \mathrm{O%
}$ (this would be impossible in the classical case). This is a rigorous
logico-mathematical proof of Bohr's complementarity.

As an example, we can consider the proposition that a quantum system is in a
stable energy state $E$, and an incompatible proposition $F$, that it
collapses at a given time $t$, say. The incompatibility $E\wedge F=\mathrm{O}
$ follows from the fact that there is no state in which the system would
collapse preserving its energy, however these two propositions are not
contradictory (i.e. not orthogonal, $EF\neq \mathrm{O}$): the system might
not collapse if it is in other than $E$ stationary state (remember the
Schr\"{o}dinger's earlier belief that the energy law is valid only on
average, and is violated in the process of radiation).

In 1952 Wick, Wightman, and Wigner \cite{WWW52} showed that there are
physical systems for which not every orthoprojector corresponds to an
observable event, so that not every one-dimensional orthoprojector $\rho
=P_\psi $ corresponding to a wave function $\psi $ is a pure state. This is
equivalent to the admission of some selective events which are
dispersion-free in all pure states. Jauch and Piron \cite{JaPi63}
incorporated this situation into quantum logics and proved in the context of
this most general approach that the hidden variable interpretation is only
possible if the theory is observably wrong, i.e. if incompatible events are
always contradictory.

Bell criticized this as well as the Gleason's theorem, but this time his
arguments were not based even on the classical ground of usual probability
theory. Although he explicitly used the additivity of the probability on the
orthogonal events in his counterexample for $\mathbb{H}=\mathbb{C}^{2}$, he
questioned : `That so much follows from such apparently innocent assumptions
leads us to question their innocence'. (p.8 in \cite{Bell87}). In fact this
was equivalent to questioning the additivity of classical probability on the
corresponding disjoint subsets, but he didn't suggest any other complete
system of physically reasonable axioms for introducing such peculiar
``nonclassical'' hidden variables, not even a single counterexample to the
orthogonal nonadditivity for the simplest case of quantum bit $\mathbb{H}=%
\frak{h}$. Thus Bell implicitly rejected classical probability theory in the
quantum world, but he didn't want to accept quantum probability as the only
possible theory for explaining the microworld.

\subsection{Entanglement and Quantum Causality}

\medskip\medskip

\begin{quote}
\textit{The nonvalidity of rigorous causality is necessary and not just
consistently possible. }- Heisenberg.
\end{quote}

Thus deterministic causality was questioned by Heisenberg when he analyzed
his uncertainty relations. The general consensus among quantum physicists
was that there is no positive answer to this question. Max Born even stated:

\begin{quote}
\textit{One does not get an answer to the question, what is the state after
collision? but only to the question, how probable is a given effect of the
collision?}
\end{quote}

Einstein was deeply concerned with loss of reality and causality in the
treatment of quantum measuring process by Heisenberg and Born. In \cite{EPR}
he suggested a gedanken experiment, now known as EPR paradox.
Schr\"{o}dinger's remained unhappy with Bohr's reply to the EPR paradox,
Schr\"{o}dinger's own analysis was:

\begin{quote}
\textit{It is pretty clear, if reality does not determine the measured
value, at least the measurable value determines reality.}
\end{quote}

In this section we develop this idea of Schr\"{o}dinger applying it to his
explanatory model for the EPR paradox, now is well-know as the Cat of
Schr\"{o}dinger. We shall see how the entanglement, decoherence and the
collapse problem can be derived for his cat from purely dynamical arguments
of Schr\"{o}dinger, extending his model to a semi-infinite string of
independent cats interacting with a single atom at the boundary of the
string by a unitary scattering. We shall see that in such extended system
the measurable value (cat is dead or alive) indeed determines reality by
simple inference (Bayes conditioning) in the same way as it does in usual
classical statistical theory. We shall see that this is possible only due to
the Schr\"{o}dinger's superselection rule which determines his measurable
state as reality, i.e. as a classical bit system state with the only two
values. This is the only way to keep the causality in its weakest,
statistical form Later we shall see that quantum causality in the form of a
superselection rule is in fact the new postulate of quantum theory which
does not contradict to the present formalism and resolves the paradoxes of
its statistical interpretation.

\subsubsection{Spooky action at distance}

After his defeat on uncertainty relations Einstein seemed to have become
resigned to the statistical interpretation of quantum theory, and at the
1933 Solvay Congress he listened to Bohr's paper on completeness of quantum
theory without objections. In 1935, he launched a brilliant and subtle new
attack in a paper \cite{EPR} with two young co-authors, Podolski and Rosen,
which has become of major importance to the world view of physics. They
stated the following requirement for a complete theory as a seemingly
necessary one:

\begin{quote}
\textit{Every element of physical reality must have a counterpart in the
physical theory.}
\end{quote}

The question of completeness is thus easily answered as soon as soon as we
are able to decide what are the elements of the physical reality. \ EPR then
proposed a sufficient condition for an element of physical reality:

\begin{quote}
\textit{If, without in any way disturbing the system, we can predict with
certainty the value of a physical quantity, then there exists an element of
physical reality corresponding to this quantity.}
\end{quote}

Then they designed a thought experiment the essence of which is that two
quantum ``bits'', particle spins of two electrons say, are brought together
to interact, and after separation an experiment is made to measure the spin
orientation of one of them. The state after interaction is such that the
measurement value $\upsilon =\pm \frac{1}{2}$ of one particle uniquely
determines the spin $z$-orientation $\sigma =\mp \frac{1}{2}$ of the other
particle indipendently of its initial state. EPR apply their criterion of
local reality: since the value of $\sigma $ can be predicted by measuring $%
\upsilon $ without in any way disturbing $\sigma $, it must correspond to an
existing element of physical reality determining the state. Yet the
conclusion contradicts a fundamental postulate of quantum mechanics,
according to which the sign of spin is not an intrinsic property of a
complete description of the spin by state but is evoked only by a process of
measurement. Therefore, EPR conclude, quantum mechanics must be incomplete,
there must be hidden variables not yet discovered, which determine the spin
as an intrinsic property. It seems Einstein was unaware of the von Neumann's
hidden variable theorem, although they both had positions at the Institute
for Advanced Studies at Princeton (being among the original six mathematics
professors appointed there in 1933).

Bohr carefully replied to this challenge by rejecting the assumption of
local physical realism as stated by EPR \cite{Bohr35}: `There is no question
of a mechanical disturbance of the system under investigation during the
last critical stage of the measuring procedure. But even at this stage there
is essentially a question of \textit{an influence on the very conditions
which define the possible types of predictions regarding the future behavior
of the system}'. This influence became notoriously famous as Bohr's \textit{%
spooky action at a distance}. He had obviously meant the semi-classical
model of measurement, when one can statistically infer the state of one
(quantum) part of a system immediately after observing the other (classical)
part, whatever the distance between them. In fact, there is no paradox of
``spooky action at distance'' in the classical case. The statistical
inference, playing the role of such immediate action, is simply based on the
Bayesian selection rule of a posterior state from the prior mixture of all
such states, corresponding to the possible results of the measurement. Bohr
always emphasized that one must treat the measuring instrument classically
(the measured spin, or another bit interacting with this spin, as a
classical bit), although the classical-quantum interaction should be
regarded as quantum. The latter follows from non-existence of semi-classical
Poisson bracket (i.e. classical-quantum potential interaction) for finite
systems. Schr\"{o}dinger clarified this point more precisely then Bohr, and
he followed in fact the mathematical pattern of von Neumann measurement
theory.

\subsubsection{Releasing Schr\"{o}dinger's cat}

Motivated by EPR paper, in 1935 Schr\"{o}dinger published a three part essay 
\cite{Schr35} on `The Present Situation in Quantum Mechanics'. He turns to
EPR paradox and analyses completeness of the description by the wave
function for the entangled parts of the system. (The word \emph{entangled}
was introduced by Schr\"{o}dinger for the description of nonseparable
states.) He notes that if one has pure states $\psi \left( \sigma \right) $
and $\chi \left( \upsilon \right) $ for each of two completely separated
bodies, one has maximal knowledge, $\psi _{1}\left( \sigma ,\upsilon \right)
=\psi \left( \sigma \right) \chi \left( \upsilon \right) $, for two taken
together. But the converse is not true for the entangled bodies, described
by a non-separable wave function $\psi _{1}\left( \sigma ,\upsilon \right)
\neq \psi \left( \sigma \right) \chi \left( \upsilon \right) $:

\begin{quote}
\textit{Maximal knowledge of a total system does not necessary imply maximal
knowledge of all its parts, not even when these are completely separated one
from another, and at the time can not influence one another at all.}
\end{quote}

To make absurdity of the EPR argument even more evident he constructed his
famous burlesque example in quite a sardonic style. A cat is shut up in a
steel chamber equipped with a camera, with an atomic mechanism in a pure
state $\rho _{0}=P_{\psi }$ which triggers the release of a phial of cyanide
if an atom disintegrates spontaneously, and this proposition is represented
by a one-dimensional projector $F$. It is assumed that it might not
disintegrate in a course of an hour $t=1$ with probability $\mathrm{Tr}%
\left( EP_{\psi }\right) =1/2$, where $E=\mathrm{I}-F$. If the cyanide is
released, the cat dies, if not, the cat lives. Because the entire system is
regarded as quantum and closed, after one hour, without looking into the
camera, one can say that the entire system is still in a pure state in which
the living and the dead cat are smeared out in equal parts.

Schr\"{o}dinger resolves this paradox by noting that the cat is a
macroscopic object, the states of which (alive or dead) could be
distinguished by a macroscopic observation as distinct from each other
whether observed or not. He calls this `the principle of state distinction'
for macroscopic objects, which is in fact the postulate that the directly
measurable system (consisting of the cat) must be classical:

\begin{quote}
\textit{It is typical in such a case that an uncertainty initially
restricted to an atomic domain has become transformed into a macroscopic
uncertainty which can be resolved through direct observation.}
\end{quote}

The dynamical problem of the transformation of the atomic, or ``coherent''
uncertainty, corresponding to a probability amplitude $\psi \left( \sigma
\right) $, into a macroscopic uncertainty, corresponding to a mixed state $%
\rho $, is called quantum \emph{decoherence }problem. Thus he suggested that
the solution of EPR paradox is in non-equivalence of two spins in this
thought experiment, one being observed and thus must be open macroscopic
subsystem, and the other, nonobserved, can stay microscopic and closed. This
was the true reason why he replaced the observed spin by the classical two
state cat, and the other by an unstable atom as a quantum model of a two
level closed system. The only problem was to construct the corresponding
classical-quantum transformation in a consistent way.

In order to make this idea clear, let us formulate the dynamical
Schr\"{o}dinger's cat problem in the purely mathematical way. For the
notational simplicity instead of the values $\pm 1/2$ for the spin-variables 
$\sigma $ and $\upsilon $ we shall use the indexing values $\tau ,\upsilon
\in $ $\left\{ 0,1\right\} $ describing the states of a ``bit'', the
''atomic'' system of the classical information theory.

Consider the atomic mechanism as a quantum ``bit'' with Hilbert space $\frak{%
h}=\mathbb{C}^{2}$, the pure states of which are described by $\psi $%
-functions of the variable $\tau \in \left\{ 0,1\right\} $, i.e. by
2-columns with scalar (complex) entries $\psi \left( \tau \right) =\langle
\tau |\psi $ defining the probabilities $\left| \psi \left( \tau \right)
\right| ^{2}$ of the quantum elementary propositions corresponding to $\tau
=0,1$. If atom is disintegrated, $\psi =|1\rangle $ corresponding to $\tau
=1 $, if not, $\psi =|0\rangle $ corresponding to $\tau =0$. The
Schr\"{o}dinger's cat is a classical bit with only two pure states $\upsilon
\in \left\{ 0,1\right\} $ which can be identified with the Kr\"{o}nicker
delta probability distributions $\delta _{0}\left( \upsilon \right) $ when
alive $\left( \upsilon =0\right) $ and $\delta _{1}\left( \upsilon \right) $
when dead $\left( \upsilon =1\right) $. These pure and even mixed states of
the cat can also be described by the complex amplitudes $\chi \left(
\upsilon \right) =\langle \upsilon |\chi $ as it were initially quantum bit.
However the 2-columns $\chi $ are uniquely defined by the probabilities $%
\left| \chi \left( \upsilon \right) \right| ^{2}$ not just up to a phase
constant as in the case of the atom (only constants commute with all atomic
observables $\mathrm{B}\in \frak{L}$ on the Hilbert space $\frak{h}$), but
up to a phase function of $\upsilon $ (the phase multiplier of $\chi \in 
\frak{g}$, $\frak{g}=\mathbb{C}^{2}$ commuting with all cat observables $%
c\left( \upsilon \right) $) if the cat is considered as being classical.
Initially the cat is alive, so its amplitude defined as $\chi _{0}=|0\rangle 
$ up to a phase function by the probability distribution $\delta _{0}\equiv
\delta $ on $\left\{ 0,1\right\} $, is equal $1$ if $\upsilon =0$, and $0$
if $\upsilon =1 $ as $\langle \upsilon |0\rangle =\delta _{0}\left( \upsilon
\right) $.

The dynamical interaction in this semiclassical system can be described by
the unitary transformation 
\begin{equation}
\mathrm{S}=E\otimes \hat{1}+F\otimes \sigma _{1}=\sigma _{1}^{\mathrm{X}%
\otimes \hat{1}}  \label{3.1}
\end{equation}
in $\frak{h}\otimes \frak{g}$ as it was purely quantum composed system. Here 
$\sigma _{1}$ is the unitary flip-operator $\sigma _{1}\chi \left( \upsilon
\right) =\chi \left( \upsilon \bigtriangleup 1\right) $ in $\frak{g}$, where 
$\upsilon \bigtriangleup \tau =\left| \upsilon -\tau \right| =\tau
\bigtriangleup \upsilon $ is the difference ($\func{mod}2$) on $\left\{
0,1\right\} $, and $\mathrm{X}=0E+1F$ is the orthoprojector $F$ in $\frak{h}$%
. This is the only meaningful interaction affecting the cat but not the atom
after the hour in a way suggested by Schr\"{o}dinger, 
\begin{equation*}
\mathrm{S}\left[ \psi \otimes \chi \right] \left( \tau ,\upsilon \right)
:=\langle \tau ,\upsilon |\mathrm{S}\left( \psi \otimes \chi \right) =\psi
\left( \tau \right) \chi \left( \left( \upsilon \bigtriangleup \tau \right)
\right) ,
\end{equation*}
where $\langle \tau ,\upsilon |=\langle \tau |\otimes \langle \upsilon |$.
Applied to the initial product-state $\psi _{0}=\psi \otimes \delta $
corresponding to $\chi _{0}=\delta $ it has the resulting probability
amplitude 
\begin{equation}
\psi _{1}\left( \tau ,\upsilon \right) =\psi \left( \tau \right) \delta
\left( \upsilon \bigtriangleup \tau \right) =0\quad \mathrm{if}\quad \tau
\neq \upsilon .  \label{3.2}
\end{equation}
Because the initial state $\delta $ is pure for the cat considered either as
classical bit or quantum, the initial composed state $\psi _{0}=\psi \otimes
\delta $ is also pure even if this system is considered as semiquantum,
corresponding to the Cartesian product $\left( \psi ,0\right) $ of the
initial pure classical $\upsilon =0$ and quantum states $\psi \in \frak{h}$.
Despite this fact one can easily see that the unitary operator $\mathrm{S}$
induces in $\mathbb{H}=\frak{h}\otimes \frak{g}$ the mixed state for the
quantum-classical system, although it is still described by the vector $\psi
_{1}=\mathrm{S}\psi _{0}\in \mathbb{H}$ as the wave function $\psi
_{1}\left( \tau ,\upsilon \right) $ of the ``atom+cat'' corresponding to $%
\psi _{0}=\psi \otimes \delta $.

Indeed, the potential observables of such a system at the time of
observation $t=1$ are all operator-functions $X$ of $\upsilon $ with values $%
X\left( \upsilon \right) $ in Hermitian 2$\times $2-matrices, represented as
block-diagonal $\left( \tau ,\upsilon \right) $-matrices $\hat{X}=\left[
X\left( \upsilon \right) \delta _{\upsilon ^{\prime }}^{\upsilon }\right] $
of the multiplication $X\left( \upsilon \right) \psi _{1}\left( \cdot
,\upsilon \right) $ at each point $\upsilon \in \left\{ 0,1\right\} $. This
means that the amplitude $\psi _{1}$ (and its density matrix $\omega
=P_{\psi _{1}}$) induces the same expectations 
\begin{equation}
\left\langle \hat{X}\right\rangle =\sum_{\upsilon }\psi _{1}\left( \upsilon
\right) ^{\dagger }X\left( \upsilon \right) \psi _{1}\left( \upsilon \right)
=\sum_{\upsilon }\mathrm{Tr}X\left( \upsilon \right) \varrho \left( \upsilon
\right) =\mathrm{Tr}\hat{X}\hat{\varrho}  \label{3.3}
\end{equation}
as the block-diagonal density matrix $\hat{\varrho}=\left[ \varrho \left(
\upsilon \right) \delta _{\upsilon ^{\prime }}^{\upsilon }\right] $ of the
multiplication by 
\begin{equation*}
\varrho \left( \upsilon \right) =F\left( \upsilon \right) P_{\psi }F\left(
\upsilon \right) =\pi \left( \upsilon \right) P_{F\left( \upsilon \right)
\psi }
\end{equation*}
where $\pi \left( \upsilon \right) =\left| \psi \left( \upsilon \right)
\right| ^{2}$, $F\left( \upsilon \right) =P_{\delta _{\upsilon }}$ is the
projection operator $E$ if $\upsilon =0$ and $F$ if $\upsilon =1$: 
\begin{equation}
\left[ F\left( \upsilon \right) \psi \right] \left( \tau \right) =\delta
\left( \upsilon \bigtriangleup \tau \right) \psi \left( \tau \right) =\psi
\left( \upsilon \right) \delta _{\upsilon }\left( \tau \right) ,  \label{3.4}
\end{equation}
and $P_{F\left( \upsilon \right) \psi }=P_{\delta _{\upsilon }}$ is also
projector onto $\delta _{\upsilon }\left( \cdot \right) =\delta \left( \cdot
\bigtriangleup \upsilon \right) $. The $4\times 4$-matrix $\hat{\varrho}$ is
a mixture of two orthogonal projectors $P_{\delta _{\upsilon }}\otimes
P_{\delta _{\upsilon }}$, $\upsilon =0,1$: 
\begin{equation*}
\hat{\varrho}=\left[ P_{\delta _{\upsilon }}\delta _{\upsilon ^{\prime
}}^{\upsilon }\pi \left( \upsilon \right) \right] =\sum_{\upsilon =0}^{1}\pi
\left( \upsilon \right) P_{\delta _{\upsilon }}\otimes P_{\delta _{\upsilon
}}.
\end{equation*}

The only remaining problem is to explain how the cat, initially interacting
with atom as a quantum bit described by the algebra $\mathcal{A}=\mathcal{B}%
\left( \frak{g}\right) $ of all operators on $\frak{g}$, after the
measurement becomes classical, described by the commutative subalgebra $%
\mathcal{C}=\mathcal{D}\left( \frak{g}\right) $ of all diagonal operators on 
$\frak{g}$. As will be shown in the next section even, this can be done in
purely dynamical terms if the system ''atom plus cat'' is extended to an
infinite system by adding a quantum string of ''incoming cats'' and a
classical string of ''outgoing cats'' with a potential interaction (\ref{3.1}%
) with the atom at the boundary. The free dynamics in the strings is modeled
by the simple shift which replaces the algebra $\mathcal{A}$ of the quantum
cat at the boundary by the algebra $\mathcal{C}$ of the classical one, and
the total discrete-time dynamics of this extended system is induced on the
infinite semi-classical algebra of the ''atom plus strings'' observables by
a unitary dynamics on the extended Hilbert space $\mathcal{H}=\mathbb{%
H\otimes }\mathcal{H}_{0}$. Here $\mathbb{H}=\frak{h}\otimes \frak{g}$, and $%
\mathcal{H}_{0}$ is generated by the orthonormal infinite products $|\tau
_{0}^{\infty },\upsilon _{0}^{\infty }\rangle =\otimes |\tau _{i},\upsilon
_{i}\rangle $ for all the strings of quantum $\tau _{0}^{\infty }=\left(
\tau _{1},\tau _{2},\ldots \right) $ and classical $\upsilon _{0}^{\infty
}=\left( \upsilon _{1},\upsilon _{2},\ldots \right) $ bits with almost all
(but finite number) of $\tau _{n}$ and $\upsilon _{m}$ being zero. In this
space the total dynamics is described by the single-step unitary
transformation 
\begin{equation}
U:|\tau _{0},\upsilon \rangle \otimes |\tau \sqcup \tau _{0}^{\infty
},\upsilon _{0}^{\infty }\rangle \mapsto |\tau _{0},\tau _{0}+\tau \rangle
\otimes |\tau _{0}^{\infty },\upsilon \sqcup \upsilon _{0}^{\infty }\rangle ,
\label{3.5}
\end{equation}
incorporating the shift and the scattering $\mathrm{S}$, where $\tau
+\upsilon $ is the sum $\func{mod}2$ (which coincides with $\tau
\bigtriangleup \upsilon =\left| \tau -\upsilon \right| $), and 
\begin{equation*}
|\tau _{0}^{\infty },\upsilon \sqcup \upsilon _{0}^{\infty }\rangle =|\tau
_{1},\tau _{2},\ldots \rangle \otimes |\upsilon ,\upsilon _{1},\upsilon
_{2},\ldots \rangle ,\quad \tau ,\upsilon \in \left\{ 0,1\right\}
\end{equation*}
\ are the shifted orthogonal vectors which span the whole infinite Hilbert
product space $\mathcal{H}_{0}=\otimes _{r>0}\mathbb{H}_{r}$ of $\mathbb{H}%
_{r}=\frak{h}_{r}\otimes \frak{g}_{r}$ (the copies of the four-dimensional
Hilbert space $\mathbb{H}$). Thus the states $|\tau _{0}^{\infty },\upsilon
_{0}^{\infty }\rangle =|\tau _{0}^{\infty }\rangle \otimes |\upsilon
_{0}^{\infty }\rangle $ can be interpreted as the products of two discrete
waves interacting only at the boundary via the atom. The incoming wave $%
|\tau _{0}^{\infty }\rangle $ is the quantum probability amplitude wave
describing the state of ''input quantum cats''. The outgoing wave $|\upsilon
_{0}^{\infty }\rangle $ is the classical probability amplitude wave
describing the states of ''output classical cats''.

\subsubsection{The measurement problem}

Inspired by Bohr's complementarity principle, von Neumann proposed even
earlier the idea that every quantum measuring process involves an
unanalysable element. He postulated \cite{Neum32} that, in addition to the
continuous causal propagation $\psi \left( 0\right) \mapsto \psi \left(
t\right) $ of the wave function generated by the Schr\"{o}dinger equation,
the function $\psi $ undergoes a discontinuous, irreversible instantaneous
change due to an action of the observer on the object preparing the
measurement at the time $t$. Just prior to the reading of measurement result
of an event $F$, disintegration of the atom, say, the quantum pure state $%
\sigma =P_{\psi }$ changes to the mixed one 
\begin{equation}
\rho =\lambda P_{E\psi }+\mu P_{F\psi }=E\sigma E+F\sigma F,  \label{3.6}
\end{equation}
where $E=I-F$ is the orthocomplement event, and 
\begin{equation*}
\lambda =\left\| E\psi \right\| ^{2}=\mathrm{Tr}E\rho ,\quad \mu =\left\|
F\psi \right\| ^{2}=\mathrm{Tr}F\rho .
\end{equation*}
are the probabilities of $E$ and $F$. Such change is projective as shown in
the second part of this equation, and it is called the von Neumann
projection postulate.

This linear irreversible decoherence process should be completed by the
nonlinear, acausal random jump to one of the pure states 
\begin{equation}
\rho \mapsto P_{E\psi }\text{, or }\rho \mapsto P_{F\psi }  \label{3.7}
\end{equation}
depending on whether the tested event $F$ is false (the cat is alive, $\psi
_{0}=\lambda ^{-1/2}E\psi $), or true (the cat is dead, $\psi _{1}=\mu
^{-1/2}F\psi $). This final step is the posterior prediction, called \emph{%
filtering} of the decoherent mixture of $\psi _{0}$ and $\psi _{1}$ by
selection of only one result of the measurement, and is an unavoidable
element in every measurement process relating the state of the pointer of
the measurement (in this case the cat) to the state of the whole system.
This assures that the same result would be obtained in case of immediate
subsequent measurement of the same event $F$. The resulting change of the
prior wave-function $\psi $ is described up to normalization by one of the
projections 
\begin{equation*}
\psi \mapsto E\psi ,\quad \psi \mapsto F\psi
\end{equation*}
and is sometimes called the L\"{u}ders projection postulate \cite{Lud51}.

Although unobjectionable from the purely logical point of view the von
Neumann theory of measurement soon became the target of severe criticisms.
Firstly it seams radically subjective, postulating the spooky action at
distance in a purely quantum system instead of deriving it. Secondly the von
Neumann analysis is applicable to only the idealized situation of discrete
instantaneous measurements.

The first objection can be regarded as a result of misinterpretation of the
projection postulate $\sigma \mapsto \rho $ which can be avoided by its more
adequate formulation in the ''Heisenberg picture'' for any atomic observable 
$\mathrm{B}$ as the transformation $\mathrm{B}\mapsto \mathrm{A}$, 
\begin{equation}
\mathrm{A}=E\mathrm{B}E+F\mathrm{B}F,\quad \mathrm{B}\in \mathcal{B}\left( 
\frak{h}\right) ,  \label{3.8}
\end{equation}
where $\mathcal{B}\left( \frak{h}\right) $ is the algebra of operators on $%
\frak{h}$. Note that this transformation is irreversible, so the
''Heisenberg'' and ''Schr\"{o}dinger'' pictures for measurements are no
longer unitary equivalent as the matter of choice of the mathematically
equivalent representations as it is in the case of the conservative
(reversible) dynamical transformations. However these two pictures are
physically (statistically) equivalent as they give the same prediction of
the expectations just prior the measurement: 
\begin{equation*}
\mathrm{TrB}\sigma =\left\langle \psi |E\mathrm{A}E+F\mathrm{A}F|\psi
\right\rangle =\mathrm{TrA}\rho .
\end{equation*}

Indeed, thus reformulated, the projection postulate can be interpreted as
the reduction of the set of all potential observables to only those
compatible with the measurement in the result of the preparation of this
measurement. There is nothing subjective in this, the states $\sigma
=P_{\psi }$ describing the reality by wave functions $\psi $ prior the
measurement are not changed, and the reduction of potential observables is
merely a rule to satisfy the Bohrs complementarity for the given
measurement. However after the reduction of quantum potentialities to only
those which are compatible with the measurement, the pure state $\sigma $
becomes mixed even without the change of $\psi $, and this state can be
described not only by the prior density operator $P_{\psi }$ but by the
mixture $\rho $ in (\ref{3.6}) of the posterior $P_{E\psi }$ and $P_{F\psi }$
such that $\mathrm{TrA}P_{\psi }=\mathrm{TrA}\rho $ for all reduced
observables $\mathrm{A}$.

As we already mentioned when discussing the EPR paradox, the process of
filtering $\rho \mapsto P_{F\left( \upsilon \right) \psi }$ is free from any
conceptual difficulty if it is understood as the statistical inference about
a mixed state in an extended stochastic representation of the quantum system
as a part of a semiclassical one, based upon the results $\upsilon $ of
observation in its classical part. As we shall see in the next section this
is as simple as the transition from the prior to posterior classical
probabilities by the conditioning upon the results of statistical inference.
Note that in classical statistics due to complete commutativity such
conditioning is always possible, and this is why there is no measurements
problem in classical physics.

In the previous section we mentioned that the amount of commutativity which
is necessary to derive the projection postulate as the result of inference,
can be dynamically achieved by extending the system to the infinity. In the
next section we shall show how to do this in the general case, but here let
us demonstrate this for the dynamical model of ''cat'', identifying the
quantum system in question with the Schr\"{o}dinger's atom. The event $E$
(the atom exists) will correspond then to $\upsilon =0$ (the cat is alive), $%
E=F\left( 0\right) $, and the complementary event will be $F=F\left(
1\right) $.

Consider the semi-classical string of ''incoming and outgoing cats'' as the
quantum and classical bits moving freely in the opposite directions along
the discrete coordinate $r\in \mathbb{N}$. The Hamiltonian interaction of
the quantum (incoming) cats with the atom at the boundary $r=0$ is described
by the unitary scattering (\ref{3.1}). The whole system is described by the
unitary transformation (\ref{3.5}) which induces an injective endomorphism $%
\vartheta \left( A\right) =U^{\dagger }AU$ on the infinite product algebra $%
\frak{A}=\mathcal{A}\otimes \frak{A}_{0}$ of the atom-cat observables $\hat{X%
}\in \mathcal{A}$ at $r=0$ and other quantum-classical cats $\frak{A}%
_{0}=\otimes _{r>0}\mathcal{A}_{r}$. Here $\mathcal{A}=\mathcal{B}\left( 
\frak{h}\right) \otimes \mathcal{C}$ is the block-diagonal algebra of
operator-valued functions $\left\{ 0,1\right\} \ni \upsilon \mapsto X\left(
\upsilon \right) $ describing the observables of the string boundary on the
Hilbert space $\frak{h}\otimes \frak{g}$, where $\frak{h}=\mathbb{C}^{2}=%
\frak{g}$, and $\mathcal{A}_{r}=\mathcal{B}\left( \frak{h}_{r}\right)
\otimes \mathcal{C}_{r}$ are copies of $\mathcal{A}_{0}$ represented on
tensor products $\frak{h}_{r}\otimes \frak{g}_{r}$ of the copies $\frak{h}%
_{r}=\mathbb{C}^{2}=\frak{g}_{r}$ at $r>0$.

The input quantum probability waves $|\tau _{0}^{\infty }\rangle =\otimes
_{r>0}|\tau _{r}\rangle $ describe initially disentangled pure states on the
noncommutative algebra $\mathcal{B}\left( \mathcal{H}_{0}\right) =\otimes
_{r>0}\mathcal{B}\left( \frak{h}_{r}\right) $ of ''incoming quantum cats''
in $\mathcal{H}_{0}=\otimes _{r>0}\frak{h}_{r}$, and the output classical
probability waves $|\upsilon _{0}^{\infty }\rangle =\otimes _{r>0}|\upsilon
_{r}\rangle $ describe initially pure states on the commutative algebra $%
\frak{C}_{0}=\otimes _{r>0}\mathcal{C}_{r}$ of ''outgoing classical cats''
in $\mathcal{G}_{0}=\otimes _{r>0}\frak{g}_{r}$. At the boundary $r=0$ there
is a transmission of information from the quantum algebra $\mathcal{B}\left( 
\mathcal{H}\right) $ on $\mathcal{H}=\frak{h}\otimes \mathcal{H}_{0}$ to the
classical one $\frak{C}=\mathcal{C}\otimes \frak{C}_{0}$ on $\mathcal{G}=%
\frak{g}\otimes \mathcal{G}_{0}$ which is induced by the Heisenberg
transformation $\vartheta :\frak{A}\mapsto \frak{A}$. Note that although the
Schr\"{o}dinger transformation $U$ is reversible on $\mathbb{H}=\mathcal{H}%
\otimes \mathcal{G}$, $U^{-1}=U^{\dagger }$, and thus the Heisenberg
endomorphism $\vartheta $\ is one-to-one on the semi-commutative algebra $%
\frak{A}=\mathcal{B}\left( \mathcal{H}\right) \otimes \frak{C}$ , it
describes an irreversible dynamics because the image subalgebra $\vartheta
\left( \frak{A}\right) =U^{\dagger }\frak{A}U$ of the algebra $\frak{A}$
does not coincide with $\frak{A}\subset \mathcal{B}\left( \mathbb{H}\right) $%
. The initially distinguishable pure states on $\frak{A}$ may become
identical and mixed on the smaller algebra $U^{\dagger }\frak{A}U$, and this
explains the decoherence.

Thus, this dynamical model explains that the origin of the von Neumann
irreversible decoherence $\sigma =P_{\psi }\mapsto \rho $ of the atomic
state is in the ignorance of the result of the measurement described by the
partial tracing over the cat's Hilbert space $\frak{g}=\mathbb{C}^{2}$: 
\begin{equation}
\rho =\mathrm{Tr}_{\frak{g}}\hat{\varrho}=\sum_{\upsilon =0}^{1}\pi \left(
\upsilon \right) P_{\delta _{\upsilon }}=\varrho \left( 0\right) +\varrho
\left( 1\right) ,  \label{3.9}
\end{equation}
where $\varrho \left( \upsilon \right) =\left| \psi \left( \upsilon \right)
\right| ^{2}P_{\delta _{\upsilon }}$. It has entropy $\mathsf{S}\left( \rho
\right) =-\mathrm{Tr}\rho \log \rho $ of the compound state $\hat{\varrho}$
of the combined semi-classical system prepared for the indirect measurement
of the disintegration of atom by means of cat's death: 
\begin{equation*}
\mathsf{S}\left( \rho \right) =-\sum_{\upsilon =0}^{1}\left| \psi \left(
\upsilon \right) \right| ^{2}\log \left| \psi \left( \upsilon \right)
\right| ^{2}=\mathsf{S}\left( \hat{\varrho}\right)
\end{equation*}
It is the initial coherent uncertainty in the pure quantum state of the atom
described by the wave-function $\psi $ which is equal to one bit in the case 
$\left| \psi \left( 0\right) \right| ^{2}=1/2=\left| \psi \left( 1\right)
\right| ^{2}$. Each step of the unitary dynamics adds this entropy to the
total entropy of the state on $\frak{A}$ at the time $t\in \mathbb{N}$, so
the total entropy produced by this dynamical decoherence model is equal
exactly $t$.

The described dynamical model of the measurement interprets filtering $\rho
\mapsto \sigma _{\upsilon }$ simply as the conditioning 
\begin{equation}
\sigma _{\upsilon }=\varrho \left( \upsilon \right) /\pi \left( \upsilon
\right) =P_{\delta _{\upsilon }}  \label{3.10}
\end{equation}
of the joint classical-quantum state $\varrho \left( \cdot \right) $ with
respect to the events $F\left( \upsilon \right) $ by the Bayes formula which
is applicable due to the commutativity of actually measured observables $%
C\in \mathcal{C}$ (the life observables of cat at the time $t=1$) with any
other potential observable of the combined semi-classical system.

Thus the atomic decoherence is derived from the unitary interaction of the
quantum atom with the cat which should be treated as classical due to the
projection superselection rule in the ''Heisenberg'' picture of von Neumann
measurement. The spooky action at distance, affecting the atomic state by
measuring $\upsilon $, is simply the result of the statistical inference
(prediction after the measurement) of the atomic posterior state $\sigma
_{\upsilon }=P_{\delta _{\upsilon }}$: the atom disintegrates if and only if
the cat is dead.

A formal derivation of the von-Neumann-L\"{u}ders projection postulate and
the decoherence in the case of more general (discrete and continuous)
spectra by explicit construction of unitary transformation in the extended
semi-classical system as outlined in \cite{Be94, StBe96} is given in the
next section. An extension of this analysis to quantum continuous time and
spectra will also be considered in the last\ sections.

\section{The Dynamical Solution for Quantum Measurement}

\medskip\medskip

\begin{quote}
\textit{How wonderful we have met with a paradox, now we have some hope of
making progress} - Niels Bohr.
\end{quote}

\subsection{Instantaneous Quantum Measurements}

\medskip\medskip

\begin{quote}
\textit{If we have to go on with these dammed quantum jumps, then I'm sorry
that I ever got involved} - Schr\"{o}dinger.
\end{quote}

In this Chapter we present the main ideas of modern quantum measurement
theory and the author's views on the quantum measurement problem which might
not coincide with the present scientific consensus that this problem is
unsolvable in the standard framework, or at least unsolved \cite{BuLa91}. It
will be shown that there exists such solution along the line suggested by
the great founders of quantum theory Schr\"{o}dinger, Heisenberg and Bohr.
We shall see that von Neumann only partially solved this problem which he
studied in his Mathematical Foundation of Quantum Theory \cite{Neum32}, and
that the direction in which the solution might be found was envisaged by the
modern quantum philosopher J Bell \cite{Bell87}.

Here we develope the approach suggested in the previous section for solving
the famous Schr\"{o}dinger's cat paradox. We shall see that even the most
general quantum decoherence and wave packet reduction problem for an
instantaneous or even sequential measurements can be solved in a canonical
way which corresponds to adding a single initial cat's state. This resolves
also the other paradoxes of quantum measurement theory in a constructive
way, giving exact nontrivial models for the statistical analysis of quantum
observation processes determining the reality underlying these paradoxes.
Conceptually it is based upon a new idea of quantum causality as a
superselection rule called the Nondemolition Principle \cite{Be94} which
divides the world into the classical past, forming the consistent histories,
and the quantum future, the state of which is predictable for each such
history. This new postulate of quantum theory making \ the solution of
quantum measurement possible can not be contradicted by any experiment as we
prove that any sequence of usual, ``demolition'' measurements based on the
projection postulate or any other phenomenological measurement theories is
statistically equivalent, and in fact can be dynamically realized as a
simultaneous nondemolition measurement in a canonically extended infinite
semi-quantum system. The nondemolition models give exactly the same
predictions as the orthodox, ``demolition'' theories, but they do not
require the projection or any other postulate except the causality
(nondemolition) principle.

\subsubsection{Generalized reduction and its dilation}

Von Neumann's projection postulate, even reformulated in Heisenberg picture,
is only a phenomenological reduction principle which requires a dynamical
justification. Before formulating this quantum measurement problem in the
most interesting time-continuous case, let us consider how the reduction
principle can be generalized to include not only discrete but also
continuous measurement spectra for a single time $t$.

The generalized reduction of the wave function $\psi \left( x\right) $,
corresponding to a complete measurement with discrete or even continuous
data $y$, is described by a function $V\left( y\right) $ whose values are
linear operators $\frak{h}\ni \psi \mapsto V\left( y\right) \psi $ for each $%
y$ which are not assumed to be isometric on the quantum system Hilbert space 
$\frak{h}$, $V\left( y\right) ^{\dagger }V\left( y\right) \neq I$, but have
the following normalization condition. The resulting wave-function 
\begin{equation*}
\psi _{1}\left( x,y\right) =\left[ V\left( y\right) \psi \right] \left(
x\right)
\end{equation*}
is normalized with respect to a given measure $\mu $ on $y$ in the sense 
\begin{equation*}
\iint \left| \left[ V\left( y\right) \psi \right] \left( x\right) \right|
^{2}\mathrm{d}\lambda _{x}\mathrm{d}\mu _{y}=\int \left| \psi \left(
x\right) \right| ^{2}\mathrm{d}\lambda _{x}
\end{equation*}
for any probability amplitude $\psi $ normalized with respect to a measure $%
\lambda $ on $x$. This can be written as the isometry condition $\mathrm{V}%
^{\dagger }\mathrm{V}=\mathrm{I}$ of the operator $\mathrm{V}:\psi \mapsto
V\left( \cdot \right) \psi $ in terms of the integral 
\begin{equation}
\int_{y}V\left( y\right) ^{\dagger }V\left( y\right) \mathrm{d}\mu _{y}=%
\mathrm{I},\quad \mathrm{or}\quad \sum_{y}V\left( y\right) ^{\dagger
}V\left( y\right) =\mathrm{I}.  \label{4.1}
\end{equation}
with respect to the base measure $\mu $ which is usually the counting
measure, $\mathrm{d}\mu _{y}=1$ in the discrete case, e.g. in the case of
two-point variables $y=\upsilon $ (EPR paradox, or Schr\"{o}dinger cat with
the projection-valued $V\left( \upsilon \right) =F\left( \upsilon \right) $%
). The general case of orthoprojectors $V\left( y\right) =F\left( y\right) $
corresponds to the Kr\"{o}nicker $\delta $-function $V\left( y\right)
=\delta _{y}^{\mathrm{X}}$ of a self-adjoint operator $\mathrm{X}$ on $\frak{%
h}$ with the discrete spectrum coinciding with the measured values $y$.

As in the simple example of the Schr\"{o}dinger's cat, the unitary
realization of such $V$ can always be constructed in terms of a unitary
transformation on an extended Hilbert space $\frak{h}\otimes \frak{g}$ and a
normalized wave function $\chi ^{\circ }\in \frak{g}$. It is easy to find
such unitary dilation of any reduction family $V$ of the form 
\begin{equation}
V\left( y\right) =\mathrm{e}^{-i\mathrm{E}/\hbar }\mathrm{\exp }\left[ -%
\mathrm{X}\frac{\mathrm{d}}{\mathrm{d}y}\right] \varphi \left( y\right) =%
\mathrm{e}^{-i\mathrm{E}/\hbar }F\left( y\right) ,  \label{4.2}
\end{equation}
given by a normalized wave-function $\varphi \in L^{2}\left( \mathbb{G}%
\right) $ on a cyclic group $\mathbb{G}\ni y$ (e.g. $\mathbb{G}=\mathbb{R}$
or $\mathbb{G}=\mathbb{Z}$). Here the shift $F\left( y\right) =\varphi
\left( y-\mathrm{X}\right) $ of $\chi ^{\circ }=\varphi $ by a measured
operator $\mathrm{X}$ in $\frak{h}$ is well-defined by the unitary shifts $%
\mathrm{\exp }\left[ -x\frac{\mathrm{d}}{\mathrm{d}y}\right] $ in $\frak{g}%
=L^{2}\left( \mathbb{G}\right) $ in the eigen-representation of\ any
selfadjoint $\mathrm{X}$ having the spectral values $x\in \mathbb{G}$, and $%
\mathrm{E}=\mathrm{E}^{\dagger }$ is any free evolution action after the
measurement. As was noted by von Neumann for the case $\mathbb{G}=\mathbb{R}$%
\ in \cite{Neum32}, the operator $\mathrm{S}=\mathrm{\exp }\left[ -\mathrm{X}%
\frac{\mathrm{d}}{\mathrm{d}y}\right] $ is unitary in $\frak{h}\otimes \frak{%
g}$, and it coincides on $\psi \otimes \varphi $ with the isometry $\mathrm{F%
}=\mathrm{S}\left( \mathrm{I}\otimes \varphi \right) $ on each $\psi \in 
\frak{h}$ such that the unitary operator $\mathrm{W}=\mathrm{e}^{-i\mathrm{E}%
/\hbar }\mathrm{S}$ \ dilates the isometry $\mathrm{V}=\mathrm{e}^{-i\mathrm{%
E}/\hbar }\mathrm{F}$ in the sense 
\begin{equation*}
\mathrm{W}\left( \psi \otimes \chi ^{\circ }\right) =\mathrm{e}^{-i\mathrm{E}%
/\hbar }\mathrm{S}\left( \psi \otimes \varphi \right) =\mathrm{e}^{-i\mathrm{%
E}/\hbar }\mathrm{F}\psi ,\quad \forall \psi \in \frak{h}.
\end{equation*}
The wave function $\chi ^{\circ }=\varphi $ defines the initial probability
distribution $\left| \varphi \left( y\right) \right| ^{2}$ of the pointer
coordinate $y$ which can be dispersionless only if $\varphi $ is an
eigen-function of the pointer operator $Y=\hat{y}$ (multiplication operator
by $y$ in $\frak{g}$) corresponding to a discrete spectral value $y^{\circ }$
as a predetermined initial value of the pointer, $y^{\circ }=0$ say. This
corresponds to ortho-projectors $V\left( y\right) =\delta _{y}^{\mathrm{X}%
}=F\left( y\right) $ ($\mathrm{E}=\mathrm{O}$) indexed by $y$ from a
discrete cycle group, $y\in \mathbb{Z}$ for the discrete $\mathrm{X}$ having
eigenvalues $x\in \mathbb{Z}$ say. Thus the projection postulate is always
dilated by such shift operator $\mathrm{S}$ with $\chi ^{\circ }\left(
y\right) =\delta _{y}^{0}$ given as the eigen-function $\varphi \left(
y\right) =\delta _{y}^{0}$ corresponding to the initial value $y=0$ for the
pointer operator $Y=\hat{y}$ in $\frak{g}=L^{2}\left( \mathbb{Z}\right) $ \
(In the case of the Schr\"{o}dinger's cat $\mathrm{U}$ was simply the shift $%
\mathrm{W}$ (mod 2) \ in $\frak{g}=L^{2}\left( 0,1\right) :=\mathbb{C}^{2}$).

There exist another, canonical construction of the unitary operator $\mathrm{%
W}$ with the eigen-vector $\chi ^{\circ }\in \frak{g}$ for a `pointer
observable' $Y$ in an extended Hilbert space $\frak{g}$ even if $y$ is a
continuous variable of the general family $V\left( y\right) $. More
precisely, it can always be represented on the tensor product of the system
space $\frak{h}$ and the space $\frak{g}=\mathbb{C}\oplus L_{\mu }^{2}$ of
square-integrable functions $\chi \left( y\right) $ defining also the values 
$\chi \left( y^{\circ }\right) \in \mathbb{C}$ at an additional point $%
y^{\circ }\neq y$ corresponding to the absence of a result $y$ and $\chi
^{\circ }=1\oplus 0$ such that 
\begin{equation}
\langle x|V\left( y\right) \psi =\left( \langle x|\otimes \langle y|\right) 
\mathrm{W}\left( \psi \otimes \chi ^{\circ }\right) ,\quad \forall \psi \in 
\frak{h}  \label{4.3}
\end{equation}
for each measured value $y\neq y^{\circ }$.

Now we prove this unitary dilation theorem for the general $V\left( y\right) 
$ by the explicit construction of the matrix elements $\mathrm{W}_{y^{\prime
}}^{y}$ in the unitary block-operator $\mathrm{W}=\left[ \mathrm{W}%
_{y^{\prime }}^{y}\right] $ defined as $\left( \mathrm{I}\otimes \langle
y|\right) \mathrm{W}\left( \mathrm{I}\otimes |y^{\prime }\rangle \right) $
by 
\begin{equation*}
\psi ^{\dagger }\mathrm{W}_{y^{\prime }}^{y}\psi ^{\prime }=\left( \psi
^{\dagger }\otimes \langle y|\right) \mathrm{W}\left( \psi ^{\prime }\otimes
|y^{\prime }\rangle \right) ,
\end{equation*}
identifying $y^{\circ }$ with $0$ (assuming that $y\neq 0$, e.g. $y=1,\ldots
,n$). We shall use the short notation $\frak{f}=L_{\mu }^{2}$ for the
functional Hilbert space on the measured values $y$ and $\chi ^{\circ
}=|y^{\circ }\rangle $ (=$|0\rangle $ if $y^{\circ }=0$) for the additional
state-vector $\chi ^{\circ }\in \frak{g}$, identifying the extended Holbert
space $\frak{g}=\mathbb{C}\oplus \frak{f}$ \ with the space $L_{\mu \oplus
1}^{2}$ of square-integrable functions of all $y$ by the extention $\mu
\oplus 1$ of the measure $\mu $ at $y^{\circ }$ as \textrm{d}$\mu _{y^{\circ
}}=1$.

Indeed, we can always assume that $V\left( y\right) =\mathrm{e}^{-i\mathrm{E}%
/\hbar }F\left( y\right) $ where the family $F$ is viewed as an isometry $%
\mathrm{F}:\frak{h}\rightarrow \frak{h}\otimes \frak{f}$ corresponding $%
\mathrm{F}^{\dagger }\mathrm{F}=\mathrm{I}$ (not necessarily of the form $%
F\left( y\right) =\chi ^{\circ }\left( y-\mathrm{X}\right) $ as in (\ref{4.2}%
)). Denoting $\mathrm{e}^{-i\mathrm{E}/\hbar }\mathrm{F}$ as the column of $%
\mathrm{W}_{0}^{y}$, $y\neq 0$, and $\mathrm{e}^{-i\mathrm{E}/\hbar }\mathrm{%
F}^{\dagger }$ as the raw of $\mathrm{W}_{y}^{0}$, $y\neq 0$, we can compose
the unitary block-matrix 
\begin{equation}
\left[ \mathrm{W}_{y^{\prime }}^{y}\right] :=\mathrm{e}^{-i\mathrm{E}/\hbar }%
\left[ 
\begin{tabular}{rr}
$\mathrm{O}$ & $\mathrm{F}^{\dagger }$ \\ 
$\mathrm{F}$ & $\mathrm{I}\otimes \hat{1}-\mathrm{FF}^{\dagger }$%
\end{tabular}
\right] ,\quad \mathrm{I}\otimes \hat{1}=\left[ \mathrm{I}\delta _{y^{\prime
}}^{y}\right] _{y^{\prime }\neq 0}^{y\neq 0}  \label{4.4}
\end{equation}
describing an operator $\mathrm{W}=\left[ \mathrm{W}_{y^{\prime }}^{y}\right]
$\ on the product $\frak{h}\otimes \frak{g}$, where $\frak{g}=\mathbb{%
C\oplus }\frak{f}$, represented as$\frak{h}\oplus \left( \frak{h}\otimes 
\frak{f}\right) $, $\frak{f}=L_{\mu }^{2}$. It has the adjoint $\mathrm{W}%
^{\dagger }=\mathrm{e}^{i\mathrm{E}/\hbar }\mathrm{We}^{-i\mathrm{E}/\hbar }$%
, and obviously 
\begin{equation*}
\left( \mathrm{I}\otimes \langle y|\right) \mathrm{W}\left( \mathrm{I}%
\otimes |0\rangle \right) =V\left( y\right) ,\quad \forall y\neq 0.
\end{equation*}
The unitarity $\mathrm{W}^{-1}=\mathrm{W}^{\dagger }$ of the constructed
operator $\mathrm{W}$ is the consequence of the isometricity $\mathrm{F}%
^{\dagger }\mathrm{F}=\mathrm{I}$ and thus the projectivity $\left( \mathrm{%
FF}^{\dagger }\right) ^{2}=\mathrm{FF}^{\dagger }$ of $\mathrm{FF}^{\dagger
} $ and of $\mathrm{I}\otimes \hat{1}-\mathrm{FF}^{\dagger }$: 
\begin{equation*}
\mathrm{W}^{\dagger }\mathrm{W}=\left[ 
\begin{array}{ll}
\mathrm{F}^{\dagger }\mathrm{F} & \mathrm{F}^{\dagger }\left( \mathrm{I}%
\otimes \hat{1}-\mathrm{FF}^{\dagger }\right) \\ 
\left( \mathrm{I}\otimes \hat{1}-\mathrm{FF}^{\dagger }\right) \mathrm{F} & 
\mathrm{FF}^{\dagger }+\mathrm{I}\otimes \hat{1}-\mathrm{FF}^{\dagger }
\end{array}
\right] =\left[ 
\begin{array}{ll}
\mathrm{I} & \mathrm{O} \\ 
\mathrm{O} & \mathrm{I}\otimes \hat{1}
\end{array}
\right] .
\end{equation*}

In general the observation may be incomplete: the data $y$ may be the only
observable part of a pair $\left( z,y\right) $ defining the stochastic wave
propagator $V\left( z,y\right) .$ Consider for simplicity a discrete $z$
such that 
\begin{equation*}
V^{\dagger }V:=\sum_{z}\int V\left( z,y\right) ^{\dagger }V\left( z.y\right) 
\mathrm{d}\mu _{y}=\mathrm{I}.
\end{equation*}
Then the linear unital map on the algebra $\mathcal{B}\left( \frak{h}\right)
\otimes \mathcal{C}$ of the completely positive form 
\begin{equation*}
\pi \left( \hat{g}\mathrm{B}\right) =\sum_{z}\int g\left( y\right) V\left(
z,y\right) ^{\dagger }\mathrm{B}V\left( z.y\right) \mathrm{d}\mu _{y}\equiv 
\mathsf{M}\left[ g\pi \left( \mathrm{B}\right) \right]
\end{equation*}
describes the ''Heisenberg picture'' for generalized von Neumann reduction
with an incomplete measurement results $y$. Here $\mathrm{B}\in \mathcal{B}%
\left( \frak{h}\right) $, $\hat{g}$ is the multiplication operator by a
measurable function of $y$ defining any system-pointer observable by linear
combinations of $B\left( y\right) =g\left( y\right) \mathrm{B}$, and 
\begin{equation*}
\pi \left( y,\mathrm{B}\right) =\sum_{z}V\left( z,y\right) ^{\dagger }%
\mathrm{B}V\left( z,y\right) ,\quad \mathsf{M}\left[ B\left( y\right) \right]
=\int B\left( y\right) \mathrm{d}\mu _{y}.
\end{equation*}
The function $y\mapsto \pi \left( y\right) $ with values in the completely
positive maps $\mathrm{B}\mapsto \pi \left( y,\mathrm{B}\right) $, or
operations, is the basic tool in the operational approach to quantum
measurements. Its adjoint 
\begin{equation*}
\pi ^{\ast }\left( \sigma \right) =\sum_{z}V\left( y,z\right) \sigma V\left(
y,z\right) ^{\dagger }\mathrm{d}\mu _{y}=\pi ^{\ast }\left( y,\sigma \right) 
\mathrm{d}\mu _{y},
\end{equation*}
is given by the density matrix transformation and it is called the \emph{%
instrument} in the phenomenological measurement theories. The operational
approach was introduced by Ludwig \cite{Lud68}, and the mathematical
implementation of the notion of instrument was originated by Davies and
Lewis \cite{DaLe70}.

An abstract instrument now is defined as the adjoint to a unital completely
positive map $\pi $ for which $\pi _{y}^{\ast }\left( \sigma \right) $ is a
trace-class operator for each $y$, normalized to a density operator $\rho
=\int \mathrm{d}\pi _{y}^{\ast }\left( \sigma \right) $. The quantum mixed
state described by the operator $\rho $ is called the \emph{prior} state,
i.e. the state which has been prepared for the measurement. A unitary
dilation of the generalized reduction (or ``instrumental'') map $\pi $ was
constructed by Ozawa \cite{Oz84}, but as we shall now see, this, as well as
the canonical dilation (\ref{4.4}), is only a preliminary step towards the
its quantum stochastic realization allowing the dynamical derivation of the
reduction postulate as a result of the statistical inference as it was
suggested in \cite{Be94}.

\subsubsection{The future-past boundary value problem}

The additional system of the constructed unitary dilation for the
measurement propagator $V\left( y\right) $ represents only the pointer
coordinate of the measurement apparatus $y$ with the initial value $%
y=y^{\circ }$ ($=0$ corresponding to $\chi ^{\circ }=|0\rangle $). It should
be regarded as a classical system (like the Schr\"{o}dinger's cat) at the
instants of measurement $t>0$ in order to avoid the applying of the
projection postulate for inferences in the auxiliary system. Indeed, the
actual events of the measurement can be only those propositions $E$ in the
extended system which may serve as the conditions for any other proposition $%
F$ as a potential in future event, otherwise there can't be any causality
even in the weak, statistical sense. This means that future states should be
statistically predictable in any prior state of the system in the result of
testing the measurable event $E$ by the usual conditional probability
(Bayes) formula 
\begin{equation}
\Pr \left\{ F=1|E=1\right\} =\Pr \left\{ E\wedge F=1\right\} /\Pr \left\{
E=1\right\} \text{ \quad }\forall F,  \label{4.5}
\end{equation}
and this predictability, or statistical causality means that the prior
quantum probability $\Pr \left\{ F\right\} \equiv \Pr \left\{ F=1\right\} $
must coincide with the statistical expectation of $F$ as\ the weighted sum 
\begin{equation*}
\Pr \left\{ F|E\right\} \Pr \left\{ E\right\} +\Pr \left\{ F|E^{\perp
}\right\} \Pr \left\{ E^{\perp }\right\} =\Pr \left\{ F\right\}
\end{equation*}
of this $\Pr \left\{ F|E\right\} \equiv \Pr \left\{ F=1|E=1\right\} $ and
the complementary conditional probability $\Pr \left\{ F|E^{\perp }\right\}
=\Pr \left\{ F=1|E=0\right\} $. As one can easily see, this is possible if
and only if (\ref{2.7}) holds, i.e. any other future event-orthoprojector $F$
of the extended system must be compatible with the actual
event-orthoprojector $E$.

The actual events in the measurement model obtained by the unitary dilation
are only the orthoprojectors $E=\mathrm{I}\otimes \hat{1}_{\Delta }$ on $%
\frak{h}\otimes \frak{g}$ corresponding to the propositions ''$y\in \Delta $%
'' where $\hat{1}_{\Delta }$ is the multiplication by the indicator $%
1_{\Delta }$ for a measurable on the pointer scale subset $\Delta $. Other
orthoprojectors which are not compatible with these orthoprojectors, are
simply not admissible as the questions by the choice of time arrow. This
choice restores the quantum causality as statistical predictability, i.e.
the statistical inference made upon the sample data. And the actual
observables in question are only the measurable functions $g\left( y\right) $
of $y\neq y^{\circ }$ represented on $\frak{f}=L^{2}\left( \mu \right) $ by
the commuting operators $\hat{g}$ of multiplication by these functions, $%
\langle y|\hat{g}\chi =g\left( y\right) \chi \left( y\right) $. As follows
from $\mathrm{W}_{0}^{0}=\mathrm{O}$, the initial value $y^{\circ }=0$ is
never observed at the time $t=1$: 
\begin{equation*}
\left\| \psi _{1}\left( 0\right) \right\| ^{2}=\left\| \left( \mathrm{I}%
\otimes \langle 0|\right) \mathrm{W}\left( \psi \otimes |0\rangle \right)
\right\| ^{2}=\left\| \mathrm{W}_{0}^{0}\psi \right\| ^{2}=0,\quad \forall
\psi \in \frak{h}
\end{equation*}
(that is a measurable value $y\neq y^{\circ }$ is certainly observed at $t=1$%
). These are the only appropriate candidates for Bell's ''beables'', \cite
{Bell87}, p.174. Indeed, such commuting observables, extended to the quantum
counterpart as $\mathrm{G}_{0}=\mathrm{I}\otimes \hat{g}$ on $\frak{h}%
\otimes \frak{f}$, are compatible with any admissible question or observable 
$\mathrm{B}$ on $\frak{h}$ represented with respect to the output states $%
\psi _{1}=\mathrm{W}\psi _{0}$ at the time of measurement $t=1$ by an
operator $\mathrm{B}_{1}=\mathrm{B}\otimes \hat{1}$ on $\frak{h}\otimes 
\frak{f}$. The probabilities (or, it is better to say, the propensities) of
all such questions are the same in all states whether an observable $\mathrm{%
G}_{0}$ was measured but the result not read, or it was not measured at all.
In this sense the measurement of $\mathrm{G}_{0}$ is called \textit{%
nondemolition} with respect to the system observables $\mathrm{B}_{1}$, they
do not demolish the propensities, or prior expectations of $\mathrm{B}$.
However as we shall show now they are not necessary compatible with the same
operators $\mathrm{B}$ of the quantum system at the initial stage and\
currently represented as $\mathrm{WB}_{0}\mathrm{W}^{\dagger }$ on $\psi
_{1} $, where $\mathrm{B}_{0}=\mathrm{B}\otimes \hat{1}$ is the$\,$\
Schr\"{o}dinger representation of $\mathrm{B}$ at the time $t=0$ on the
corresponding input states $\psi _{0}=\mathrm{W}^{\dagger }\psi _{1}$ in $%
\frak{h}\otimes \frak{g}$ $.$

Indeed, we can see this on the example of the Schr\"{o}dinger cat, where $%
\mathrm{W}$ is the flip $\mathrm{S}$ in $\frak{g}=\mathbb{C}^{2}$ (shift $%
\func{mod}2$). In this case the operators the operators $\hat{g}_{1}$ in the
Heisenberg picture $\mathrm{G}=\mathrm{S}^{\dagger }\mathrm{G}_{0}\mathrm{S}$
are represented on $\frak{h}\otimes \frak{g}$ as the diagonal operators $%
\mathrm{G}=\left[ g\left( \tau +\upsilon \right) \delta _{\tau ^{\prime
}}^{\tau }\delta _{\upsilon ^{\prime }}^{\upsilon }\right] $ of
multiplication by $g\left( \tau +\upsilon \right) $, where the sum $\tau
+\upsilon =\left| \tau -\upsilon \right| $ is modulo 2. Obviously they do
not commute with $\mathrm{B}_{0}$ unless $\mathrm{B}$ is also a diagonal
operator $\hat{f}$ of multiplication by a function $f\left( \tau \right) $,
in which case 
\begin{equation*}
\left[ \mathrm{B}_{0},\mathrm{G}\right] \psi _{0}\left( \tau ,\upsilon
\right) =\left[ f\left( \tau \right) ,g\left( \tau +\upsilon \right) \right]
\psi _{0}\left( \tau ,\upsilon \right) =0,\quad \forall \psi _{0}\in \frak{h}%
\otimes \frak{g}\text{.}
\end{equation*}
The restriction of the possibilities in a quantum system to only the
diagonal operators $\mathrm{B}=\hat{f}$ of the atom which would eliminate
the time arrow in the nondemolition condition, amounts to the redundancy of
the quantum consideration: all such (possible and actual) observables can be
simultaneously represented as classical observables by the measurable
functions of $\left( \tau ,\upsilon \right) $.

Thus the constructed semiclassical algebra $\mathcal{B}_{-}=\mathcal{B}%
\left( \frak{h}\right) \otimes \mathcal{C}$ of the Schr\"{o}dinger's atom
and the pointer (dead or alive cat) is not dynamically invariant in the
sense that transformed algebra $\mathrm{W}^{\dagger }\mathcal{B}_{-}\mathrm{W%
}$ does not coincide and is not a part of $\mathcal{B}_{-}$ but of $\mathcal{%
B}_{+}=\mathcal{B}\left( \frak{h}\right) \otimes \mathcal{B}\left( \frak{g}%
\right) $. This is also true in the general case, unless all the
system-pointer observables in the Heisenberg picture are still decomposable, 
\begin{equation*}
\mathrm{W}^{\dagger }\left( \mathrm{B}\otimes \hat{g}\right) \mathrm{W}=\int
|y\rangle g\left( y\right) B\left( y\right) \langle y|\mathrm{d}\mu _{y},
\end{equation*}
which would imply $\mathrm{W}^{\dagger }\mathcal{B}\mathrm{W}\subseteq 
\mathcal{B}$. (Such dynamical invariance of the decomposable algebra , given
by the operator-valued functions $B\left( y\right) $, can be achieved by
this unitary dilations only in trivial cases.) This is why the von Neumann
type dilation (\ref{3.1}), and even more general dilations (\ref{4.4}), or 
\cite{Oz84, Be94} cannot yet be considered as the dynamical solution of the
instantaneous quantum measurement problem which we formulate in the
following way.

\emph{Given a reduction postulate defined by an isometry }$V$ \emph{on} $%
\frak{h}$ \emph{into }$\frak{h}\otimes \frak{g}$\emph{, find a triple }$%
\left( \mathcal{G},\frak{A},\Phi ^{\circ }\right) $ \emph{consisting of
Hilbert space} $\mathcal{G=G}_{-}\otimes \mathcal{G}_{+}$ \emph{embedding
the Hilbert spaces }$\frak{f=}L_{\mu }^{2}$\emph{\ by an isometry into }$%
\mathcal{G}_{+}$\emph{, an algebra }$\frak{A=A}_{-}\otimes \frak{A}_{+}$ 
\emph{on $\mathcal{G}$ with an Abelian subalgebra }$\frak{A}_{-}=\frak{C}$%
\emph{\ generated by an observable (beable) }$Y$\emph{\ on $\mathcal{G}_{-}$%
, and a state-vector }$\Phi ^{\circ }=\Phi _{-}^{\circ }\otimes \Phi
_{+}^{\circ }\in \mathcal{G}$\emph{\ such that there exist a unitary
operator }$U$\emph{\ on }$\mathcal{H=}\frak{h}\otimes \mathcal{G}$\emph{\
which induces an endomorphism on the product algebra $\frak{B=}\mathcal{B}%
\left( \frak{h}\right) \otimes \frak{A}$ in the sense }$U^{\dagger }AU\in 
\frak{B}$ \emph{for all }$B\in \frak{B}$\emph{, with } 
\begin{equation*}
\pi \left( \hat{g}\otimes \mathrm{B}\right) :=\left( \mathrm{I}\otimes \Phi
^{\circ }\right) ^{\dagger }U^{\dagger }\left( \mathrm{B}\otimes g\left(
Y\right) \right) U\left( \mathrm{I}\otimes \Phi ^{\circ }\right) =\mathsf{M}%
\left[ gV^{\dagger }\mathrm{B}V\right]
\end{equation*}
\emph{\ for all }$\mathrm{B}\in \mathcal{B}\left( \frak{h}\right) $ \emph{%
and measurable functions }$g$\emph{\ of }$Y$\emph{, where }$\mathsf{M}\left[
B\right] =\int B\left( y\right) \mu _{y}$.

As it was pointed out in \cite{Be79}, it is always possible to achieve this
dynamical invariance by extending the classical measurement apparatus' to an
infinite auxiliary semi-classical system. It has been done in the previous
section for the Schr\"{o}dinger's atom and cat, and here we sketch this
construction for the general unitary dilation (\ref{4.4}). (The full proof
is given in the Appendix 2.)

It consists of five steps. The first, preliminary step of a unitary dilation
for the isometry $\mathrm{V}$ has been already described in the previous
subsection.

Second, we construct the triple $\left( \mathcal{G},\frak{A},\Phi ^{\circ
}\right) $. Denote by $\frak{g}_{s}$, $s=\pm 0,\pm 1,\ldots $ (the indices $%
\pm 0$ are distinct and ordered as $-0<+0$) the copies of the Hilbert space $%
\frak{g}=\mathbb{C}\oplus \frak{f}$ in the dilation (\ref{4.4}) represented
as the functional space $L_{\mu }^{2}$ on the values of $y$ including $%
y^{\circ }=0$, and $\mathbb{G}_{n}=\frak{g}_{-n}\otimes \frak{g}_{+n}$, $%
n\geq 0$. We define the Hilbert space of the past $\mathcal{G}_{-}$ and the
future $\mathcal{G}_{+}$ as the state-vector spaces of semifinite discrete
strings generated by the infinite tensor products $\Phi _{-}=\chi
_{-0}\otimes \chi _{-1}\otimes \ldots $ and $\Phi _{+}=\chi _{+0}\otimes
\chi _{+1}\otimes \ldots $ with all but finite number of $\chi _{s}\in \frak{%
g}_{s}$ equal to the initial state $\chi _{s}^{\circ }$, the copies of $\chi
^{\circ }=$ $|0\rangle \in \frak{g}$. Denoting by $\mathcal{A}_{s}$ the
copies of the algebra $\mathcal{B}\left( \frak{g}\right) $ of bounded
operators if $s\geq +0$, of the diagonal subalgebra $\mathcal{D}\left( \frak{%
g}\right) $ on $\frak{g}$ if $s\leq -0$, \ and $\mathcal{A}_{n}=\mathcal{A}%
_{-n}\otimes \mathcal{A}_{+n}$ we construct the algebras of the past $\frak{A%
}_{-}$ and the future $\frak{A}_{+}$ and the whole algebra $\frak{A}$. $%
\frak{A}_{\pm }$ are generated on $\mathcal{G}_{\pm }$ respectively by the
diagonal operators $\hat{f}_{-0}\otimes \hat{f}_{-1}\otimes \ldots $ and by $%
\mathrm{X}_{+0}\otimes \mathrm{X}_{+1}\otimes \ldots $ with all but finite
number of $\hat{f}_{s}\in \mathcal{A}_{s}$, $s<0$ and $\mathrm{X}_{s}\in 
\mathcal{A}_{s}$, $s>0$ equal the identity operator $\hat{1}$ in $\frak{g}$.
Here $\hat{f}$ stands for the multiplication operator by a function $f$ of $%
y\in \mathbb{R}$, in particular, $\hat{y}$ is the multiplication by $y$,
with the eigen--vector $\chi ^{\circ }=|0\rangle $ corresponding to the
eigen-value $y^{\circ }=0$. The Hilbert space $\mathcal{G}_{-}\otimes 
\mathcal{G}_{+}$ identified with $\mathcal{G}=\otimes \mathbb{G}_{n}$, the
decomposable algebra $\frak{A}_{-}\otimes \frak{A}_{+}$ identified with $%
\frak{A}=\otimes \mathcal{A}_{n}$ and the product vector $\Phi _{-}\otimes
\Phi _{+}$ identified with $\Phi =\otimes \phi _{n}\in \mathcal{G}$, where $%
\phi _{n}=\chi _{-n}\otimes \chi _{+n}\equiv \chi _{-n}\chi _{+n}$ with all $%
\chi _{s}=\chi ^{\circ }$ stand as candidates for the triple $\left( 
\mathcal{G},\frak{A},\Phi \right) $. Note that the eigen-vector $\Phi
^{\circ }=\otimes \phi _{n}^{\circ }$ with all $\phi _{n}^{\circ }=\chi
^{\circ }\otimes \chi ^{\circ }$ corresponds to the initial eigen-state $%
y^{\circ }=0$ of all observables $Y_{\pm n}=\hat{1}_{0}\otimes \ldots
\otimes \hat{1}_{n-1}\otimes \hat{y}_{\pm }\otimes \hat{1}_{n+1}\otimes $ in 
$\mathcal{G}$, where $\hat{1}=\hat{1}_{-}\otimes \hat{1}_{+}$, $\hat{y}_{-}=%
\hat{y}\otimes \hat{1}_{+}$, $\hat{y}_{+}=\hat{1}_{-}\otimes \hat{y}$ and $%
\hat{1}_{\pm n}$ are the identity operators in $\frak{g}_{\pm n}$.

Third, we define the unitary evolution on the product space $\frak{h}\otimes 
\mathcal{G}$ of the total system by 
\begin{equation}
U:\psi \otimes \chi _{-}\chi _{+}\otimes \chi _{-1}\chi _{+1}\cdots \mapsto 
\mathrm{W}\left( \psi \otimes \chi _{+}\right) \chi _{+1}\otimes \chi
_{-}\chi _{+2}\cdots ,  \label{4.6}
\end{equation}
incorporating the right shift in $\mathcal{G}_{-}$, the left shift in $%
\mathcal{G}_{+}$ and the conservative boundary condition $\mathrm{W}:\frak{h}%
\otimes \frak{g}_{+}\rightarrow \frak{h}\otimes \frak{g}_{-}$ given by the
unitary dilation (\ref{4.4}). We have obviously 
\begin{equation*}
\left( \mathrm{I}\otimes \langle y_{-},y_{+},y_{-1},y_{+1}\ldots |\right)
U\left( \mathrm{I}\otimes |y_{-}^{0},0,y_{-1}^{0},0\ldots \rangle \right)
=\cdots \delta _{0}^{y_{+1}}\delta _{0}^{y_{+}}V\left( y_{-}\right) \delta
_{y_{-}^{0}}^{y_{-1}}\delta _{y_{-1}^{0}}^{y_{-2}}\cdots
\end{equation*}
so that the extended unitary operator $U$ still reproduces the reduction $%
V\left( y\right) $ in the result $y\neq y^{\circ }$ of the measurement $%
Y=Y_{-0}$ in a sequence $\left( Y_{-0},Y_{+0},Y_{-1}Y_{+1},\ldots \right) $
with all other $y_{s}$ being zero $y^{\circ }=0$ with the probability one
for the initial ground state $\Phi ^{\circ }$ of the connected string.

Fourth, we prove the dynamical invariance $U^{\dagger }\left( \mathcal{B}%
\left( \frak{h}\right) \otimes \frak{A}\right) U\subseteq \mathcal{B}\left( 
\frak{h}\right) \otimes \frak{A}$ of the decomposable algebra of the total
system, incorporating the measured quantum system $\mathcal{B}\left( \frak{h}%
\right) $ as the boundary between the quantum future (the right string
considered as quantum, $\frak{A}_{+}=\mathcal{B}\left( \mathcal{G}%
_{+}\right) $) with the classical past (the left string considered as
classical, $\frak{A}_{-}=\mathcal{D}\left( \mathcal{G}_{-}\right) $ ). This
follows straightforward from the definition of $U$%
\begin{equation*}
U^{\dagger }\left( \mathrm{B}\otimes \hat{g}_{-}\mathrm{X}_{+}\otimes \hat{g}%
_{-1}\mathrm{X}_{+1}\cdots \right) U=\hat{g}_{-1}\mathrm{W}^{\dagger }\left( 
\mathrm{B}\otimes \hat{g}_{-}\right) \mathrm{W}\otimes \hat{g}_{-2}\mathrm{X}%
_{+}\cdots
\end{equation*}
due to $\mathrm{W}^{\dagger }\left( \mathrm{B}\otimes \hat{g}\right) \mathrm{%
W}\in \mathcal{B}\left( \frak{h}\right) \otimes \mathcal{B}\left( \frak{g}%
\right) $ for all $\hat{g}\in \mathcal{D}\left( \frak{g}\right) $. However
this algebra representing the total algebra $\mathcal{B}\left( \frak{h}%
\right) \otimes \frak{A}$ on $\frak{h}\otimes \mathcal{G}$ is not invariant
under the inverse transformation, and there in no way to achieve the inverse
invariance keeping $\frak{A}$ decomposable as the requirement for
statistical causality of quantum measurement if $\mathrm{W}\left( \mathrm{B}%
\otimes \mathrm{X}\right) \mathrm{W}^{\dagger }\notin \mathcal{B}\left( 
\frak{h}\right) \otimes \mathcal{D}\left( \frak{g}\right) $ for some $%
\mathrm{B}\in \mathcal{B}\left( \frak{h}\right) $ and $\mathrm{X}\in 
\mathcal{B}\left( \frak{g}\right) $: 
\begin{equation*}
U\left( \mathrm{B}\otimes \hat{g}_{-}\mathrm{X}_{+}\otimes \hat{g}_{-1}%
\mathrm{X}_{+1}\cdots \right) U^{\dagger }=\mathrm{W}\left( \mathrm{B}%
\otimes \mathrm{X}_{+}\right) \mathrm{W}^{\dagger }\mathrm{X}_{+1}\otimes 
\hat{g}_{-}\mathrm{X}_{+2}\cdots .
\end{equation*}

And the fifth step is to explain on this dynamical model the decoherence
phenomenon, irreversibility and causality by giving a constructive scheme in
terms of equation for quantum predictions as statistical inferences by
virtue of gaining the measurement information.

Because of the crucial importance of these realizations for developing
understanding of the mathematical structure and interpretation of modern
quantum theory, we need to analyze the mathematical consequences which can
be drawn from such schemes.

\subsubsection{Decoherence and quantum prediction}

The analysis above shows that the dynamical realization of a quantum
instantaneous measurement is possible in an infinitely extended system, but
the discrete unitary group of unitary transformations $U^{t}$, $t\in \mathbb{%
N}$ with $U^{1}=U$ induces not a group of Heisenberg authomorphisms but an
injective irreversible semigroup of endomorphisms on the decomposable
algebra $\frak{B}=\mathcal{B}\left( \frak{h}\right) \otimes \frak{A}$ of
this system. However it is locally invertible on the center of the algebra $%
\frak{A}$ in the sense that it reverses the shift dynamics on $\frak{A}^{0]}$%
: 
\begin{equation}
T_{-t}\left( \mathrm{I}\otimes Y_{s}\right) T_{t}:=\mathrm{I}\otimes
Y_{s-t}=U^{t}\left( \mathrm{I}\otimes Y_{s}\right) U^{-t},\quad \forall
s\leq -0,t\in \mathbb{N}.  \label{4.7}
\end{equation}
Here $Y_{-n}=\hat{1}^{\otimes n}\otimes \hat{y}_{-}\otimes I_{n}$, where $%
I_{n}=\otimes _{k>n}\hat{1}_{k}$, and $T_{-t}=\left( T\right) ^{t}$ is the
power of the isometric shift $T:\Phi _{-}\mapsto \chi ^{\circ }\otimes \Phi
_{-}$ on $\mathcal{G}_{-}$ extended to the free unitary dynamics of the
whole system as 
\begin{equation*}
T:\psi \otimes \chi _{-}\chi _{+}\otimes \chi _{-1}\chi _{+1}\cdots \mapsto
\psi \otimes \chi _{+}\chi _{+1}\otimes \chi _{-}\chi _{+2}\cdots .
\end{equation*}

The extended algebra $\frak{B}$ is the minimal algebra containing all
consistent events of the history and all admissible questions about the
future of the open system under observation initially described by $\mathcal{%
B}\left( \frak{h}\right) $. Indeed, it contains all Heisenberg operators 
\begin{equation*}
B\left( t\right) =U^{-t}\left( \mathrm{B}\otimes I\right) U^{t},\quad
Y_{-}\left( t\right) =U^{-t}\left( \mathrm{I}\otimes Y_{-0}\right)
U^{t},\quad \forall t>0
\end{equation*}
of $\mathrm{B}\in \mathcal{B}\left( \frak{h}\right) $, and these operators
not only commute at each $t$, but also satisfy \emph{the nondemolition
causality condition} 
\begin{equation}
\left[ B\left( t\right) ,Y_{-}\left( r\right) \right] =0,\quad \left[
Y_{-}\left( t\right) ,Y_{-}\left( r\right) \right] =0,\quad \forall t\geq
r\geq 0.  \label{4.8}
\end{equation}
This follows from the commutativity of the Heisenberg string operators 
\begin{equation*}
Y_{r-t}\left( t\right) =U^{-t}\left( \mathrm{I}\otimes Y_{r-t}\right)
U^{t}=Y_{-}\left( r\right)
\end{equation*}
at the different points $s=r-t<0$ coinciding with $Y_{s}\left( r-s\right) $
for any $s<0$ because of (\ref{4.7}), and also from the commutativity with $%
\mathrm{B}\left( t\right) $ due to the simultaneous commutativity of all $%
Y_{s}\left( 0\right) =\mathrm{I}\otimes \mathrm{Y}_{s}$ and $B\left(
0\right) =\mathrm{B}\otimes I$. Thus all output Heisenberg operators $%
Y_{-}\left( r\right) $, $0<r\leq t$ at the boundary of the string can be
measured simultaneously as $Y_{-n}\left( t\right) =Y_{-}\left( t-n\right) $
at the different points $n<t$, or sequentially at the point $s=-0$ as the
commutative nondemolition family $Y_{0}^{t]}=\left( Y^{1},\ldots
,Y^{t}\right) $, where $Y^{r}=Y_{-}\left( r\right) $. This defines the
reduced evolution operators 
\begin{equation*}
V\left( t,y_{0}^{t]}\right) =V\left( y^{t}\right) V\left( y^{t-1}\right)
\cdots V\left( y^{1}\right) ,\quad t>0
\end{equation*}
of a \emph{sequential measurement} in the system Hilbert space $\frak{h}$
with measurement data $y_{0}^{t]}=\left\{ (0,t]\ni r\mapsto y^{r}\right\} $.
One can prove this (see the Appendix 2) using the \emph{filtering recurrency
equation} 
\begin{equation}
\psi \left( t,y_{0}^{t]}\right) =V\left( y_{t}\right) \psi \left(
t-1,y_{0}^{t-1]}\right) ,\quad \psi \left( 0\right) =\psi  \label{4.9}
\end{equation}
for $\psi \left( t,y_{0}^{t]}\right) =V\left( t,y_{0}^{t]}\right) \psi $ and
for $\Psi \left( t\right) =U^{t}\left( \psi \otimes \Phi _{-}\otimes \Phi
_{+}^{\circ }\right) $, where $\psi \in \frak{h}$, and $V\left( y_{t}\right) 
$ is defined by 
\begin{equation*}
\left( \mathrm{I}\otimes \langle y_{-\infty }^{t]}|\otimes \langle
y_{t}^{\infty }|\right) U\Psi \left( t-1\right) =V\left( y_{t}\right) \psi
\left( t-1,y_{0}^{t-1]}\right) \langle \delta _{0}^{y_{t}^{\infty
}}y_{-\infty }^{0]}|\Phi _{-}.
\end{equation*}

Moreover, any future expectations in the system, say the probabilities of
the questions $F\left( t\right) =U^{-t}\left( F\otimes I\right) U^{t},t\geq
s $ \ given by orthoprojectors $F$ on $\frak{h}$, can be statistically
predicted upon the results of the past measurements of $Y_{-}\left( r\right) 
$, $0<r\leq t$ and initial state $\psi $ by\ the simple conditioning 
\begin{equation*}
\Pr \left\{ F\left( t\right) |E\left( \mathrm{d}y^{1}\times \cdots \times 
\mathrm{d}y^{t}\right) \right\} =\frac{\Pr \left\{ F\left( t\right) \wedge
E\left( \mathrm{d}y^{1}\times \cdots \times \mathrm{d}y^{t}\right) \right\} 
}{\Pr \left\{ E\left( \mathrm{d}y^{1}\times \cdots \times \mathrm{d}%
y^{t}\right) \right\} }.
\end{equation*}
Here $E$ is the joint spectral measure for $Y^{1},\ldots ,Y^{t}$, and the
probabilities in the numerator (and denominator) are defined as 
\begin{equation*}
\left\| F\left( t\right) E\left( \mathrm{d}y^{1}\times \cdots \times \mathrm{%
d}y^{t}\right) \left( \psi \otimes \Phi ^{\circ }\right) \right\|
^{2}=\left\| F\psi \left( t,y_{0}^{t]}\right) \right\| ^{2}\mathrm{d}\mu
_{y^{1}}\cdots \mathrm{d}\mu _{y^{t}}
\end{equation*}
(and for $F=\mathrm{I}$) due to the commutativity of $F\left( t\right) $
with $E_{-}$. This implies the usual sequential instrumental formula 
\begin{equation*}
\left\langle \mathrm{B}\right\rangle \left( t,y_{0}^{r]}\right) =\frac{\psi
^{\dagger }\pi \left( t,y_{0}^{r]},\mathrm{B}\right) \psi }{\psi ^{\dagger
}\pi \left( t,y_{0}^{r]},\mathrm{I}\right) \psi }=\mathsf{M}\left[ \psi
_{y_{0}^{t]}}^{\dagger }\left( t\right) \mathrm{B}\psi _{y_{0}^{t]}}\left(
t\right) |y_{0}^{r]}\right]
\end{equation*}
for the future expectations of $\mathrm{B}\left( t\right) $ conditioned by $%
Y_{-}\left( 1\right) =y^{1},\ldots ,Y_{-}\left( s\right) =y^{s}$ for any $%
t>r $. Here $\psi _{y_{0}^{t]}}\left( t\right) =\psi \left(
t,y_{0}^{t]}\right) /\left\| \psi \left( t,y_{0}^{t]}\right) \right\| $, and 
\begin{equation*}
\pi \left( t,y_{0}^{r]},\mathrm{B}\right) =\idotsint V\left(
t,y_{0}^{r]},y_{r}^{t]}\right) ^{\dagger }\mathrm{B}V\left(
t,y_{0}^{r]},y_{r}^{t]}\right) \mathrm{d}\mu _{y^{_{r+1}}}\cdots \mathrm{d}%
\mu _{y^{t}}
\end{equation*}
is the sequential reduction map $V\left( t,y_{0}^{t]}\right) ^{\dagger }%
\mathrm{B}V\left( t,y_{0}^{t]}\right) $ defining the prior probability
distribution 
\begin{equation*}
\mathsf{P}\left( \mathrm{d}y_{0}^{t]}\right) =\psi ^{\dagger }\pi \left(
t,y_{0}^{t]},\mathrm{I}\right) \psi \mathrm{d}\mu _{y_{0}^{t]}}=\left\| \psi
\left( t,y_{0}^{t]}\right) \right\| ^{2}\mathrm{d}\mu _{y^{1}}\cdots \mathrm{%
d}\mu _{y^{t}}
\end{equation*}
integrated over $y_{r}^{t]}$ if these data are ignored for the quantum
prediction of the state at the time $t>r$.\ 

Note that the stochastic vector $\psi \left( t,y_{0}^{t]}\right) $,
normalized as 
\begin{equation*}
\idotsint \left\| \psi \left( t,y_{0}^{t]}\right) \right\| ^{2}\mathrm{d}\mu
_{y^{1}}\cdots \mathrm{d}\mu _{y^{t}}=1
\end{equation*}
depends linearly on the initial state vector $\psi \in \frak{h}$. However 
\emph{the posterior state vector} $\psi _{y_{0}^{t]}}\left( t\right) $ is
nonlinear, satisfying the \emph{nonlinear stochastic recurrency equation} 
\begin{equation}
\psi _{y_{0}^{t]}}\left( t\right) =V_{y_{0}^{t-1]}}\left( t,y^{t}\right)
\psi _{y_{0}^{t-1]}}\left( t-1\right) ,\quad \psi \left( 0\right) =\psi ,
\label{4.10}
\end{equation}
where $V_{y_{0}^{t-1]}}\left( t,y^{t}\right) =\left\| V\left(
t-1,y_{0}^{t-1]}\right) \psi \right\| V\left( y^{t-1]}\right) /\left\|
V\left( t,y_{0}^{t]}\right) \psi \right\| $.

In particular one can always realize in this way any sequential observation
of the noncommuting operators $\mathrm{B}_{t}=\mathrm{e}^{i\mathrm{E}/\hbar }%
\mathrm{B}_{0}\mathrm{e}^{-i\mathrm{E}/\hbar }$ given by a selfadjoint
operator $\mathrm{B}_{0}$ with discrete spectrum and the energy operator $%
\mathrm{E}$ in $\frak{h}$. It corresponds to the sequential collapse given
by $V\left( y\right) =\delta _{0}^{\mathrm{B}_{0}}\mathrm{e}^{-i\mathrm{E}%
/\hbar }$. Our construction suggests that any demolition sequential
measurement can be realized as the nondemolition by the commutative family $%
Y_{-}\left( t\right) $, $t>0$ with a common eigenvector $\Phi ^{\circ }$ as
the pointers initial state, satisfying the causality condition (\ref{4.8})
with respect to all future Heisenberg operators $\mathrm{B}\left( t\right) .$
And the sequential collapse (\ref{4.10}) follows from the usual Bayes
formula for conditioning of the compatible observables due to the classical
inference in the extended system. Thus, we have solved the \emph{sequential
quantum measurement problem} which can rigorously be formulated as

\emph{Given a sequential reduction family }$V\left( t,y_{0}^{t]}\right)
,t\in \mathbb{N}$ \emph{of isometries resolving the filtering equation (\ref
{4.9}) on} $\frak{h}$\emph{\ into $\frak{h}\otimes \frak{f}^{\otimes t}$,
find a triple }$\left( \mathcal{G},\frak{A},\Phi \right) $ \emph{consisting
of a Hilbert space} $\mathcal{G=G}_{-}\otimes \mathcal{G}_{+}$ \emph{%
embedding all tensor products }$\frak{f}^{\otimes t}$\emph{\ of the Hilbert
spaces }$\frak{f=}L_{\mu }^{2}$\emph{\ by an isometry into }$\mathcal{G}_{+}$%
\emph{, an algebra }$\frak{A=A}_{-}\otimes \frak{A}_{+}$ \emph{on $\mathcal{G%
}$ with an Abelian subalgebra }$\frak{A}_{-}=\frak{C}$\emph{\ generated by a
compatible discrete family }$Y_{-\infty }^{0]}=\left\{ Y_{s}\text{ }s\leq
0\right\} $\emph{\ } \emph{of the observables (beables) }$Y_{s}$\emph{\ on $%
\mathcal{G}_{-}$, and a state-vector }$\Phi ^{\circ }=\Phi _{-}^{\circ
}\otimes \Phi _{+}^{\circ }\in \mathcal{G}$\emph{\ such that there exist a
unitary group }$U^{t}$\emph{\ on }$\mathcal{H=}\frak{h}\otimes \mathcal{G}$%
\emph{\ inducing a semigroup of endomorphisms }$\frak{B}\ni B\mapsto
U^{-t}BU^{t}\in \frak{B}$ \emph{on the product algebra $\frak{B=}\mathcal{B}%
\left( \frak{h}\right) \otimes \frak{A}$ (\ref{4.7}on }$\frak{A}$\emph{,
with } 
\begin{equation*}
\pi ^{t}\left( \hat{g}_{-t}\otimes \mathrm{B}\right) =\left( \mathrm{I}%
\otimes \Phi ^{\circ }\right) ^{\dagger }U^{-t}\left( g_{-t}\left(
Y_{-t}^{0]}\right) \otimes \mathrm{B}\right) U^{t}\left( \mathrm{I}\otimes
\Phi ^{\circ }\right) =\mathsf{M}\left[ gV\left( t\right) ^{\dagger }\mathrm{%
B}V\left( t\right) \right]
\end{equation*}
\emph{\ for any }$\mathrm{B}\in \mathcal{B}\left( \frak{h}\right) $ \emph{%
and any operator }$\hat{g}_{-t}=\hat{g}_{-t}\left( Y_{-t}^{0]}\right) \in 
\frak{C}$ $\ $\emph{represented as the shifted function }$\hat{g}_{-t}\left(
y_{-t}^{0]}\right) =g\left( y_{0}^{t]}\right) $\emph{\ of }$%
Y_{-t}^{-0]}=\left( Y_{1-t},\ldots ,Y_{0}\right) $ \emph{on $\mathcal{G}$ by
any measurable function }$g$\emph{\ of }$y_{0}^{t]}=\left( y_{1},\ldots
,y_{t}\right) $\emph{\ with arbitrary }$t>0$\emph{, where} 
\begin{equation*}
\mathsf{M}\left[ gV\left( t\right) ^{\dagger }\mathrm{B}V\left( t\right) %
\right] =\idotsint g\left( y_{0}^{t]}\right) V\left( t,y_{0}^{t]}\right)
^{\dagger }\mathrm{B}V\left( t,y_{0}^{t]}\right) \mathrm{d}\mu
_{y^{1}}\cdots \mathrm{d}\mu _{y^{t}}.
\end{equation*}

Note that our construction of the solution to this problem admits also the
time reversed representation of the sequential measurement process described
by the isometry $\mathrm{V}$. The reversed system leaves in the same Hilbert
space, with the same initial state-vector $\Phi ^{\circ }$ in the auxiliary
space G, however the reversed auxiliary system is described by the reflected
algebra $\widetilde{\frak{A}}=R\frak{A}R$ where the reflection $R$ is
described by the unitary flip-operator $R:\Phi _{-}\otimes \Phi _{+}\mapsto
\Phi _{+}\otimes \Phi _{-}$ on $\mathcal{G}=\mathcal{G}_{-}\otimes \mathcal{G%
}_{+}$. The past and future in the reflected algebra $\widetilde{\frak{A}}=%
\frak{A}_{+}\otimes \frak{A}_{-}$ are flipped such that its left subalgebra
consists now of all operators on $\mathcal{G}_{-}$, $\widetilde{\frak{A}}%
_{-}=\mathcal{B}\left( \mathcal{G}\right) \supset \frak{A}_{-}$ and its
right subalgebra is the diagonal algebra $\widetilde{\frak{A}}_{+}=\mathcal{D%
}\left( \mathcal{G}_{+}\right) \subset \frak{A}_{+}$ on $\mathcal{G}_{+}$.
The inverse operators $U^{t},t<0$ induce the reversed dynamical semigroup of
the injective endomorphisms $B\mapsto U^{-t}BU^{t}$ which leaves invariant
the algebra $\widetilde{\frak{B}}=\mathcal{B}\left( \frak{h}\right) \otimes 
\widetilde{\frak{A}}$ but not $\frak{B}$. The reversed canonical measurement
process is described by another family $Y_{[+0}^{\infty }=\left(
Y_{+t}\right) $ of commuting operators $Y_{+t}=RY_{-t}R$ in $\widetilde{%
\frak{A}}_{+}$, and the Heisenberg operators 
\begin{equation*}
Y_{+}\left( t\right) =Y_{s}\left( t-s\right) =RY_{-}\left( -t\right) R,\quad
t<0,s>0,
\end{equation*}
are compatible and satisfy the reversed causality condition 
\begin{equation*}
\left[ B\left( t\right) ,Y_{+}\left( r\right) \right] =0,,\quad \left[
Y_{+}\left( t\right) ,Y_{+}\left( r\right) \right] =0,\quad \forall t\leq
r\leq 0.
\end{equation*}
It reproduces another, reversed sequence of the successive measurements 
\begin{equation*}
V^{\ast }\left( t,y_{[t}^{0}\right) =V^{\ast }\left( y_{t}\right) V^{\ast
}\left( y_{t+1}\right) \cdots V^{\ast }\left( y_{-1}\right) ,\quad t<0,
\end{equation*}
where $V^{\ast }\left( y\right) =\left( \mathrm{I}\otimes \langle y|\right) 
\mathrm{W}^{-1}\left( \mathrm{I}\otimes |0\rangle \right) $ depends on the
choice of the unitary dilation $\mathrm{W}$ of $\mathrm{V}$. In the case of
the canonical dilation (\ref{4.4}) uniquely defined up to the system
evolution between the measurements, we obtain $V^{\ast }\left( y\right)
=F\left( y\right) \mathrm{e}^{i\mathrm{E}/\hbar }$. If the system the
Hamiltonian is time-symmetric, i.e. $\overline{\mathrm{E}}=\mathrm{E}$ in
the sense $\mathrm{E}\bar{\psi}=\overline{\mathrm{E}\psi }$ with respect to
the complex (or another) conjugation in $\frak{h}$, and if $\overline{F}%
\left( y\right) =\mathrm{e}^{-i\mathrm{E}/\hbar }F\left( \tilde{y}\right) 
\mathrm{e}^{i\mathrm{E}/\hbar }$, where $y\mapsto \tilde{y}$ is a covariant
flip, $\widetilde{\tilde{y}}=y$ (e.g. $\tilde{y}=y$, or reflection of the
measurement data under the time reflection $t\mapsto -t$), then $V^{\ast
}\left( y\right) =\overline{V}\left( \tilde{y}\right) $. This means that the
reversed measurement process can be described as time-reflected direct
measurement process under the $\ast $-conjugation $\psi ^{\ast }\left(
y\right) =\bar{\psi}\left( \tilde{y}\right) $ in the space $\frak{h}\otimes 
\frak{f}$. And it can be modelled as the time reflected direct nondemolition
process under the involution $J\left( \psi \otimes \Phi \right) =\bar{\psi}%
\otimes R\Phi ^{\ast }$ induced by $\chi ^{\ast }\left( y\right) =\bar{\chi}%
\left( \tilde{y}\right) $ in $\frak{g}$ with the flip-invariant eigen-value $%
y^{\circ }=0$ and $|0\rangle ^{\ast }=|0\rangle $ corresponding to the real
ground state $\chi ^{\circ }\left( y\right) =\delta _{y}^{0}$.

Thus, the choice of time arrow, which is absolutely necessary for restoring
statistical causality in quantum theory, is equivalent to a superselection
rule. This corresponds to a choice of the minimal algebra $\frak{B}\subset 
\mathcal{B}\left( \mathcal{H}\right) $ generated by all admissible questions
on a suitable Hilbert space $\mathcal{H}$ of the nondemolition
representation for a process of the successive measurements. All consistent
events should be drown from the center of $\frak{B}$: the events must be
compatible with the questions, otherwise the propensities for the future
cannot be inferred from the testing of the past. The decoherence is
dynamically induced by a unitary evolution from any pure state on the
algebra $\frak{B}$ \ corresponding to the initial eigen-state for the
measurement apparatus pointer which is described by the center of $\frak{B}$%
. Moreover, the reversion of the time arrow corresponds to another choice of
the admissible algebra. It can be implemented by a complex conjugation $J$
on $\mathcal{H}$ on the transposed algebra $\widetilde{\frak{B}}=J\frak{B}J$
. Note that the direct and reversed dynamics respectively on $\frak{B}$ and
on $\widetilde{\frak{B}}$ are only endomorphic, and that the invertible
authomorphic dynamics induced on the total algebra $\mathcal{B}\left( 
\mathcal{H}\right) =\frak{B}\vee \widetilde{\frak{B}}$ does not reproduce
the decoherence due to the redundancy of one of its part for a given time
arrow $t$.

As Lawrence Bragg, another Nobel prize winner, once said, everything in the
future is a wave, everything in the past is a particle.

\subsection{Quantum Jumps as a Boundary Value Problem}

\medskip\medskip

\begin{quote}
\textit{Have the `jump' in the equations and not just the talk} - J Bell.
\end{quote}

Perhaps the closest to the truth was Bohr when he said that it `must be
possible so to describe the extraphysical process of the subjective
perception as\ if it were in reality in the physical world'. He regarded the
measurement apparatus, or meter, as a semiclassical object which interacts
with the world in a quantum mechanical way but is essentially classical: it
has only commuting observables - pointers. It relates the reality to a
subjective observer as the classical part of the classical-quantum closed
mechanical system. Thus Bohr accepted that \textit{not all the world is
quantum mechanical, there is a classical part of the physical world, and we
belong partly to this classical world. }

In the previous section we have already shown how to realize this program as
a discrete time boundary value problem of unitary interaction of the
classical past with the quantum future. This however gives little for
explanation of quantum jumps because any, even the Schr\"{o}dinger unitary
evolution in discrete time is described by jumps. Schr\"{o}dinger himself
tried unsuccessfully to derive the time continuous jumps from a boundary
value problem between past and future for a maybe more general than his
equation which would be relativistic and with infinite degrees of freedom.
Here we shall see that in order to realize this program one should indeed
consider a quantum field Dirac type boundary value problem, and the quantum
stochastic models of jumps correspond to its ultrarelativistic limit.

In realizing this program I will start along the line suggested by John Bell 
\cite{Bell87} that the ``development towards greater physical precision
would be to have the `jump' in the equations and not just the talk -- so
that it would come about as a dynamical process in dynamically defined
conditions.''

\subsubsection{Stochastic decoherence equation}

The generalized wave mechanics which enables us to treat the quantum
spontaneous events, unstable systems and processes of time-continuous
observation, or in other words, quantum mechanics with trajectories $\omega
=\left( x_{t}\right) $, was discovered only quite recently, in \cite{Be88,
Be89a, Be89b}. The basic idea of the theory is to replace the deterministic
unitary Schr\"{o}dinger propagation $\psi \mapsto \psi \left( t\right) $ by
a linear causal stochastic one $\psi \mapsto \psi \left( t,\omega \right) $
which is not necessarily unitary for each history $\omega $, but unitary in
the mean square sense, $\mathsf{M}\left[ \left\| \psi \left( t\right)
\right\| ^{2}\right] =1$, with respect to a standard probability measure $%
\mu \left( \mathrm{d}\omega \right) $ for the measurable history subsets $%
\mathrm{d}\omega $. The unstable quantum systems can also be treated in the
stochastic formalism by relaxing this condition by allowing the decreasing
survival probabilities $\mathsf{M}\left[ \left\| \psi \left( t\right)
\right\| ^{2}\right] \leq 1$. Due to this the positive measures 
\begin{equation*}
\mathsf{P}\left( t,\mathrm{d}\omega \right) =\left\| \psi \left( t,\omega
\right) \right\| ^{2}\mu \left( \mathrm{d}\omega \right) ,\quad \tilde{\mu}%
\left( \mathrm{d}\omega \right) =\lim_{t\rightarrow \infty }\mathsf{P}\left(
t,\mathrm{d}\omega \right)
\end{equation*}
are normalized (if $\left\| \psi \right\| =1$) for each $t$, and are
interpreted as the probability measure for the histories $\omega
_{t}=\left\{ (0,t]\ni r\mapsto x^{r}\right\} =x_{0}^{t]}$.of the output
stochastic process $x^{t}$ with respect to the measure $\tilde{\mu}$. In the
same way as the abstract Schr\"{o}dinger equation can be derived from only
unitarity of propagation, the abstract decoherence wave equation can be
derived from the mean square unitarity in the form of a linear stochastic
differential equation. The reason that Bohr and Schr\"{o}dinger didn't
derive such an equation despite their firm belief that the measurement
process can be described `as\ if it were in reality in the physical world'
is that the appropriate (stochastic and quantum stochastic) differential
calculus had not been yet developed early in that century. As classical
differential calculus has its origin in classical mechanics, quantum
stochastic calculus has its origin in quantum stochastic mechanics. A formal
algebraic approach to this new calculus, which was developed in \cite{Be88,
Be92a}, is presented in the Appendix 2.

Assuming that the superposition principle also holds for the stochastic
waves such that $\psi \left( t,\omega \right) $ is given by a linear
stochastic propagator $V\left( t,\omega \right) $, let us derive the general
linear stochastic wave equation which preserves the mean-square
normalization of these waves. Note that the abstract Schr\"{o}dinger
equation $i\hbar \partial _{t}\psi =\mathrm{E}\psi $ can also be derived as
the general linear deterministic equation which preserves the normalization
in a Hilbert space $\frak{h}$. For the notational simplicity we shall
consider here only the a finite-dimensional maybe complex trajectories $%
x^{t}=\left( x_{k}^{t}\right) $, $k=1,\ldots ,d$, the infinite-dimensional
trajectories (fields) with even continuous index $k$ can be found elsewhere
(e.g. in \cite{Be88, Be92a}). It is usually assumed that the these $%
x_{k}^{t} $ as \emph{input} stochastic processes have stationary independent
increment $\mathrm{d}x_{k}^{t}=x_{k}^{t+\mathrm{d}t}-x_{k}^{t}$ with given
expectations $\mathsf{M}\left[ \mathrm{d}x_{k}^{t}\right] =\lambda _{k}%
\mathrm{d}t$. The abstract linear stochastic decoherence wave equation is
written then as 
\begin{equation*}
\mathrm{d}\psi \left( t\right) +\left( \frac{\lambda ^{2}}{2}\mathrm{R}+%
\frac{i}{\hbar }\mathrm{E}\right) \psi \left( t\right) \mathrm{d}t=\mathrm{L}%
^{k}\psi \left( t\right) \mathrm{d}x_{k}^{t},\quad \psi \left( 0\right)
=\psi .
\end{equation*}
Here $\mathrm{E}$ is the system energy operator (the Hamiltonian of free
evolution of the system), $\mathrm{R}=\mathrm{R}^{\dagger }$ is a
selfadjoint operator describing a relaxation process in the system, $\mathrm{%
L}^{k}$ are any operators coupling the system to the trajectories $x_{k}$,
and we use the Einstein summation rule $\mathrm{L}^{k}x_{k}=\sum \mathrm{L}%
^{k}x_{k}\equiv \mathrm{L}x$, $\lambda ^{2}=\lambda ^{k}\lambda _{k}$ with $%
\lambda ^{k}=\bar{\lambda}_{k}$. In order to derive the relations between
these operators which will imply the mean-square normalization of $\psi
\left( t,\omega \right) $, let us rewrite this equation in the standard form 
\begin{equation}
\mathrm{d}\psi \left( t\right) +\mathrm{K}\psi \left( t\right) \mathrm{d}t=%
\mathrm{L}^{k}\psi \left( t\right) \mathrm{d}y_{k}^{t},\quad \mathrm{K}=%
\frac{\lambda ^{2}}{2}\mathrm{R}+\frac{i}{\hbar }\mathrm{E}-\mathrm{L}%
\lambda ,  \label{5.1}
\end{equation}
where $y_{k}^{t}=x_{k}^{t}-t\lambda _{k}$ are input noises as zero mean
value independent increment processes with respect to the input probability
measure $\mu $. Note that these noises will become the output information
processes which will have dependent increments and correlations with the
system with respect to the output probability measure $\tilde{\mu}=\mathsf{P}%
\left( \infty ,\mathrm{d}\omega \right) $. If the Hilbert space valued
stochastic process $\psi \left( t,\omega \right) $ is normalized in the mean
square sense for each $t$, it represents a stochastic probability amplitude $%
\psi \left( t\right) $ as an element of an extended Hilbert space $\mathcal{H%
}_{0}=\frak{h}\otimes L_{\mu }^{2}$. The stochastic process $t\mapsto \psi
\left( t\right) $ describes a process of continual decoherence of the
initial pure state $\rho \left( 0\right) =P_{\psi }$ into the mixture 
\begin{equation*}
\rho \left( t\right) =\int P_{\psi _{\omega }\left( t\right) }\mathsf{P}%
\left( t,\mathrm{d}\omega \right) =\mathsf{M}\left[ \psi \left( t\right)
\psi \left( t\right) ^{\dagger }\right]
\end{equation*}
of the posterior states corresponding to $\psi _{\omega }\left( t\right)
=\psi \left( t,\omega \right) /\left\| \psi \left( t,\omega \right) \right\| 
$, where $\mathsf{M}$ denotes mean with respect to the measure $\mu $.
Assuming that the conditional expectation $\left\langle \mathrm{d}\bar{y}%
_{k}^{t}\mathrm{d}y_{k}^{t}\right\rangle _{t}$ in 
\begin{align*}
\left\langle \mathrm{d}\left( \psi ^{\dagger }\psi \right) \right\rangle
_{t}& =\left\langle \mathrm{d}\psi ^{\dagger }\mathrm{d}\psi +\psi ^{\dagger
}\mathrm{d}\psi +\mathrm{d}\psi ^{\dagger }\psi \right\rangle _{t} \\
& =\psi ^{\dagger }\left( \mathrm{L}^{k\dagger }\left\langle \mathrm{d}\bar{y%
}_{k}\mathrm{d}y_{k}\right\rangle _{t}\mathrm{L}^{k}-\left( \mathrm{K}+%
\mathrm{K}^{\dagger }\right) \mathrm{d}t\right) \psi
\end{align*}
is $\mathrm{d}t$ (as in the case of the standard independent increment
processes with $\bar{y}=y$ and $\left( \mathrm{d}y\right) ^{2}=\mathrm{d}%
t+\varepsilon \mathrm{d}y_{t}$, see the Appendix 3), the mean square
normalization in the differential form $\left\langle \mathrm{d}\left( \psi
^{\dagger }\psi \right) \right\rangle _{t}=0$ (or $\left\langle \mathrm{d}%
\left( \psi ^{\dagger }\psi \right) \right\rangle _{t}\leq 0$ for the
unstable systems) can be expressed \cite{Be89a, Be90b} as $\mathrm{K}+%
\mathrm{K}^{\dagger }\geq \mathrm{L}^{\dagger }\mathrm{L}$. In the stable
case this defines the self-adjoint part of $\mathrm{K}$ as half of $\mathrm{L%
}^{\dagger }\mathrm{L}$,\ i.e. 
\begin{equation*}
\mathrm{K}=\frac{1}{2}\mathrm{L}^{\dagger }\mathrm{L}+\frac{i}{\hbar }%
\mathrm{H},\quad \mathrm{L}^{\dagger }\mathrm{L}=\sum_{k}\mathrm{L}%
^{k\dagger }\mathrm{L}^{k}\equiv \mathrm{L}_{k}\mathrm{L}^{k}
\end{equation*}
where $\mathrm{H}=\mathrm{H}^{\dagger }$ is the Schr\"{o}dinger Hamiltonian
in this equation when $\mathrm{L}=0$. One can also derive the corresponding
Master equation 
\begin{equation*}
\frac{\mathrm{d}}{\mathrm{d}t}\rho \left( t\right) +\mathrm{K}\rho \left(
t\right) +\rho \left( t\right) \mathrm{K}^{\dagger }=\mathrm{L}^{k}\rho
\left( t\right) \mathrm{L}_{k}
\end{equation*}
for mixing decoherence of the initially pure state $\rho \left( 0\right)
=\psi \psi ^{\dagger }$, as well as a stochastic nonlinear wave equation for
the dynamical prediction of the posterior state vector $\psi _{\omega
}\left( t\right) $, the normalization of $\psi \left( t,\omega \right) $ at
each $\omega $.

\subsubsection{Quantum jumps and unstable systems}

Actually, there are two basic standard forms \cite{Be89b, Be90a} of such
stochastic wave equations, corresponding to two basic types of stochastic
integrators with independent increments: the Brownian standard type $%
x_{k}^{t}\simeq b_{k}^{t}$, and the Poisson standard type $x_{k}^{t}\simeq
n_{k}^{t}$ with respect to the basic measure $\mu $, see the Appendix 3. We
shall start with the Poisson case of the identical $n_{k}^{t}$ having all
the expectations $\mathsf{M}n_{k}^{t}=\nu t$ and characterized by a very
simple differential multiplication table 
\begin{equation*}
\mathrm{d}n_{k}^{t}\left( \omega \right) \mathrm{d}n_{l}^{t}\left( \omega
\right) =\delta _{l}^{k}\mathrm{d}n_{l}^{t}\left( \omega \right)
\end{equation*}
as it is for the only possible values $\mathrm{d}n_{k}^{t}=0,1$ of the
counting increments at each time $t$. By taking all $x_{k}^{t}=n_{k}^{t}/\nu
^{1/2}$ such that they have the expected rates $\lambda _{k}=\nu ^{1/2}$ we
can get the standard Poisson noises $y_{k}^{t}=x_{k}^{t}-\nu ^{1/2}t\equiv
m_{k}^{t}$ with respect to the input Poisson probability measure $\mu =%
\mathsf{P}_{m}$, described by the multiplication table 
\begin{equation*}
\mathrm{d}m_{k}\mathrm{d}m_{l}=\delta _{k}^{l}\left( \mathrm{d}t+\nu ^{-1/2}%
\mathrm{d}m_{k}\right) ,\quad \mathrm{d}m_{k}\mathrm{d}t=0=\mathrm{d}t%
\mathrm{d}m_{k},
\end{equation*}
Let us set now in our basic equation (\ref{5.1}) the Hamiltonian $\mathrm{H}%
=\hbar \left( \mathrm{K}-\mathrm{K}^{\dagger }\right) /2i$ and the coupling
operators $\mathrm{L}^{k}$ of the form 
\begin{equation*}
\mathrm{L}^{k}=\lambda (\mathrm{C}^{k}-\mathrm{I}),\quad \mathrm{H}=\mathrm{E%
}+i\frac{\nu }{2}\left( \mathrm{C}^{k}-\mathrm{C}_{k}\right) ,
\end{equation*}
with the coupling constant $\lambda =\nu ^{1/2}$ and $\mathrm{C}^{k}\equiv 
\mathrm{C}_{k}^{\dagger }$ given by the collapse operators $\mathrm{C}_{k}$
(e.g. orthoprojectors, or contractions, $\mathrm{C}_{k}^{\dagger }$\textrm{C}%
$_{k}\leq \mathrm{I}$). This corresponds to the stochastic decoherence
equation of the form 
\begin{equation}
\mathrm{d}\psi \left( t\right) +\left( \frac{\nu }{2}\mathrm{R}+\frac{i}{%
\hbar }\mathrm{E}\right) \psi \left( t\right) \mathrm{d}t=\left( \mathrm{C}%
^{k}-\mathrm{I}\right) \psi \left( t\right) \mathrm{d}n_{k}^{t},\quad \psi
\left( 0\right) =\psi ,  \label{5.2}
\end{equation}
where $\mathrm{R}\geq \mathrm{C}^{\dagger }\mathrm{C}-\mathrm{I}$, or in the
standard form (\ref{5.1}) with $y_{k}^{t}=m_{k}^{t}$. In the stable case
when $\mathrm{R}=\mathrm{C}^{\dagger }\mathrm{C}-\mathrm{I}$ this was
derived from a unitary quantum jump model for counting nondemolition
observation in \cite{Be89a, BaBe}. It correspond to the linear stochastic
decoherence Master-equation 
\begin{equation*}
\mathrm{d}\varrho \left( t\right) +\left[ \mathrm{G}\varrho \left( t\right)
+\varrho \left( t\right) \mathrm{G}^{\dagger }-\nu \varrho \left( t\right) %
\right] \mathrm{d}t=\left[ \mathrm{C}_{k}\varrho \left( t\right) \mathrm{C}%
^{k}-\varrho \left( t\right) \right] \mathrm{d}n_{k}^{t},\quad \varrho
\left( 0\right) =\rho ,
\end{equation*}
for the not normalized (but normalized in the mean) density matrix $\varrho
\left( t,\omega \right) $, where $\mathrm{G}=\frac{\nu }{2}\mathrm{C}_{k}%
\mathrm{C}^{k}+\frac{i}{\hbar }\mathrm{E}$ (it has the form $\psi \left(
t,\omega \right) \psi \left( t,\omega \right) ^{\dagger }$ in the case of a
pure initial state $\rho =\psi \psi ^{\dagger }$).

The nonlinear filtering equation for the posterior state vector 
\begin{equation*}
\psi _{\omega }\left( t\right) =\psi \left( t,\omega \right) /\left\| \psi
\left( t,\omega \right) \right\|
\end{equation*}
has in this case the following form \cite{Be90a} 
\begin{equation}
\mathrm{d}\psi _{\omega }+\left( \frac{\nu }{2}\left( \mathrm{C}_{k}\mathrm{C%
}^{k}-\sum \left\| \mathrm{C}_{k}^{\dagger }\psi _{\omega }\right\|
^{2}\right) +\frac{i}{\hbar }\mathrm{E}\right) \psi _{\omega }\mathrm{d}%
t=\left( \mathrm{C}_{k}^{\dagger }/\left\| \mathrm{C}_{k}^{\dagger }\psi
_{\omega }\right\| -\mathrm{I}\right) \psi _{\omega }\mathrm{d}n_{k,\omega
}^{t,\rho },  \label{5.3}
\end{equation}
where $\left\| \psi \right\| =\left\langle \psi |\psi \right\rangle ^{1/2}$
(see also \cite{BeMe96} for the infinite-dimensional case). It corresponds
to the nonlinear stochastic Master-equation 
\begin{equation*}
\mathrm{d}\rho _{\omega }+\left[ \mathrm{G}\rho _{\omega }+\rho _{\omega }%
\mathrm{G}^{\dagger }-\nu \rho _{\omega }\mathrm{TrC}^{k}\rho _{\omega }%
\mathrm{C}_{k}\right] \mathrm{d}t=\left[ \mathrm{C}^{k}\rho _{\omega }%
\mathrm{C}_{k}/\mathrm{TrC}^{k}\rho _{\omega }\mathrm{C}_{k}-\rho \right] 
\mathrm{d}n_{k,\omega }^{t,\rho }
\end{equation*}
for the posterior density matrix $\rho _{\omega }\left( t\right) $ which is
the projector $P_{\omega }\left( t\right) =\psi _{\omega }\left( t\right)
\psi _{\omega }\left( t\right) ^{\dagger }$ for the pure initial state $\rho
_{\omega }\left( 0\right) =P_{\psi }$. Here $n_{k}^{\rho }\left( t,\omega
\right) =n_{k,\omega }^{t,\rho \left( t\right) }$ are the output counting
processes which are described by the history probability measure 
\begin{equation*}
\mathsf{P}\left( t,\mathrm{d}\omega \right) =\pi \left( t,\omega \right) \mu
\left( \mathrm{d}\omega \right) ,\quad \pi \left( t,\omega \right) =\mathrm{%
Tr}\varrho \left( t,\omega \right)
\end{equation*}
with the increment $\mathrm{d}n_{k}^{\rho }\left( t\right) $ independents of 
$n_{k}^{\rho }\left( t\right) $ under the condition $\rho _{\omega }\left(
t\right) =\rho $ and the conditional expectations 
\begin{equation*}
\mathsf{M}\left[ \mathrm{d}n_{k}^{\rho }\left( t\right) |\rho _{\omega
}\left( t\right) =\rho \right] =\nu \mathrm{TrC}_{k}^{\dagger }\rho \mathrm{C%
}_{k}\mathrm{d}t
\end{equation*}
which are $\nu \left\| \mathrm{C}_{k}^{\dagger }\psi \right\| ^{2}\mathrm{d}%
t $ for $\rho =P_{\psi }$. The derivation and solution of this equation was
also considered in \cite{BeSt91}, and its solution was applied in quantum
optics in \cite{Car93, Car94}.

This nonlinear quantum jump equation can be written also in the quasi-linear
form \cite{Be89b, Be90a} 
\begin{equation}
\mathrm{d}\psi _{\omega }\left( t\right) +\widetilde{\mathrm{K}}\left(
t\right) \psi _{\omega }\left( t\right) \mathrm{d}t=\mathrm{L}^{k}\left(
t\right) \psi _{\omega }\left( t\right) \mathrm{d}\tilde{m}_{k,\omega
}^{t,\rho },  \label{5.4}
\end{equation}
where $\tilde{m}_{k}^{\rho }\left( t,\omega \right) =\tilde{m}_{k,\omega
}^{t,\rho \left( t,\omega \right) }$ are the innovating martingales with
respect to the output measure which is described by the differential 
\begin{equation*}
\mathrm{d}\tilde{m}_{k}^{\rho }\left( t\right) =\nu ^{-1/2}\left\| \mathrm{C}%
_{k}^{\dagger }\psi _{\omega }\left( t\right) \right\| ^{-1}\mathrm{d}%
n_{k}^{\rho }\left( t\right) -\nu ^{1/2}\left\| \mathrm{C}_{k}^{\dagger
}\psi _{\omega }\left( t\right) \right\| \mathrm{d}t
\end{equation*}
with $\rho =P_{\psi }$ for $\psi =\psi _{\omega }\left( t\right) $ and the
initial $\tilde{m}_{k}^{\rho }\left( 0\right) =0$, the operator $\widetilde{%
\mathrm{K}}\left( t\right) $ similar to $\mathrm{K}$ has the form 
\begin{equation*}
\widetilde{\mathrm{K}}\left( t\right) =\frac{1}{2}\widetilde{\mathrm{L}}%
\left( t\right) ^{\dagger }\widetilde{\mathrm{L}}\left( t\right) +\frac{i}{%
\hbar }\widetilde{\mathrm{H}}\left( t\right) ,
\end{equation*}
and $\widetilde{\mathrm{H}}\left( t\right) ,\widetilde{\mathrm{L}}\left(
t\right) $ depend on $t$ (and $\omega $) through the dependence on $\psi
=\psi _{\omega }\left( t\right) $: 
\begin{equation*}
\widetilde{\mathrm{L}}^{k}=\nu ^{1/2}\left( \mathrm{C}_{k}^{\dagger
}-\left\| \mathrm{C}_{k}^{\dagger }\psi \right\| \right) ,\quad \widetilde{%
\mathrm{H}}=\mathrm{E}+\frac{\nu }{2i}\left( \mathrm{C}_{k}-\mathrm{C}%
_{k}^{\dagger }\right) \left\| \mathrm{C}_{k}^{\dagger }\psi \right\| .
\end{equation*}

The latter form of the nonlinear filtering equation admits the central limit 
$\nu \rightarrow \infty $ corresponding to the standard Wiener case when $%
y_{k}^{t}=w_{k}^{t}$, 
\begin{equation*}
\mathrm{d}w_{k}\mathrm{d}w_{l}=\delta _{k}^{l}\mathrm{d}t,\quad \mathrm{d}%
w_{k}\mathrm{d}t=0=\mathrm{d}t\mathrm{d}w_{k},
\end{equation*}
with respect to the limiting input Wiener measure $\mu $. If $\mathrm{L}^{k}$
and $\mathrm{H}$ do not depend on $\nu $, i.e. $\mathrm{C}_{k}$ and $\mathrm{%
E}$ depend on $\nu $ as 
\begin{equation*}
\mathrm{C}_{k}=\mathrm{I}+\nu ^{-1/2}\mathrm{L}_{k},\;\mathrm{E}=\mathrm{H}+%
\frac{\nu ^{1/2}}{2i}\left( \mathrm{L}_{k}^{\dagger }-\mathrm{L}_{k}\right) ,
\end{equation*}
then $\tilde{m}_{k}^{\rho }\left( t\right) \rightarrow \tilde{w}_{k}^{t}$,
where the innovating diffusion process $\tilde{w}^{t}$ defined as 
\begin{equation*}
\mathrm{d}\tilde{w}_{k}^{t}\left( \omega \right) =\mathrm{d}w_{k}^{t}\left(
\omega \right) -2\func{Re}\left\langle \psi _{\omega }\left( t\right) |%
\mathrm{L}_{k}^{\dagger }\psi _{\omega }\left( t\right) \right\rangle 
\mathrm{d}t,
\end{equation*}
are also standard Wiener processes but with respect to the output
probability measure $\tilde{\mu}\left( \mathrm{d}\omega \right) =\mu \left( 
\mathrm{d}\tilde{\omega}\right) $ due to 
\begin{equation*}
\mathrm{d}\tilde{w}_{k}\mathrm{d}\tilde{w}_{l}=\delta _{k}^{l}\mathrm{d}%
t,\quad \mathrm{d}\tilde{w}_{k}\mathrm{d}t=0=\mathrm{d}t\mathrm{d}\tilde{w}%
_{k}.
\end{equation*}
.If $\left\| \psi _{\omega }\left( t\right) \right\| =1$ (which follows from
the initial condition $\left\| \psi \right\| =1$), the stochastic
operator-functions $\widetilde{\mathrm{L}}^{k}\left( t\right) $, $\widetilde{%
\mathrm{H}}\left( t\right) $ defining the nonlinear filtering equation have
the limits 
\begin{equation*}
\widetilde{\mathrm{L}}^{k}=\mathrm{L}^{k}-\func{Re}\left\langle \psi |%
\mathrm{L}^{k}\psi \right\rangle ,\quad \widetilde{\mathrm{H}}=\mathrm{H}+%
\frac{i}{2}\left( \mathrm{L}_{k}^{\dagger }-\mathrm{L}_{k}\right) \func{Re}%
\left\langle \psi |\mathrm{L}^{k}\psi \right\rangle .
\end{equation*}
The corresponding nonlinear stochastic diffusion equation 
\begin{equation*}
\mathrm{d}\psi _{\omega }\left( t\right) +\widetilde{\mathrm{K}}\left(
t\right) \psi _{\omega }\left( t\right) \mathrm{d}t=\widetilde{\mathrm{L}}%
^{k}\left( t\right) \psi _{\omega }\left( t\right) \mathrm{d}\tilde{w}%
_{k}^{t}
\end{equation*}
was first derived in the general multi-dimensional density-matrix form 
\begin{equation*}
\mathrm{d}\rho _{\omega }+\left[ \mathrm{K}\rho _{\omega }+\rho _{\omega }%
\mathrm{K}^{\dagger }-\mathrm{L}^{k}\rho _{\omega }\mathrm{L}_{k}\right] 
\mathrm{d}t=\left[ \mathrm{L}^{k}\rho _{\omega }+\rho _{\omega }\mathrm{L}%
^{k\dagger }-\rho _{\omega }\mathrm{Tr}\left( \mathrm{L}^{k}+\mathrm{L}%
^{k\dagger }\right) \rho _{\omega }\right] \mathrm{d}\tilde{w}_{k}^{t}
\end{equation*}
for the renormalized density matrix $\rho _{\omega }=\rho \left( \omega
\right) /\mathrm{Tr}\rho \left( \omega \right) $ in \cite{Be88, Be90c} from
the microscopic reversible quantum stochastic unitary evolution models by
the quantum filtering method. The general microscopic derivation for the
case of multi-dimensional complete and incomplete measurements and solution
in the linear-Gaussian case is given in \cite{Be92b}. It has been recently
applied in quantum optics \cite{WiMi93, GoGr93, WiMi94, GoGr94} for the
description of counting, homodyne and heterodyne time-continuous
measurements introduced in \cite{BaBe}. It has been shown in \cite{ChSt92,
Kol95} that the nondemolition observation of such a particle is described by
filtering of the quantum noise which results in the continual collapse of
any initial wave packet to the Gaussian stationary one localized at the
position posterior expectation.

The connection between the above diffusive nonlinear filtering equation and
our linear decoherence Master-equation 
\begin{equation*}
\mathrm{d}\varrho \left( t\right) +\left[ \mathrm{K}\varrho \left( t\right)
+\varrho \left( t\right) \mathrm{K}^{\dagger }-\mathrm{L}^{k}\varrho \left(
t\right) \mathrm{L}_{k}\right] \mathrm{d}t=\left[ \mathrm{L}^{k}\varrho
\left( t\right) +\varrho \left( t\right) \mathrm{L}^{k\dagger }\right] 
\mathrm{d}w_{k}^{t},\quad \varrho \left( 0\right) =\rho ,
\end{equation*}
for the stochastic density operator $\varrho \left( t,\omega \right) $,
defining the output probability density $\mathrm{Tr}\varrho \left( t,\omega
\right) $, was well understood and presented in \cite{GPR90, GoGr94, GGH95}.
However it has also found an incorrect mathematical treatment in recent
Quantum State Diffusion theory \cite{Perc99} based on the case $\varepsilon
=0$ of our filtering equation (this particular nonlinear filtering equation
is empirically postulated as the `primary quantum state diffusion', and its
more fundamental linear version $\mathrm{d}\psi +\mathrm{K}\psi \mathrm{d}t=%
\mathrm{L}^{k}\psi \mathrm{d}w_{k}$ is `derived' in \cite{Perc99} simply by
dropping the non-linear terms without appropriate change of the probability
measures for the processes $\tilde{y}_{k}=\tilde{w}_{k}$ and $y_{k}=w_{k}$).
The most general stochastic decoherence Master equation is given in the
Appendix 3.

\subsubsection{The derivation of jumps and localizations.}

Here we give the solution of the quantum jump problem for the stochastic
model described by the equation (\ref{5.2}) in the case $\mathrm{C}^{\dagger
}\mathrm{C}\leq \mathrm{I}$, $\mathrm{R}=0$ which corresponds to the
Hamiltonian evolution between the jumps with energy operator $\mathrm{E}$,
and the jumps are cased only by the spontaneous decays or measurements. When 
$\mathrm{C}^{\dagger }\mathrm{C}=\mathrm{I}$, the quantum system certainly
decays at the random moment of the jump $\mathrm{d}n_{k}^{t}=1$ to one of
the $m$ products ending in the state $\mathrm{C}^{k}\psi /\left\| \mathrm{C}%
^{k}\psi \right\| $ from any state $\psi \in \frak{h}$ with the probability $%
\left\| \mathrm{C}^{k}\psi \right\| ^{2}$, or one of the measurement results 
$k=1,\ldots ,m$ localizing the product is gained at the random moment of the
spontaneous disintegration. The spontaneous evolution and its unitary
quantum stochastic dilation was studied in details in \cite{Be95}. When $%
\mathrm{C}^{\dagger }\mathrm{C}<\mathrm{I}$, the unstable system does not
decay to one of the measurable products with the probability $\left\| \psi
\right\| ^{2}-\left\| \mathrm{C}\psi \right\| ^{2}$, or no result is gained
at the jump. This unstable spontaneous evolution and its unitary quantum
stochastic dilation was considered in details for one dimensional case $m=1$
in \cite{BeSt00}.

First, we consider the operator $\mathrm{C}$ as a construction (or isometry
if $\mathrm{C}^{\dagger }\mathrm{C}=I$) from $\frak{h}$ into $\frak{h}%
\otimes \frak{f}$, where $\frak{f}=\mathbb{C}^{m}$. We dilate this $\mathrm{C%
}$ in the canonical way to the selfadjoint scattering operator 
\begin{equation}
\mathrm{S}=\left[ 
\begin{tabular}{ll}
$-\left( \mathrm{I}-\mathrm{C}^{\dagger }\mathrm{C}\right) ^{1/2}$ & $%
\mathrm{C}^{\dagger }$ \\ 
$\mathrm{C}$ & $\left( \mathrm{I}\otimes \hat{1}-\mathrm{CC}^{\dagger
}\right) ^{1/2}$%
\end{tabular}
\right] ,  \label{5.5}
\end{equation}
where $\mathrm{I}\otimes \hat{1}$ is the identity operator in $\frak{h}%
\otimes \frak{f}$, and $\mathrm{C}^{\dagger }$ is the adjoint construction $%
\frak{h}\otimes \frak{f}\rightarrow \frak{h}$ and $\mathrm{CC}^{\dagger }$
is a positive construction (orthoprojector $\mathrm{C}^{\dagger }\mathrm{C}=%
\mathrm{I}$) in this space. The unitarity $\mathrm{S}^{\dagger }=\mathrm{S}%
^{-1}$ of the operator $\mathrm{S}=\mathrm{S}^{\dagger }$ in the space $%
\frak{g}=\mathbb{C}\oplus \frak{f}=\mathbb{C}^{1+m}$ is easily proved as 
\begin{equation*}
\mathrm{S}^{2}=\left[ 
\begin{tabular}{ll}
$\left( \mathrm{I}-\mathrm{C}^{\dagger }\mathrm{C}\right) +\mathrm{C}%
^{\dagger }\mathrm{C}$ & $\mathrm{O}$ \\ 
$\mathrm{O}$ & \textrm{CC}$^{\dagger }+\left( \mathrm{I}\otimes \hat{1}-%
\mathrm{CC}^{\dagger }\right) $%
\end{tabular}
\right] =\left[ 
\begin{tabular}{ll}
\textrm{I} & \textrm{O} \\ 
\textrm{O} & $\mathrm{I}\otimes \hat{1}$%
\end{tabular}
\right]
\end{equation*}
by use of the identity $\left( \mathrm{I}\otimes \hat{1}-\mathrm{CC}%
^{\dagger }\right) ^{1/2}\mathrm{C}=\mathrm{C}\left( \mathrm{I}-\mathrm{C}%
^{\dagger }\mathrm{C}\right) ^{1/2}$. The operators $\mathrm{C}^{k}=\mathrm{S%
}_{0}^{k}$ are obtained from $\mathrm{S}$ as the partial matrix elements $%
\left( \mathrm{I}\otimes \langle k|\right) \mathrm{S}\left( \mathrm{I}%
\otimes |0\rangle \right) $ corresponding to the transition of the auxiliary
system (pointer) form the initial state $|0\rangle =1\oplus 0$ to one of the
measured orthogonal states $0\oplus |k\rangle $, $k=1,\ldots ,m$ in the
extended space $\frak{g}$.

Second, we consider two continuous seminfinite strings indexed by $s=\pm r$,
where $r>0$ is any real positive number (one can think that $s$ is the
coordinate on the right or left semistrings on the real line without the
point $s=0$). Let us denote by $\frak{g}^{\otimes }=\frak{g}_{1}\otimes 
\frak{g}_{2}\otimes \ldots \equiv \otimes \frak{g}_{n}$ the infinite tensor
product generated by $\otimes \chi _{n}$ with almost all components $\chi
_{n}\in \frak{g}_{n}=\frak{g}$ $\ $ equal to $\varphi =|0\rangle $ as they
were defined in the Section 5.2.\ We shall consider right-continuous
amplitudes $\Phi \left( \upsilon \right) $ with values in $\frak{g}^{\otimes
}$ for all infinite increasing sequences $\upsilon =\left\{
r_{1},r_{2},\ldots \right\} $, $r_{n-1}<r_{n}$ having a finite number $%
n^{t}\left( \upsilon \right) =\left| \upsilon \cap \lbrack 0,t)\right| $ of
elements $r\in \upsilon ^{t}$ in the finite intervals $[0,t)$ for all $t>0$
such that $f\left( r_{n}\right) \in \frak{g}_{n}$ if $\Phi \left( \upsilon
\right) =\otimes f\left( r_{n}\right) $. Let us also define the Hilbert
space $L_{\mu }^{2}\otimes \frak{g}^{\otimes }$ of the square-integrable
functions $\upsilon \mapsto \Phi \left( \upsilon \right) $ in the sense 
\begin{equation*}
\left\| \Phi \right\| ^{2}=\int \langle \Phi \left( \upsilon \right) |\Phi
\left( \upsilon \right) \rangle \mathrm{d}\mu _{\upsilon }^{\nu }\equiv 
\mathsf{M}\left( \left\| \Phi \left( \cdot \right) \right\| ^{2}\right)
<\infty
\end{equation*}
with respect to the standard Poisson measure $\mu _{\upsilon }^{\nu }=%
\mathsf{P}_{m}$ with the constant intensity $\nu $ on $\mathbb{R}_{+}$. In
other words we consider a countable number of the similar auxiliary systems
(Schr\"{o}dinger cats, bubbles or other pointers) described by the identical
state spaces $\frak{g}_{n}$ as independent randomly distributed on $\mathbb{R%
}_{+}$ with the average number $\nu $ on any unit interval of the string.
Let $\mathcal{G}_{\pm }=L_{\mu }^{2}\otimes \frak{g}_{\pm }^{\otimes }$ be
two copies of such space, one for the past, another for the future, and let $%
\mathcal{G}=L_{\mu }^{2}\otimes \mathbb{G}^{\otimes }$ with $\mathbb{G}=%
\frak{g}_{-}\otimes \frak{g}_{+}$ be canonically identified with the space $%
\mathcal{G}_{-}\otimes \mathcal{G}_{+}$ of square-integrable with $\mu
_{-}^{\nu }\otimes \mu _{+}^{\nu }$ functions $\Phi \left( \upsilon
_{-},\upsilon _{+}\right) $ having the values in $\frak{g}_{-}^{\otimes
}\otimes \frak{g}_{+}^{\otimes }$. In particular, the ground state described
by the constant function $\Phi ^{\circ }\left( \upsilon \right) =\otimes
\phi _{n}^{\circ }$, where $\phi ^{\circ }=\varphi \otimes \varphi $, is
identified with $\Phi _{-}^{\circ }\otimes \Phi _{+}^{\circ }$, where $\Phi
_{\pm }^{\circ }\left( \upsilon \right) =\otimes \varphi _{n}$. In order to
maintain the quantum causality we shall select a decomposable algebra $\frak{%
A}_{-}\otimes \frak{A}_{+}$ of the string with left string being classical,
described by the commutative algebra $\frak{A}_{-}=\mathcal{D}\left( 
\mathcal{G}_{-}\right) $, and the right string being quantum, described by
the commutant $\frak{A}_{+}$ of the Abelian algebra $L_{\mu }^{\infty }$ of
random scalar functions $f:\upsilon _{+}\mapsto \mathbb{C}$ represented by
multiplication operators $\hat{f}$ \ on $\mathcal{G}_{+}$. More precisely, $%
\frak{A}_{\pm }$ are the von Neumann algebras generated respectively on $%
\mathcal{G}_{\pm }$ by the functions $\upsilon _{\pm }\mapsto A\left(
\upsilon _{\pm }\right) $ with operator values $A\left( \upsilon _{+}\right)
\in \mathcal{A}_{+}^{\otimes }$ and $A\left( \upsilon _{-}\right) \in 
\mathcal{A}_{-}^{\otimes }$, where $\mathcal{A}_{+}^{\otimes }=\otimes 
\mathcal{B}\left( \frak{g}_{n}\right) $ is the algebra of all bounded
operators $\mathcal{B}\left( \frak{g}^{\otimes }\right) $ corresponding to $%
\mathcal{A}_{+}=\mathcal{B}\left( \mathbb{C}^{1+m}\right) $, and $\mathcal{A}%
_{-}^{\otimes }=\otimes \mathcal{D}\left( \frak{g}_{n}\right) $ is its
diagonal subalgebra $\mathcal{D}\left( \frak{g}^{\otimes }\right) $
corresponding to $\mathcal{A}_{-}=\mathcal{D}\left( \mathbb{C}^{1+m}\right) $%
. The canonical triple $\left( \mathcal{G},\frak{A},\Phi ^{\circ }\right) $
is the appropriate candidate for the dynamical dilation of quantum jumps in
the unstable system described by our spontaneous localization equation.

Third, we construct the time-continuous unitary group evolution which will
dynamically induce the spontaneous jumps in the interaction representation
at the boundary of the string. Let $T_{t}$ be the one parametric continuous
unitary group on $\mathcal{G}=\mathcal{G}_{-}\otimes \mathcal{G}_{+}$
describing the free evolution by right shifts $\Phi _{t}\left( \omega
\right) =\Phi \left( \omega -t\right) $ when $\Phi \left( \upsilon
_{-},\upsilon _{+}\right) $ is represented as $\Phi \left( \omega \right) $
with $\left( -\upsilon _{-}\right) \cup \left( +\upsilon _{+}\right) \subset 
\mathbb{R}$ for the two sided string parametrized by $\mathbb{R\supset
\omega }$. (As in the discrete time case the corresponding Hilbert space $%
\mathcal{G}$ for such $\Phi $ will be denote as $\mathcal{G}^{0]}\otimes 
\mathcal{G}_{0}$ with $\mathcal{G}_{0}=\mathcal{G}_{+}$ and $\mathcal{G}%
^{0]} $ obtained by the reflection of $\mathcal{G}_{-}$). This can be
written as 
\begin{equation*}
T_{t}\Phi \left( \upsilon _{-},\upsilon _{+}\right) =\Phi \left( \upsilon
_{-}^{t},\upsilon _{+}^{t}\right) ,
\end{equation*}
where $\upsilon _{\pm }^{t}=\pm \left[ \left( \left[ \left( -\upsilon
_{-}\right) \cup \left( +\upsilon _{+}\right) \right] -t\right) \cap \mathbb{%
R}_{\pm }\right] $. Let us denote the selfadjoint generator of this free
evolution on $\mathcal{G}$ by $P$ such that $T_{t}=\mathrm{e}^{-iPt/\hbar }$%
. This $P$ is the first order operator 
\begin{equation}
P\Phi \left( \upsilon _{-},\upsilon _{+}\right) =\frac{\hbar }{i}\left(
\sum_{r_{+}\in \upsilon _{+}}\frac{\partial }{\partial r_{+}}-\sum_{r_{-}\in
\upsilon _{-}}\frac{\partial }{\partial r_{-}}\right) \Phi \left( \upsilon
_{-},\upsilon _{+}\right)  \label{5.6}
\end{equation}
on $\mathbb{R}_{+}$ which is well defined on the differentiable functions $%
\Phi \left( \upsilon \right) $ which are constants for almost all $r\in
\upsilon $. It is selfadjoint on a natural domain $\mathcal{D}_{0}$ in $%
\mathcal{G}$ corresponding to the boundary condition $\Phi \left( 0\sqcup
\upsilon _{-},\upsilon _{+}\right) =\Phi \left( \upsilon _{-},0\sqcup
\upsilon _{+}\right) $ for all $\upsilon _{\pm }>0$, where 
\begin{equation*}
\quad \Phi \left( 0\sqcup \upsilon _{-},\upsilon _{+}\right)
=\lim_{r\searrow 0}\Phi \left( r\sqcup \upsilon _{-},\upsilon _{+}\right)
,\quad \Phi \left( \upsilon _{-},0\sqcup \upsilon _{+}\right)
=\lim_{r\searrow 0}\Phi \left( \upsilon _{-},r\sqcup \upsilon _{+}\right)
\end{equation*}
as induced by the continuity condition at $s=0$ on the whole $\mathbb{R}$.
Here $r\sqcup \upsilon $ is adding a point $r\notin \upsilon $ to the
ordered sequence $\upsilon =\left\{ 0,r_{1},\ldots \right\} $, and $0\sqcup
\upsilon _{\pm }$ can be formally treated as adding two zeroth $s=\pm 0$ to $%
\left\{ \pm r_{1},\pm r_{2},\ldots \right\} $ such that $-r<-0<+0<+r$ for
all $r>0$. Note that the Hamiltonian $-P$, not $+P$ corresponds to the right
free evolution in positive arrow of time in which the states $\Phi \left(
\upsilon _{+}\right) $ describe the incoming ``from the future'' quantized
waves, and the states $\Phi \left( \upsilon _{-}\right) $ describe the
outgoing ``to the past'' classical particles. This is a relativistic many
particle Dirac type Hamiltonian on the half of the line $\mathbb{R}$
corresponding to zero particle mass and the orientation of spin along $%
\mathbb{R}$, and it has unbounded from below spectrum. However as we showed
in \cite{Be00, Be01a}, the Heisenberg free evolution corresponding to this
shift can be obtained as a WKB approximation in the ultra-relativistic limit 
$\langle p\rangle \rightarrow \infty $ of any free evolution with a positive
single particle Hamiltonian, $\varepsilon \left( p\right) =\left| p\right| $
say. The unitary group evolution $U^{t}$ corresponding to the scattering
interaction at the boundary with the unstable system which has its own free
evolution described by the energy operator $\mathrm{E}$ can be obtained by
resolving the following generalized Schr\"{o}dinger equation 
\begin{equation}
\frac{\partial }{\partial t}\Psi ^{t}\left( \upsilon _{-},\upsilon
_{+}\right) +\frac{i}{\hbar }\left( \mathrm{E}\otimes I\right) \Psi
^{t}\left( \upsilon _{-},\upsilon _{+}\right) =\left( \sum_{r\in \upsilon
_{+}}\frac{\partial }{\partial r}-\sum_{r\in \upsilon _{-}}\frac{\partial }{%
\partial r}\right) \Psi ^{t}\left( t,\upsilon _{-},\upsilon _{+}\right)
\label{5.7}
\end{equation}
in the Hilbert space $\mathcal{H}=\frak{h}\otimes \mathcal{G}$, with the
following boundary condition 
\begin{equation}
\Psi ^{t}\left( 0\sqcup \upsilon _{-},\upsilon _{+}\right) =\mathrm{S}%
_{0}\Psi ^{t}\left( \upsilon _{-},0\sqcup \upsilon _{+}\right) ,\quad
\forall t>0,\upsilon _{\pm }>0.  \label{5.8}
\end{equation}
Here $\mathrm{S}_{0}$ is the boundary action of $\mathrm{S}$ defined by $%
\mathrm{S}_{0}\Psi \left( 0\sqcup \upsilon \right) =\mathrm{S}\left[ \psi
\otimes \varphi \right] \otimes \Phi \left( \upsilon \right) $ on the
products $\Psi \left( 0\sqcup \upsilon \right) =\psi \otimes \varphi \otimes
\Phi \left( \upsilon \right) $. Due to the unitarity of the scattering
matrix $\mathrm{S}$ this simply means $\left\| \Psi \left( 0\sqcup \upsilon
_{-},\upsilon _{+}\right) \right\| =\left\| \Psi \left( \upsilon
_{-},0\sqcup \upsilon _{+}\right) \right\| $, that is Dirac current has zero
value at the boundary\ $r=0$. Note that this natural Dirac boundary
condition corresponds to an unphysical discontinuity condition $\Psi \left(
-0\sqcup \omega \right) =\left( \mathrm{S}\otimes I\right) \Psi \left(
+0\sqcup \omega \right) $ at the origin $s=0$ when the doubled semi-string
is represented as the two-sided string on $\mathbb{R}$.

Fourth, we have to solve this equation, or at least to prove that it has a
unitary solution which induces the injective Heisenberg dynamics on the
algebra $\frak{B}=\mathcal{B}\left( \frak{h}\right) \otimes \frak{A}$ of the
combined system with the required properties $U^{-t}\frak{B}U^{t}\subseteq 
\frak{B}$. The latter can be done by proof that the Hamiltonian in the
equation (\ref{5.7}) is selfadjoint on a natural domain corresponding to the
perturbed boundary condition (\ref{5.8}). It has been done by finding the
appropriate domain for the perturbed Hamiltonian in the recent paper \cite
{BeKo01}. However we need a more explicit construction of the time
continuous resolving operators $U^{t}:\Psi \mapsto \Psi ^{t}$ for the
equation (\ref{5.7}). To obtain this we note that the this equation apart
from the boundary condition coincides with the free evolution equation given
by the Hamiltonian $-P$ up to the free unitary transformation $\mathrm{e}^{-i%
\mathrm{E}t/\hslash }$ in the initial space $\frak{h}$. This implies that
apart from the boundary the evolution $U^{t}$ coincides with $\mathrm{e}^{-i%
\mathrm{E}t/\hslash }\otimes T_{-t}$ such that the interaction evolution $%
U\left( t\right) =T_{t}U^{t}$, where $T_{t}$ is the shift extended trivially
onto the component $\frak{h}$ of $\mathcal{H}=\mathcal{G}^{0]}\otimes \frak{h%
}\otimes \mathcal{G}_{0}$, is adapted in the sense 
\begin{equation*}
U\left( t\right) \left( \Phi ^{0]}\otimes \psi \otimes \varphi
_{0}^{t]}\otimes \Phi _{t}\right) =\Phi ^{0]}\otimes \psi \left( t\right)
\otimes \Phi _{t},
\end{equation*}
where $\psi \left( t\right) =\mathrm{W}_{0}^{t]}\left( \psi \otimes
f_{0}^{t]}\right) \in \frak{h}\otimes \mathcal{G}_{0}^{t]}$ for all $\psi
\in \frak{h}$ and $f_{0}^{t]}\in \mathcal{F}_{0}^{t]}$. Here we use the
tensor product decomposition $\mathcal{G}_{0}=\mathcal{F}_{0}^{t]}\otimes 
\mathcal{G}_{t}$, where $\mathcal{F}_{0}^{t]}$ is Fock space generated by
the products $f_{0}^{t]}\left( \upsilon \right) =\otimes _{r\in \upsilon
}f\left( r\right) $ with finite $\upsilon \subset \lbrack 0,t)$ and $f\left(
r\right) \in \frak{g}$ (we use the representation of $\mathcal{H}$ as the
Hilbert space $\mathcal{G}^{0]}\otimes \frak{h}\otimes \mathcal{G}_{0}$ for
the two sided string on $\mathbb{R}$ with the measured system inserted at
the origin $s=0$, and the notations $\mathcal{G}_{0}=\mathcal{G}_{+}$ , $%
\mathcal{G}^{0]}=\mathcal{G}_{-}$ , identifying $s=+r$ with $z=\left|
s\right| $ for all $r>0$, and $s=-r$ with $z=-\left| s\right| $ including $%
r=0$). Moreover, in this interaction picture $U\left( t\right) $ is
decomposable, 
\begin{equation*}
\left[ U\left( t\right) \Psi \right] \left( \omega \right) =U\left( t,\omega
\right) \Psi \left( \omega \right) ,\quad U\left( t,\omega \right)
=I^{0]}\otimes \mathrm{W}_{0}^{t]}\left( \omega _{0}^{t]}\right) \otimes
I_{t}
\end{equation*}
as decomposable is each $\mathrm{W}_{0}^{t]}$ into the unitary operators $%
\mathrm{W}_{0}^{t]}\left( \upsilon \right) $ in $\frak{h}\otimes \mathcal{F}%
_{0}^{t]}\left( \upsilon \right) $ with $\mathcal{F}_{0}^{t]}\left( \upsilon
\right) =\otimes _{n=1}^{\left| \upsilon \right| }\frak{g}_{n}$ for any
finite $\upsilon \subset \lbrack 0,t)$. The dynamical invariance $%
U^{-t}BU^{t}\in \frak{B}$ of the algebra $\frak{B}=\mathcal{B}\left( \frak{h}%
\right) \otimes \frak{A}$ for any positive $t>0$ under the Heisenberg
transformations of the operators $B\in \frak{B}$ induced by $%
U^{t}=T_{-t}U\left( t\right) $ simply follows from the right shift
invariance $T_{t}\mathcal{B}T_{-t}\subseteq \frak{B}$ of this algebra and $%
U\left( t,\omega \right) ^{\dagger }\frak{B}\left( \omega \right) U\left(
t,\omega \right) =\frak{B}\left( \omega \right) $ for each $\omega $ due to
unitarity of $\mathrm{W}_{0}^{t]}\left( \upsilon \right) $ and the
simplicity of the local algebras of future $\mathcal{B}_{0}^{t]}\left(
\upsilon \right) =\mathcal{B}\left( \frak{h}\right) \otimes \mathcal{A}%
_{+}^{\otimes \left| \upsilon \right| }$ corresponding to $\mathcal{A}_{+}=%
\mathcal{B}\left( \frak{g}\right) $.

And finally we can find the nondemolition processes $N_{k}^{t}$, $k=1,\ldots
,m$ which count the spontaneous disintegrations of the unstable system to
one of the measured products $k=1,\ldots ,m$, and with $N_{0}^{t}$
corresponding to the unobserved jumps. These are given on the space $%
\mathcal{G}_{-}$ as the sums of the orthoprojectors $|k\rangle \langle k|\in 
\frak{g}$ corresponding to the arriving particles at the times $r_{n}\in
\upsilon _{-}$ up to the time $t:$ 
\begin{equation}
N_{k}^{t}\left( \upsilon _{-},\upsilon _{+}\right) =\mathrm{I}\otimes
\sum_{n=1}^{n^{t}\left( \upsilon _{-}\right) }\mathrm{I}_{0}^{n-1]}\otimes
\left( |k\rangle \langle k|\otimes \mathrm{I}_{+}\right) \otimes I_{n}.
\label{5.10}
\end{equation}
Due to commutativity of this compatible family with all operators $\mathrm{B}%
\otimes I$ of the unstable system, the Heisenberg processes $N_{k}\left(
t\right) =U^{-t}N_{k}^{t}U^{t}$ satisfies the nondemolition causality
condition. The independent increment quantum nondemolition process with zero
initial expectations corresponding to $y_{k}^{t}=m_{k}^{t}$ in the standard
jump-decoherence equation (\ref{5.1}) then is given by 
\begin{equation*}
Y_{k}^{t}=\nu ^{-1/2}\left( N_{k}^{t}-\nu t\right) =X_{k}^{t}-\lambda t,
\end{equation*}
where $X_{k}^{t}=\lambda ^{-1/2}N_{k}^{t}$ with the coupling constant $%
\lambda =\nu ^{1/2}$. Hence the quantum jumps, decoherence and spontaneous
localization are simply derived from this dynamical model as the results of
inference, or quantum filtering without the projection postulate by simple
conditioning. The statistical equivalence of the nondemolition countings $%
N_{k}^{t}$ in the Schr\"{o}dinger picture $\Psi ^{t}=U^{t}\left( \psi
\otimes \Phi ^{\circ }\right) $ of this continuous unitary evolution model
with fixed initial ground state $\Phi ^{\circ }=\Phi _{-}^{\circ }\otimes
\Phi _{+}^{\circ }$ and the stochastic quantum reduction model based on the
spontaneous jump equation (\ref{5.2}) for an unstable system corresponding
to $\mathrm{R}=0$ and $\mathrm{C}^{\dagger }\mathrm{C}\leq \mathrm{I}$ was
proved in the interaction representation picture in \cite{Be95, BeMe96}.

Note that before the interaction the probability to measure any of these
products is zero in the initial states $\Phi ^{\circ }\left( \upsilon
_{-},\upsilon _{+}\right) =\otimes \phi _{n}^{\circ }$ as all $\phi
_{n}^{\circ }=|0\rangle \otimes |0\rangle $. The maximal decreasing
orthoprojector 
\begin{equation*}
E_{t}\left( \upsilon _{-},\upsilon _{+}\right) =\mathrm{I}\otimes
E_{1}\left( 0\right) \otimes \ldots \otimes E_{n^{t}\left( \upsilon
_{-}\right) }\left( 0\right) \otimes I_{n^{t}\left( \upsilon _{-}\right)
},\quad E_{n}=|0\rangle \langle 0|\otimes \mathrm{I}_{+},
\end{equation*}
which is orthogonal to all product countings $N_{k}^{t}$, $k=1,\ldots ,m$,
is called the survival process for the unstable system.

Thus, the quantum jumps problem has been solved as the time-continuous
quantum boundary-value problem formulated as

\emph{Given a reduction family }$V\left( t,\omega \right) =V\left(
t,m_{0}^{t]}\right) ,t\in \mathbb{R}_{+}$ \emph{of isometries on} $\frak{h}$%
\ \emph{into} $\frak{h}\otimes L_{\mu }^{2}$\emph{\ resolving the quantum
jump equation (\ref{5.2}) with respect to the input probability measure }$%
\mu =\mathsf{P}_{m}$\emph{\ for the standard Poisson noises }$m_{k}^{t}$%
\emph{\ find a triple }$\left( \mathcal{G},\frak{A},\Phi \right) $ \emph{%
consisting of a Hilbert space} $\mathcal{G=G}_{-}\otimes \mathcal{G}_{+}$ 
\emph{embedding the Poisson Hilbert space }$L_{\mu }^{2}$\emph{\ by an
isometry into }$\mathcal{G}_{+}$\emph{, an algebra }$\frak{A=A}_{-}\otimes 
\frak{A}_{+}$ \emph{on $\mathcal{G}$ with an Abelian subalgebra }$\frak{A}%
_{-}$\emph{\ generated by a compatible continuous family }$Y_{-\infty
}^{0]}=\left\{ Y_{k}^{s},k=1,\ldots ,d,s\leq 0\right\} $\emph{\ } \emph{of
observables (beables) on $\mathcal{G}_{-}$, and a state-vector }$\Phi
^{\circ }=\Phi _{-}^{\circ }\otimes \Phi _{+}^{\circ }\in \mathcal{G}$\emph{%
\ such that there exist a time continuous unitary group }$U^{t}$\emph{\ on }$%
\mathcal{H=}\frak{h}\otimes \mathcal{G}$\emph{\ inducing a semigroup of
endomorphisms }$\frak{B}\ni B\mapsto U^{-t}BU^{t}\in \frak{B}$ \emph{on the
product algebra $\frak{B=}\mathcal{B}\left( \frak{h}\right) \otimes \frak{A}$%
, with } 
\begin{eqnarray*}
\pi ^{t}\left( \hat{g}_{-t}\otimes \mathrm{B}\right) &:&=\left( \mathrm{I}%
\otimes \Phi ^{\circ }\right) ^{\dagger }U^{-t}\left( \mathrm{B}\otimes
g_{-t}\left( Y_{-t}^{0]}\right) \right) U^{t}\left( \mathrm{I}\otimes \Phi
^{\circ }\right) \\
&=&\int g\left( y_{0}^{t]}\right) V\left( t,y_{0}^{t]}\right) ^{\dagger }%
\mathrm{B}V\left( t,y_{0}^{t]}\right) \mathrm{d}\mathsf{P}_{m}\equiv \mathsf{%
M}\left[ gV\left( t\right) ^{\dagger }\mathrm{B}V\left( t\right) \right]
\end{eqnarray*}
\emph{\ for any }$\mathrm{B}\in \mathcal{B}\left( \frak{h}\right) $ \emph{%
and any operator }$\hat{g}_{-t}=\hat{g}_{-t}\left( Y_{-t}^{0]}\right) \in 
\frak{C}$ \emph{represented as the shifted functional }$\hat{g}_{-t}\left(
y_{-t}^{0]}\right) =g\left( y_{0}^{t]}\right) $\emph{\ of }$%
Y_{-t}^{-0]}=\left\{ Y_{\cdot }^{s}:s\in (-t,0]\right\} $ \emph{on $\mathcal{%
G}$ by any measurable functional }$g$\emph{\ of }$y_{0}^{t]}=\left\{
y_{\cdot }^{r}:r\in (0,t]\right\} $\emph{\ with arbitrary }$t>0.$

Note that despite strong continuity of the unitary group evolutions $T_{t}$
and $U^{t}$, the interaction evolution $U\left( t,\omega \right) $ is
time-discontinuous for each $\omega $. It is defined as $U\left( t\right)
=I^{0]}\otimes W\left( t\right) $ for any positive $t$ by the stochastic
evolution $W\left( t\right) =\mathrm{W}_{0}^{t]}\otimes I_{t}$ resolving the
Schr\"{o}dinger unitary jump equation \cite{Be95, BeMe96} 
\begin{equation}
\mathrm{d}\Psi _{0}\left( t,\upsilon \right) +\frac{i}{\hbar }\left( \mathrm{%
E}\otimes I\right) \Psi _{0}\left( t,\upsilon \right) \mathrm{d}t=\left(
S-I\right) _{t}\left( \upsilon \right) \Psi _{0}\left( t,\upsilon \right) 
\mathrm{d}n^{t}\left( \upsilon \right)  \label{5.9}
\end{equation}
on $\mathcal{H}_{0}=\frak{h}\otimes \mathcal{G}_{0}$ as $\Psi _{0}\left(
t,\upsilon \right) =W\left( t,\upsilon \right) \Psi _{0}$. Here $L_{t}\left(
\upsilon \right) =\mathrm{L}_{n^{t}\left( \upsilon \right) }\otimes I_{t}$
is the adapted generator $\mathrm{L}=\mathrm{S}-\mathrm{I}$ which is applied
only to the system and the $n$-th particle with the number $n=n^{t}\left(
\upsilon \right) $ on the right semistring as the operator $\mathrm{L}_{n}=%
\mathrm{T}_{0}^{n]\dagger }\left( \mathrm{I}_{0}^{n-1]}\otimes \mathrm{L}%
_{0}\right) \mathrm{T}_{0}^{n]}$ obtained from $\mathrm{L}_{0}=\mathrm{TLT}%
^{\dagger }$ by the transposition operator $\mathrm{T}\left( \psi \otimes
\chi \right) =\chi \otimes \psi $ generating $\mathrm{T}_{0}^{n]}\left( \psi
\otimes \chi _{0}^{n]}\right) =\chi _{0}^{n]}\otimes \psi $ on $\frak{h}%
\otimes \frak{g}_{0}^{n]}$ by the recurrency 
\begin{equation*}
\mathrm{T}_{0}^{n]}=\left( \mathrm{I}_{0}^{n-1]}\otimes \mathrm{T}\right)
\left( \mathrm{T}_{0}^{n-1]}\otimes \mathrm{I}\right) ,\quad \mathrm{T}%
_{0}^{0]}=\mathrm{I}.
\end{equation*}
The\ operator $\mathrm{W}_{0}^{t]}$ can be explicitly found from the
equivalent stochastic integral equation 
\begin{equation*}
\mathrm{W}_{0}^{t]}\left( \upsilon \right) =\left( \mathrm{e}^{-i\mathrm{E}%
t/\hbar }\otimes \mathrm{I}_{0}^{t]}\right) +\sum_{r\in \upsilon \cap
\{0,t)}\left( \mathrm{e}^{i\mathrm{E}\left( r-t\right) /\hbar }\otimes 
\mathrm{I}_{0}^{t]}\right) \left( \mathrm{S}-\mathrm{I}\right)
_{n^{r}}\left( \mathrm{W}_{0}^{r]}\otimes \mathrm{I}_{r}^{t]}\right)
\end{equation*}
with $\mathrm{W}_{0}^{0]}=\mathrm{I}$. Indeed, the solution to this integral
equation can be written for each $\upsilon $ in terms of the finite
chronological product of unitary operators as in the discrete time case
iterating the following recurrency equation 
\begin{equation*}
\mathrm{W}_{0}^{t]}\left( \upsilon \right) =\mathrm{e}^{-i\mathrm{E}t/\hbar }%
\mathrm{S}\left( t_{n}\right) \left( \mathrm{W}_{0}^{t_{n}]}\left( \upsilon
\right) \otimes \mathrm{I}\right) ,\quad \mathrm{W}_{0}^{0]}\left( \upsilon
\right) =\mathrm{I},
\end{equation*}
where $\mathrm{S}\left( t_{n}\right) =\mathrm{e}^{i\mathrm{E}t_{n}/\hbar }%
\mathrm{S}_{n}\mathrm{e}^{-i\mathrm{E}t_{n}/\hbar }$ with $n=n_{t}\left(
\upsilon \right) $. From this we can also obtain the corresponding explicit
formula \cite{Be95, BeMe96} 
\begin{equation*}
V\left( t,\upsilon _{\cdot }\right) =\mathrm{e}^{-i\mathrm{E}t/\hbar }%
\mathrm{C}^{k_{n}}\left( t_{n}\right) \left( V\left( t_{n},\upsilon \right)
\otimes \mathrm{I}\right) ,\quad \mathrm{V}\left( \upsilon _{\cdot }\right) =%
\mathrm{I},
\end{equation*}
resolving the reduced stochastic equation (\ref{5.2}) as $\psi \left(
t\right) =V\left( t\right) \psi $ for $\mathrm{R}=\mathrm{O}$ and any
sequence $\upsilon _{\cdot }$ of pairs $\left( r_{n},k_{n}\right) $ with
increasing $\left\{ r_{n}\right\} $, where $n=n^{t}\left( \upsilon _{\cdot
}\right) $ is the maximal number in $\left\{ r_{n}\right\} \cap \lbrack 0,t)$%
.

\subsection{Continuous Trajectories and State Diffusion}

\begin{quote}
\textit{Quantum mechanics itself, whatever its interpretation, does not
account for the transition from `possible to the actual'}\textrm{\ - }%
Heisenberg.
\end{quote}

Schr\"{o}dinger believed that all quantum problems including the
interpretation of measurement should be formulated in continuous time in the
form of differential equations. He thought that the measurement problem
would have been resolved if quantum mechanics had been made consistent with
relativity theory and the time had been treated appropriately. However
Einstein and Heisenberg did not believe this, each for his own reasons.

Although Schr\"{o}dinger did not succeed to find the `true Schr\"{o}dinger
equation' to formulate the boundary value problem for such transitions, the
analysis of the phenomenological stochastic models for quantum spontaneous
jumps in the unstable systems proves that Schr\"{o}dinger was right. However
Heisenberg was also right as in order to make an account for these
transitions by filtering the actual past events simply as it is done in
classical statistics, the corresponding ultrarelativistic Dirac type
boundary value problem (\ref{5.7}), (\ref{5.8}) must be supplemented by
future-past superselection rule for the total algebra as it follows from the
nondemolition causality principle \cite{Be94}. This principle cannot be
formulated in quantum mechanics as it involves infinitely many degrees of
freedom, and it has not been formulated even in the orthodox quantum field
theory.

Here we shall deal with quantum noise models which allow to formulate the
most general stochastic decoherence equation which was derived in \cite
{Be95a}. We shall start with a simple quantum noise model and show that it
allows to prove the `true Heisenberg principle' in the form of an
uncertaincy relation for measurement errors and dynamical perturbations. The
discovery of quantum thermal noise and its white-noise approximations lead
to a profound revolution not only in modern physics but also in contemporary
mathematics comparable with the discovery of differential calculus by Newton
(for a feature exposition of this, accessible for physicists, see \cite
{Gar91}, the complete theory, which was mainly developed in the 80's \cite
{Be80, HuPa84, GaCo85, Be88a}, is sketched in the Appendix 2)

\subsubsection{The true Heisenberg principle}

The first time continuous solution of the quantum measurement problem \cite
{Be80} was motivated by analogy with the classical stochastic filtering
problem which obtains the prediction of future for an unobservable dynamical
process $x\left( t\right) $ by time-continuous measuring of another,
observable process $y\left( t\right) $. Such problems were first considered
by Wiener and Kolmogorov who found the solutions in the form of a\ causal
spectral filter for a liner estimate $\hat{x}\left( t\right) $ of $x\left(
t\right) $ which is optimal only in the stationary Gaussian case. The
complete solution of this problem was obtained by Stratonovich \cite{Str66}
in 1958 who derived a stochastic filtering equation giving the posterior
expectations $\hat{x}\left( t\right) $\ of $x\left( t\right) $ in the
arbitrary Markovian pair $\left( x,y\right) $. This was really a break
through in the statistics of stochastic processes which soon found many
applications, in particular for solving the problems of stochastic control
under incomplete information (it is possible that this was one of the
reasons why the Russians were so successful in launching the rockets to the
Moon and other planets of the Solar system in 60s).

If $X\left( t\right) $ is an unobservable Heisenberg process, or vector of
such processes $X_{k}\left( t\right) $, $k=1,\ldots ,d$ which might have
even no prior trajectories as the Heisenberg coordinate processes of a
quantum particle say, and $Y\left( t\right) $ is an actual observable
quantum processes, i.e. a sort of Bell's beable describing the vector
trajectory $y\left( t\right) $ of the particle in a cloud chamber say, why
don't we find the posterior trajectories by deriving and solving a filtering
equation for the posterior expectations $\hat{x}\left( t\right) $ of $%
X\left( t\right) $ or any other \ function of $X\left( t\right) $, defining
the posterior trajectories $x\left( t,y_{0}^{t]}\right) $ in the same way as
we do it in the classical case? If we had a dynamical model in which such
beables existed as a nondemolition process, we could solve this problem
simply by conditioning as the statistical inference problem, predicting the
future knowing a history, i.e. a particular trajectory $y\left( r\right) $
up to the time $t$. This problem was first considered and solved by finding
a nontrivial quantum stochastic model for the Markovian Gaussian pair $%
\left( X,Y\right) $. It corresponds to a quantum open linear system with
linear output channel, in particular for a quantum oscillator matched to a
quantum transmission line \cite{Be80, Be85}. By studying this example, the
nondemolition condition 
\begin{equation*}
\left[ X_{k}\left( s\right) ,Y\left( r\right) \right] =0,\quad \text{ }\left[
Y\left( s\right) ,Y\left( r\right) \right] =0\quad \forall r\leq s
\end{equation*}
was first found, and this allowed the solution in the form of the causal
equation for $x\left( t,y_{0}^{t]}\right) =\left\langle X\left( t\right)
\right\rangle _{y_{0}^{t]}}$.

Let us describe this exact dynamical model of the causal nondemolition
measurement first in terms of quantum white noise for a quantum
nonrelativistic particle of mass $m$ which is conservative if not observed,
in a potential field $\phi $. But we shall assume that this particle is
under a time continuous indirect observation which is realized by measuring
of its Heisenberg position operators $Q^{k}\left( t\right) $ with additive
random errors $e^{k}\left( t\right) :$%
\begin{equation*}
Y^{k}\left( t\right) =Q^{k}\left( t\right) +e^{k}\left( t\right) ,\quad
k=1,\ldots ,d.
\end{equation*}
We take the simplest statistical model for the error process $e\left(
t\right) $, the white noise model (the worst, completely chaotic error),
assuming that it is a classical Gaussian white noise given by the first
momenta 
\begin{equation*}
\left\langle e^{k}\left( t\right) \right\rangle =0,\quad \left\langle
e^{k}\left( s\right) e^{l}\left( r\right) \right\rangle =\sigma
_{e}^{2}\delta \left( s-r\right) \delta _{l}^{k}.
\end{equation*}
The components of measurement vector-process $Y\left( t\right) $ should be
commutative, satisfying the causal nondemolition condition with respect to
the noncommutative process $Q\left( t\right) $ (and any other Heisenberg
operator-process of the particle), this can be achieved by perturbing the
particle Newton-Erenfest equation: 
\begin{equation*}
m\frac{\mathrm{d}^{2}}{\mathrm{d}t^{2}}Q\left( t\right) +\nabla \phi \left(
Q\left( t\right) \right) =f\left( t\right) .
\end{equation*}
Here $f\left( t\right) $ is vector-process of Langevin forces $f_{k}$
perturbing the dynamics due to the measurement, which are also assumed to be
independent classical white noises 
\begin{equation*}
\left\langle f_{k}\left( t\right) \right\rangle =0,\quad \left\langle
f_{k}\left( s\right) f_{l}\left( r\right) \right\rangle =\sigma
_{f}^{2}\delta \left( s-r\right) \delta _{l}^{k}.
\end{equation*}
In classical measurement and filtering theory the white noises $e\left(
t\right) ,f\left( t\right) $ are usually considered independent, and the
intensities $\sigma _{e}^{2}$ and $\sigma _{f}^{2}$ can be arbitrary, even
zeros, corresponding to the ideal case of the direct unperturbing
observation of the particle trajectory $Q\left( t\right) $. However in
quantum theory corresponding to the standard commutation relations 
\begin{equation*}
Q\left( 0\right) =\mathrm{Q},\quad \frac{\mathrm{d}}{\mathrm{d}t}Q\left(
0\right) =\frac{1}{m}\mathrm{P},\quad \left[ \mathrm{Q}^{k},\mathrm{P}_{l}%
\right] =i\hbar \delta _{l}^{k}\mathrm{I}
\end{equation*}
the particle trajectories do not exist such that the measurement error $%
e\left( t\right) $ and perturbation force $f\left( t\right) $ should satisfy
a sort of uncertainty relation. This ``true Heisenberg principle'' had never
been mathematically proved before the discovery \cite{Be80} of quantum
causality in the form of nondemolition condition of commutativity of $%
Q\left( s\right) $, as well as any other process, the momentum $P\left(
t\right) =m\dot{Q}\left( t\right) $ say, with all $Y\left( r\right) $ for $%
r\leq s$. As we showed first in the linear case \cite{Be80, Be85}, and later
even in the most general case \cite{Be92b}, these conditions are fulfilled
if and only if $e\left( t\right) $ and $f\left( t\right) $ satisfy the
canonical commutation relations 
\begin{equation*}
\left[ e^{k}\left( r\right) ,e^{l}\left( s\right) \right] =0,\;\left[
e^{k}\left( r\right) ,f_{l}\left( s\right) \right] =\frac{\hbar }{i}\delta
\left( r-s\right) \delta _{l}^{k},\;\left[ f_{k}\left( r\right) ,f_{l}\left(
s\right) \right] =0.
\end{equation*}
From this it follows, in a similar way as it was done in the Sec. 3.1, that
the pair $\left( e,f\right) $ satisfy the uncertainty relation $\sigma
_{e}\sigma _{f}\geq \hbar /2$. This inequality constitutes the precise
formulation of the true Heisenberg principle for the square roots $\sigma
_{e}$ and $\sigma _{f}$ of the intensities of error $e$ and perturbation $f$%
: they are inversely proportional with the same coefficient of
proportionality, $\hbar /2$, as for the pair $\left( \mathrm{Q},\mathrm{P}%
\right) $. Note that the canonical pair $\left( e,f\right) $ called quantum
white noise cannot be considered classically, despite the fact that each
process $e$ and $f$ separately can. This is why we need a quantum-field
representation for the pair $\left( e,f\right) $, and the corresponding
quantum stochastic calculus. Thus, a generalized matrix mechanics for the
treatment of quantum open systems under continuous nondemolition observation
and the true Heisenberg principle was discovered 20 years ago only after the
invention of quantum white noise in \cite{Be80}. The nondemolition
commutativity of $Y\left( t\right) $ with respect to the Heisenberg
operators of the open quantum system was later rediscovered for the output
of quantum stochastic fields in \cite{GaCo85}.

Let us outline the exact quantum stochastic model \cite{Be88, Be92b} for a
quantum particle of mass $m$ in a potential $\phi $ under indirect
observation of the positions $Q^{k}$ by measuring $Y_{k}$. We define the
output process as a quantum stochastic Heisenberg transformation $%
Y_{k}^{t}=W\left( t\right) ^{\dagger }\left( \mathrm{I}\otimes \hat{y}%
_{k}^{t}\right) W\left( t\right) $ for a time-continuous quantum stochastic
unitary evolution $W\left( t\right) $ in a similar way as we did in the
discrete case, see the Appendix 2. It has been shown in \cite{Be88, Be92b}
that $W\left( t\right) $ is the resolving family for an appropriate \emph{%
quantum stochastic Schr\"{o}dinger equation }\ (See the equation (\ref{6.4})
below). It induces the following quantum stochastic Heisenberg output
equation 
\begin{equation}
\text{\quad }\mathrm{d}Y_{k}^{t}=2\lambda Q^{k}\left( t\right) \mathrm{d}t+%
\mathrm{d}\hat{w}_{k}^{t}\equiv X_{k}\left( t\right) \mathrm{d}t+\mathrm{d}%
\hat{w}_{k}^{t},  \label{6.1}
\end{equation}
where $\lambda $ is a coupling constant, or a diagonal matrix $\lambda =%
\left[ \lambda _{k}\delta _{k}^{i}\right] $ defining different accuracies of
an indirect measurement it time of $\mathrm{Q}^{k}$. Here $X\left( t\right)
=W\left( t\right) ^{\dagger }\left( \mathrm{X}\otimes I_{0}\right) W\left(
t\right) $ are the system Heisenberg operators for $\mathrm{X}_{k}=2\left(
\lambda Q\right) ^{k}$, $I_{0}$ is the identity operator in the Fock space $%
\mathcal{F}_{0}$, and $\hat{w}_{k}^{t}\equiv y_{k}^{t}$, $k=1,\ldots ,d$ are
the standard independent Wiener processes $w_{k}^{t}$ represented as the
operators $\hat{w}_{k}^{t}=A_{-}^{k}\left( t\right) +A_{k}^{+}\left(
t\right) $ on the Fock vacuum vector $\delta _{\varnothing }\in \mathcal{F}%
_{0}$ such that $w_{k}^{t}\simeq \hat{w}_{k}^{t}\delta _{o}$ (See the
notations and more about the quantum stochastic calculus in Fock space in
the Appendix 3). This model coincides with the signal plus noise model given
above if 
\begin{equation*}
\hat{e}^{k}\left( t\right) =\frac{1}{2}\left( a_{k}^{+}+a_{-}^{k}\right)
\left( t\right) =\frac{1}{2\lambda _{k}}\frac{\mathrm{d}w_{k}^{t}}{\mathrm{d}%
t},
\end{equation*}
where $a_{k}^{+}\left( t\right) ,a_{-}^{k}\left( t\right) $ are the
canonical bosonic creation and annihilation field operators, 
\begin{equation*}
\left[ a_{k}^{+}\left( s\right) ,a_{l}^{+}\left( t\right) \right] =0,\;\left[
a_{-}^{k}\left( s\right) ,a_{l}^{+}\left( t\right) \right] =\delta
_{l}^{k}\delta \left( t-s\right) ,\;\left[ a_{-}^{k}\left( s\right)
,a_{-}^{l}\left( t\right) \right] =0,
\end{equation*}
defined as the generalized derivatives of the standard quantum Brownian
motions $A_{k}^{+}\left( t\right) $ and $A_{-}^{k}\left( t\right) $ in Fock
space $\mathcal{F}_{0}$. It was proved \ in \cite{Be88, Be92b} that $%
Y_{k}^{t}$ is a commutative nondemolition process with respect to the system
Heisenberg coordinate and momentum $P\left( t\right) =W\left( t\right)
^{\dagger }\left( \mathrm{P}\otimes I\right) W\left( t\right) $ processes if
they are perturbed by\ independent Langevin forces $f_{k}\left( t\right) $
of intensity $\tau _{k}^{2}=\left( \lambda _{k}\hbar \right) ^{2}$, the
generalized derivatives of $f_{k}^{t}\simeq \hat{f}_{k}^{t}\delta
_{\varnothing }$ times $\lambda _{k}$, where $\hat{f}_{k}^{t}=i\hbar \left(
A_{-}^{k}-A_{k}^{+}\right) \left( t\right) $: 
\begin{equation}
\mathrm{d}P_{k}\left( t\right) +\phi _{k}^{\prime }\left( Q\left( t\right)
\right) \mathrm{d}t=\lambda _{k}\mathrm{d}\hat{f}_{k}^{t},\quad P_{k}\left(
t\right) =m\frac{\mathrm{d}}{\mathrm{d}t}Q^{k}\left( t\right) .  \label{6.2}
\end{equation}
Note that the quantum error operators $\hat{w}_{k}^{t}$ commute, but they do
not commute with the perturbing quantum force operators $\hat{f}_{k}^{t}$ in
Fock space due to the multiplication table 
\begin{equation*}
\left( \mathrm{d}\hat{w}_{k}\right) ^{2}=I\mathrm{d}t,\quad \mathrm{d}\hat{f}%
_{k}\mathrm{d}\hat{w}_{l}=i\hbar I\delta _{l}^{k}\mathrm{d}t,\quad \mathrm{d}%
\hat{w}_{k}\mathrm{d}\hat{f}_{l}=-i\hbar I\delta _{l}^{\kappa }\mathrm{d}%
t,\quad \left( \mathrm{d}\hat{f}_{k}\right) ^{2}=\hbar ^{2}I\mathrm{d}t\text{%
.}
\end{equation*}
This corresponds to the canonical commutation relations for the renormalized
derivatives $\hat{w}_{k}\left( t\right) $ and $\hat{f}_{l}\left( t\right) $,
so that the true Heisenberg principle is fulfilled at the boundary $\sigma
_{k}\tau _{k}=\hbar /2$. Thus our quantum stochastic model of nondemolition
observation is the minimal perturbation model for the given accuracy $%
\lambda $ of the continual indirect measurement of the position operators $%
Q\left( t\right) $ (the perturbation vanishes when $\lambda =0$).

\subsubsection{Quantum state diffusion and filtering}

Let us introduce the quantum stochastic wave equation for the unitary
transformation $\Psi _{0}\left( t\right) =W\left( t\right) \Psi _{0}$
inducing Heisenberg dynamics which is decrepid by the quantum Langevin
equation (\ref{6.2}) with white noise perturbation in terms. This equation
is well understood in terms of the generalized derivatives 
\begin{equation*}
\hat{f}_{k}\left( t\right) =\lambda _{k}\frac{\hbar }{i}\left(
a_{k}^{+}-a_{-}^{k}\right) \left( t\right) =\lambda _{k}\frac{\mathrm{d}\hat{%
f}_{k}^{t}}{\mathrm{d}t}
\end{equation*}
of the standard quantum Brownian motions $A_{k}^{+}\left( t\right) $ and $%
A_{-}^{k}\left( t\right) $ defined by the commutation relations 
\begin{equation*}
\left[ A_{k}^{+}\left( s\right) ,A_{l}^{+}\left( t\right) \right] =0,\;\left[
A_{-}^{k}\left( s\right) ,A_{l}^{+}\left( t\right) \right] =\left( t\wedge
s\right) \delta _{l}^{k},\;\left[ A_{-}^{k}\left( s\right) ,A_{-}^{l}\left(
t\right) \right] =0
\end{equation*}
in Fock space $\mathcal{F}_{0}$ ($t\wedge s=\min \left\{ s,t\right\} $) \
The corresponding quantum stochastic differential equation for the
probability amplitude in $\frak{h}\otimes \mathcal{F}_{0}$ is a particular
case 
\begin{equation*}
\mathrm{L}_{k}^{-\dagger }=-\mathrm{L}_{+}^{k},\quad \mathrm{L}%
_{+}^{k}=\left( \lambda \mathrm{Q}\right) ^{k}\equiv \mathrm{L}^{k}
\end{equation*}
of the general quantum diffusion wave equation 
\begin{equation}
\mathrm{d}\Psi _{0}\left( t\right) +\left( \mathrm{K}\otimes I\right) \Psi
_{0}\left( t\right) \mathrm{d}t=\left( \mathrm{L}_{+}^{k}\otimes \mathrm{d}%
A_{k}^{+}+\mathrm{L}_{k}^{-}\otimes \mathrm{d}A_{-}^{k}\right) \left(
t\right) \Psi _{0}\left( t\right)  \label{6.4}
\end{equation}
which describes the unitary evolution in $\frak{h}\otimes \mathcal{F}_{0}$
if $\mathrm{K}=\frac{i}{\hbar }\mathrm{H}-\frac{1}{2}\mathrm{L}_{k}^{-}%
\mathrm{L}_{+}^{k}$, where $\mathrm{H}^{\dagger }=\mathrm{H}$ is the
evolution Hamiltonian for the system in $\frak{h}$. Using the quantum
It\^{o} formula, see the Appendix 3, it was proven in \cite{Be88, Be92b}
that it is equivalent to the Langevin equation 
\begin{eqnarray}
\mathrm{d}X\left( t\right) &=&\left( f\left( XL+L^{\dagger }X\right)
+L^{\dagger }XL-K^{\dagger }X-XK\right) \left( t\right) \mathrm{d}t  \notag
\\
&&+\left( fX+L^{\dagger }X-XL\right) \left( t\right) \mathrm{d}A_{-}+\left(
fX+XL-L^{\dagger }X\right) \left( t\right) \mathrm{d}A^{+}  \label{6.5}
\end{eqnarray}
for any quantum stochastic Heisenberg process 
\begin{equation*}
X\left( t,f\right) =W\left( t\right) ^{\dagger }\left( \mathrm{X}\otimes
\exp \left[ \int_{0}^{t}\left( f^{k}\left( r\right) \mathrm{d}\hat{w}%
_{k}^{r}-\frac{1}{2}f\left( r\right) ^{2}\mathrm{d}r\right) \right] \right)
W\left( t\right) ,
\end{equation*}
where $f^{k}\left( t\right) $ are a test function for the output process $%
w_{k}^{t}$ and 
\begin{equation*}
K\left( t\right) =W\left( t\right) ^{\dagger }\left( \mathrm{K}\otimes 
\mathrm{I}\right) W\left( t\right) ,\quad L^{k}\left( t\right) =W\left(
t\right) ^{\dagger }\left( \mathrm{L}^{k}\otimes \mathrm{I}\right) W\left(
t\right) .
\end{equation*}
The Langevin equation (\ref{6.2}) for the system coordinate $X\left(
t\right) =W\left( t\right) ^{\dagger }\left( \mathrm{Q}\otimes I\right)
W\left( t\right) $ and for the output processes $Y\left( t,f\right) $
corresponding to $\mathrm{X}=\mathrm{I}$ follow straightforward in the case $%
\mathrm{L}=\lambda \mathrm{Q}$, $\mathrm{H}=\mathrm{P}^{2}/2m+\phi \left( 
\mathrm{Q}\right) $.

In the next section we shall show that this unitary evolution is the
interaction picture for a unitary group evolution $U^{t}$ corresponding to a
Dirac type boundary value problem for a generalized Schr\"{o}dinger equation
in an extended product Hilbert space $\frak{h}\otimes \mathcal{G}$. Here we
prove that the quantum stochastic evolution (\ref{6.4}) in $\frak{h}\otimes 
\mathcal{F}_{0}$ coincides with the quantum state diffusion in $\frak{h}$ if
it is considered only for the initial product states $\psi \otimes \delta
_{\varnothing }$ with $\delta _{\varnothing }$ being the Fock vacuum state
vector in $\mathcal{F}_{0}$,.

\emph{Quantum state diffusion} is a nonlinear, nonunitary, irreversible
stochastic form of quantum mechanics with trajectories put forward by Gisin
and Percival \cite{GiPe92, GiPe93} in the early 90's as a new, \emph{primary}
quantum theory which includes the diffusive reduction process into the wave
equation for pure quantum states. It has been criticized, quite rightly, as
an incomplete theory which does not satisfy the linear superposition
principle for the waves, and for not explaining the origin of irreversible
dissipativity which is build into the equation `by hands'. In fact the
`primary' equation had been derived even earlier as the posterior state
diffusion equation for pure states $\psi _{\omega }=\psi \left( \omega
\right) /\left\| \psi \left( \omega \right) \right\| $ from the linear
unitary quantum diffusion equation (\ref{6.4}) by the following method as a
particular type of the general quantum filtering equation in \cite{Be88,
Be89b, BeSt92}. Here we shall show only how to derive the corresponding
stochastic linear decoherence equation \{\ref{5.1}) for $\psi \left(
t,w\right) =V\left( t,w\right) \psi $ when all the independent increment
processes $y_{k}^{t}$ are of the diffusive type $y_{k}^{t}=w_{k}^{t}$: 
\begin{equation}
\mathrm{d}\psi \left( t,\omega \right) +\mathrm{K}\psi \left( t,\omega
\right) \mathrm{d}t=\mathrm{L}^{k}\psi \left( t,\omega \right) \mathrm{d}%
w_{k}^{t},\quad \psi \left( 0\right) =\psi .  \label{6.6}
\end{equation}
Note that the resolving stochastic propagator $V\left( t,\omega \right) $
for this equation defines the isometries 
\begin{equation*}
V\left( t\right) ^{\dagger }V\left( t\right) =\int V\left( t,\omega \right)
^{\dagger }V\left( t,\omega \right) \mathrm{d}\mu =1
\end{equation*}
of the system Hilbert space $\frak{h}$ into the Wiener Hilbert space $L_{\mu
}^{2}$ of square integrable functionals of the diffusive trajectories $%
\omega =\left\{ w\left( t\right) \right\} $ with respect to the standard
Gaussian measure $\mu =\mathsf{P}_{w}$ if $\mathrm{K}+\mathrm{K}^{\dagger }=%
\mathrm{L}^{\dagger }\mathrm{L}$.

Let us represent these Wiener processes in the equation by operators $\hat{w}%
_{k}^{t}=A_{k}^{+}+A_{-}^{k}$ on the Fock space vacuum $\delta _{\varnothing
}$ using the unitary equivalence $w_{k}^{t}\simeq \hat{w}_{k}^{t}\delta
_{\varnothing }$ in the notation explained in the Appendix 3. Then the
corresponding operator equation 
\begin{equation*}
\mathrm{d}\hat{\psi}\left( t\right) +\mathrm{K}\hat{\psi}\left( t\right) 
\mathrm{d}t=\left( \mathrm{L}^{k}\mathrm{d}A_{k}^{+}+\mathrm{L}_{k}^{\dagger
}\mathrm{d}A_{-}^{k}\right) \hat{\psi}\left( t\right) ,\quad \hat{\psi}%
\left( 0\right) =\psi \otimes \delta _{\varnothing },\psi \in \frak{h},
\end{equation*}
with $\mathrm{L}_{k}^{\dagger }=\lambda \mathrm{Q}^{k}=\mathrm{L}^{k}$,
coincides with the quantum diffusion Schr\"{o}dinger equation (\ref{6.4}),
where $\mathrm{L}_{+}^{k}=\mathrm{L}^{k},\mathrm{L}_{k}^{-}=-\mathrm{L}_{k}$
on the same initial product-states $\Psi _{0}\left( 0\right) =\psi \otimes
\delta _{\varnothing }$. Indeed, as it was noted in \cite{Be92b}, due to the
adaptedness 
\begin{equation*}
\hat{\psi}\left( t\right) =\hat{\psi}^{t}\otimes \delta _{\varnothing
},\quad \Psi _{0}\left( t\right) =\Psi _{0}^{t}\otimes \delta _{\varnothing }
\end{equation*}
both right had sides of these equations coincide on future vacuum $\delta
_{\varnothing }$ if $\hat{\psi}^{t}=\Psi _{0}^{t}$ as 
\begin{eqnarray*}
\mathrm{L}^{k}\mathrm{d}\hat{w}_{k}^{t}\hat{\psi}\left( t\right) &=&\left( 
\mathrm{L}^{k}\mathrm{d}A_{k}^{+}+\mathrm{L}_{k}^{\dagger }\mathrm{d}%
A_{-}^{k}\right) \left( \hat{\psi}^{t}\otimes \delta _{\varnothing }\right) =%
\mathrm{L}^{k}\hat{\psi}^{t}\otimes \mathrm{d}A_{k}^{+}\delta _{\varnothing }
\\
\frac{i}{\hbar }\mathrm{L}^{k}\mathrm{d}\hat{f}_{k}^{t}\Psi _{0}\left(
t\right) &=&\left( \mathrm{L}^{k}\mathrm{d}A_{k}^{+}-\mathrm{L}_{k}\mathrm{d}%
A_{-}^{k}\right) \left( \Psi _{0}^{t}\otimes \delta _{\varnothing }\right) =%
\mathrm{L}^{k}\Psi _{0}^{t}\otimes \mathrm{d}A_{k}^{+}\delta _{\varnothing }
\end{eqnarray*}
(the annihilation processes $A_{-}^{k}$ are zero on the vacuum $\delta
_{\varnothing }$). By virtue of the coincidence of the initial data $\hat{%
\psi}^{0}=\psi =\Psi _{0}^{t}$ this proves that $\hat{\psi}\left( t\right)
=\Psi _{0}\left( t\right) $ for all $t>0$. Note that the quantum stochastic
evolutions $\hat{\psi}\left( t\right) $ and $\Psi _{0}\left( t\right) $,
when extended on the whole space $\frak{h}\otimes \mathcal{G}_{0}$, are
described by the different propagators $V\left( t\right) $and $W\left(
t\right) $ as $\hat{\psi}\left( t\right) =V\left( t\right) \hat{\psi}$, $%
\Psi _{0}\left( t\right) =W\left( t\right) \Psi _{0}$. The first one is
unbounded and even not well defined on the whole space $\frak{h}\otimes 
\mathcal{G}$, while the second one is unitary, resolving another stochastic
differential equation 
\begin{equation}
\mathrm{d}\psi _{0}\left( t\right) +\left( \frac{i}{\hbar }\mathrm{H}+\frac{1%
}{2}\mathrm{Q}^{k}\lambda _{k}^{2}\mathrm{Q}^{k}\right) \psi _{0}\left(
t\right) \mathrm{d}t=\frac{i}{\hbar }\lambda \mathrm{Q}^{k}\psi _{0}\left(
t\right) \mathrm{d}f_{k}^{t},\quad \psi _{0}\left( 0\right) =\psi .
\label{6.3}
\end{equation}
by the unitary propagator $\mathrm{W}\left( t,f\right) =W\left( t\right)
\left( \mathrm{I}\otimes \delta _{\varnothing }\right) $ for each $f$ in $%
\frak{h}$ as the stochastic function $\psi _{0}\left( t,f\right) =\mathrm{W}%
\left( t,f\right) \psi $ on another classical probability space.

Thus the stochastic decoherence equation 
\begin{equation*}
\mathrm{d}\psi \left( t\right) +\left( \frac{i}{\hbar }\mathrm{H}+\frac{1}{2}%
\mathrm{Q}^{k}\lambda _{k}^{2}\mathrm{Q}^{k}\right) \psi \left( t\right) 
\mathrm{d}t=\lambda _{k}\mathrm{Q}^{k}\psi \left( t\right) \mathrm{d}%
w_{k}^{t},\quad \psi \left( 0\right) =\psi .
\end{equation*}
for the continuous observation of the position of a quantum particle with $%
\mathrm{H}=\frac{1}{2m}\mathrm{P}^{2}+\phi \left( \mathrm{Q}\right) $ was
derived for the unitary quantum stochastic evolution as an example of the
general decoherence equation (\ref{5.1}) which was obtained in this way in 
\cite{Be88}. It was explicitly solved in \cite{Be88, Be89b, BeSt92} for the
case of linear and quadratic potentials $\phi $, and it was shown that this
solution coincides with the optimal quantum linear filtering solution
obtained earlier in \cite{Be80, Be85} if the initial wave packet is Gaussian.

The nonlinear stochastic posterior equation for this particular case was
derived independently by Diosi \cite{Dio88} and (as an example) in \cite
{Be88, Be89b}. It has the following form 
\begin{equation*}
\mathrm{d}\psi _{w}\left( t\right) +\left( \frac{i}{\hbar }\mathrm{H}+\frac{1%
}{2}\widetilde{\mathrm{Q}}^{k}\left( t\right) \lambda _{k}^{2}\widetilde{%
\mathrm{Q}}^{k}\left( t\right) \right) \psi _{w}\left( t\right) \mathrm{d}%
t=\lambda _{k}\widetilde{\mathrm{Q}}^{k}\left( t\right) \psi _{w}\left(
t\right) \mathrm{d}\tilde{w}_{k}^{t},
\end{equation*}
where $\widetilde{\mathrm{Q}}\left( t\right) =\mathrm{Q}-\hat{q}\left(
t\right) $ with $\hat{q}^{k}\left( t\right) $ defined as the multiplication
operators by the components $q^{k}\left( t,w\right) =\psi _{w}^{\dagger
}\left( t\right) \mathrm{Q}^{k}\left( t\right) \psi _{w}\left( t\right) $ of
the posterior expectation (statistical prediction) of the coordinate $%
\mathrm{Q}$, and 
\begin{equation*}
\mathrm{d}\tilde{w}_{k}^{t}=\mathrm{d}w_{k}^{t}-2\lambda _{k}\hat{q}%
^{k}\left( t\right) \mathrm{d}t=\mathrm{d}y_{k}^{t}-\hat{x}_{k}\left(
t\right) \mathrm{d}t,\quad \hat{x}_{k}\left( t\right) =2\left( \lambda \hat{q%
}\right) ^{k}\left( t\right) .
\end{equation*}
Note that the innovating output processes $\tilde{w}_{k}^{t}$ are also
standard Wiener processes with respect to the output probability measure $%
\mathrm{d}\tilde{\mu}=\lim \Pr \left( t,\mathrm{d}\omega \right) $, but not
with respect to the Wiener probability measure $\mu =\Pr \left( 0,\mathrm{d}%
\omega \right) $ for the input noise $w_{k}^{t}$.

Let us give the explicit solution of this stochastic wave equation for the
free particle ($\phi =0$) in one dimension and the stationary Gaussian
initial wave packet which was found in \cite{Be88, Be89b, BeSt92}. One can
show \cite{ChSt92, Kol95} that the nondemolition observation of such
particle is described by filtering of quantum noise which results in the
continual collapse of any wave packet to the Gaussian stationary one
centered at the posterior expectation $q\left( t,w\right) $ with finite
dispersion $\left\| \left( \hat{q}\left( t\right) -\mathrm{Q}\right) \psi
_{\omega }\left( t\right) \right\| ^{2}\rightarrow 2\lambda \left( \hbar
/m\right) ^{1/2}$. This center can be found from the linear Newton equation 
\begin{equation*}
\frac{\mathrm{d}^{2}}{\mathrm{d}t^{2}}z\left( t\right) +2\kappa \frac{%
\mathrm{d}}{\mathrm{d}t}z\left( t\right) +2\kappa ^{2}z\left( t\right)
=-g\left( t\right) ,
\end{equation*}
for the deviation process $z\left( t\right) =q\left( t\right) -x\left(
t\right) $, where $x\left( t\right) $ is an expected trajectory of the
output process (\ref{6.1}) with $z\left( 0\right) =q_{0}-x\left( 0\right) $, 
$z^{\prime }\left( 0\right) =v_{0}-x^{\prime }\left( 0\right) $. Here $%
\kappa =\lambda \left( \hbar /m\right) ^{1/2}$ is the decay rate which is
also the frequency of effective oscillations, $q_{0}=\left\langle \hat{x}%
\right\rangle $, $v_{0}=\left\langle \hat{p}/m\right\rangle $ are the
initial expectations and $g\left( t\right) =x^{\prime \prime }\left(
t\right) $ is the effective gravitation for the particle in the moving
framework of $x\left( t\right) $. The following figure illustrate the
continuous collapse $z\left( t\right) \rightarrow 0$ of the posterior
trajectory $q\left( t\right) $ towards a linear trajectory $x\left( t\right)
.$

\FRAME{itbpF}{13.0479cm}{5.5706cm}{0.3712cm}{}{}{}{\special{language
"Scientific Word";type "MAPLEPLOT";width 13.0479cm;height 5.5706cm;depth
0.3712cm;display "PICT";plot_snapshots TRUE;function \TEXUX{$5\tau
$};linecolor "blue";linestyle 3;linethickness 1;pointstyle "point";function
\TEXUX{$0.5\tau -1$};linecolor "red";linestyle 3;linethickness 1;pointstyle
"point";function \TEXUX{$e^{-\tau }\left( 6\sin \tau +\cos \tau \right)
+0.5\tau -1$};linecolor "green";linestyle 1;linethickness 2;pointstyle
"point";xmin "-0.011678";xmax "5.988322";xviewmin "-0.011678";xviewmax
"5.988322";yviewmin "-1";yviewmax "2.0";viewset"XY";rangeset"X";phi 45;theta
45;plottype 4;plottickdisable TRUE;num-y-ticks 1;labeloverrides 3;x-label
"t";y-label "q(t)";numpoints 50;axesstyle "normal";xis \TEXUX{v964};var1name
\TEXUX{$x$};valid_file "T";tempfilename 'H0SHIT00.wmf';tempfile-properties
"XPR";}}

The posterior position expectation $q\left( t\right) $ in the absence of
effective gravitation, $x^{\prime \prime }\left( t\right) =0$, for the
linear trajectory $x\left( t\right) =ut-q$ collapses to the expected input
trajectory $x\left( t\right) $ with the rate $\kappa =\lambda \left( \hbar
/m\right) ^{1/2}$, remaining not collapsed, $q_{0}\left( t\right) =v_{0}t$
in the framework where $q_{0}=0$, only in the classical limit $\hbar
/m\rightarrow 0$ or absence of observation $\lambda =0$. This is the graph
of 
\begin{equation*}
q_{0}\left( t\right) =v_{0}t,\quad q\left( t\right) =ut+e^{-\kappa t}\left(
q\cos \kappa t+\left( q+\kappa ^{-1}\left( v_{0}-u\right) \right) \sin
\kappa t\right) -q
\end{equation*}
obtained as $q\left( t\right) =x\left( t\right) +z\left( t\right) $ by
explicit solving of the second order linear equation for $z\left( t\right) $.

\subsubsection{The dynamical boundary-value realization}

Finally, let us set up an explicitly solvable microscopic problem underlying
all quantum diffusion and more general quantum noise Langevin models. We
shall see that all such models exactly correspond to Dirac type boundary
value problems for a Poisson flow of independent quantum particles
interacting with the quantum system under the observation at the boundary $%
r=0$ of the half line $\mathbb{R}_{+}$ in an additional dimension, exactly
as it was done for the quantum jumps in (\ref{5.1}). One can think that the
coordinate $r>0$ of this extra dimension is a physical realization of
localizable time, at least it is so for any free evolution Hamiltonian $%
\varepsilon \left( p\right) >0$ of the incoming quantum particles in the
ultrarelativistic limit $\left\langle p\right\rangle \rightarrow -\infty $
such that the average velocity in an initial state is a finite constant, $%
c=\left\langle \varepsilon ^{\prime }\left( p\right) \right\rangle
\rightarrow 1$ say. Thus we are going to solve the problem of quantum
trajectories, individual decoherence, state diffusion, or permanent
reduction problem as the following time continuous quantum measurement
problem which we already formulated and solved in the Sec. 5.3 for the time
discrete case:

\emph{Given a reduction family }$V\left( t,\omega \right) =V\left(
t,w_{0}^{t]}\right) ,t\in \mathbb{R}_{+}$ \emph{of isometries on} $\frak{h}$%
\ \emph{into} $\frak{h}\otimes L_{\mu }^{2}$\emph{\ resolving the state
diffusion equation (\ref{6.6}) with respect to the input probability measure 
}$\mu =\mathsf{P}_{w}$\emph{\ for the standard Wiener noises }$w_{k}^{t}$%
\emph{, find a triple }$\left( \mathcal{G},\frak{A},\Phi \right) $ \emph{%
consisting of a Hilbert space} $\mathcal{G=G}_{-}\otimes \mathcal{G}_{+}$ 
\emph{embedding the Wiener Hilbert space }$L_{\mu }^{2}$\emph{\ by an
isometry into }$\mathcal{G}_{+}$\emph{, an algebra }$\frak{A=A}_{-}\otimes 
\frak{A}_{+}$ \emph{on $\mathcal{G}$ with an Abelian subalgebra }$\frak{A}%
_{-}$\emph{\ generated by a compatible continuous family }$Y_{-\infty
}^{0]}=\left\{ Y_{k}^{s},k=1,\ldots ,d,s\leq 0\right\} $\emph{\ } \emph{of
observables on$\mathcal{\ }$}$\mathcal{G}_{-}$\emph{, and a state-vector }$%
\Phi ^{\circ }=\Phi _{-}^{\circ }\otimes \Phi _{+}^{\circ }\in \mathcal{G}$%
\emph{\ such that there exist a time continuous unitary group }$U^{t}$\emph{%
\ on }$\mathcal{H=}\frak{h}\otimes \mathcal{G}$\emph{\ inducing a semigroup
of endomorphisms }$\frak{B}\ni B\mapsto U^{-t}BU^{t}\in \frak{B}$ \emph{on
the product algebra $\frak{B=}\mathcal{B}\left( \frak{h}\right) \otimes 
\frak{A}$, with } 
\begin{eqnarray*}
\pi ^{t}\left( \hat{g}_{-t}\otimes \mathrm{B}\right) &:&=\left( \mathrm{I}%
\otimes \Phi ^{\circ }\right) ^{\dagger }U^{-t}\left( \mathrm{B}\otimes
g_{-t}\left( Y_{-t}^{0]}\right) \right) U^{t}\left( \mathrm{I}\otimes \Phi
^{\circ }\right) \\
&=&\int g\left( w_{0}^{t]}\right) V\left( t,w_{0}^{t]}\right) ^{\dagger }%
\mathrm{B}V\left( t,w_{0}^{t]}\right) \mathrm{d}\mathsf{P}_{w}\equiv \mathsf{%
M}\left[ gV\left( t\right) ^{\dagger }\mathrm{B}V\left( t\right) \right]
\end{eqnarray*}
\emph{\ for any }$\mathrm{B}\in \mathcal{B}\left( \frak{h}\right) $ \emph{%
and any operator }$\hat{g}_{-t}=\hat{g}_{-t}\left( Y_{-t}^{0]}\right) \in 
\frak{C}$ \emph{represented as the shifted functional }$\hat{g}_{-t}\left(
y_{-t}^{0]}\right) =g\left( y_{0}^{t]}\right) $\emph{\ of }$%
Y_{-t}^{-0]}=\left\{ Y_{\cdot }^{s}:s\in (-t,0]\right\} $ \emph{on $\mathcal{%
G}$ by any measurable functional }$g$\emph{\ of }$y_{0}^{t]}=\left\{
y_{\cdot }^{r}:r\in (0,t]\right\} $\emph{\ with arbitrary }$t>0.$

We have already dilated the state diffusion equation (\ref{6.6}) to a
quantum stochastic unitary evolution $W\left( t\right) $ resolving the
quantum stochastic Schr\"{o}dinger equation (\ref{6.4}) on the system
Hilbert $\frak{h}$ tensored the Fock space$\mathcal{F}_{0}$ such that $%
W\left( t\right) \left( \mathrm{I}\otimes \delta _{\varnothing }\right)
=V\left( t\right) $, where $\delta _{\varnothing }\in \mathcal{F}_{0}$ is
the Fock vacuum vector. In fact the state diffusion equation was first
derived \cite{Be88, Be89b} in this way from even more general quantum
stochastic unitary evolution which satisfy the equation 
\begin{equation*}
\left( \mathrm{I}\otimes \delta _{\varnothing }\right) ^{\dagger }W\left(
t\right) ^{\dagger }\left( \mathrm{B}\otimes g\left( \hat{w}_{0}^{t]}\right)
\right) W\left( t\right) \left( \mathrm{I}\otimes \delta _{\varnothing
}\right) =\mathsf{M}\left[ gV\left( t\right) ^{\dagger }\mathrm{B}V\left(
t\right) \right] .
\end{equation*}
Indeed, this equation is satisfied for the model (\ref{6.4}) as one can
easily check for 
\begin{equation*}
g\left( w_{0}^{t]}\right) =\exp \left[ \int_{0}^{t}\left( f^{k}\left(
r\right) \mathrm{d}w_{k}^{r}-\frac{1}{2}f\left( r\right) ^{2}\mathrm{d}%
r\right) \right]
\end{equation*}
given by a test vector function $f$ by conditioning the Langevin equation (%
\ref{6.5}) with respect to the vacuum vector $\delta _{\varnothing }$: 
\begin{equation*}
\left( \mathrm{I}\otimes \delta _{\varnothing }\right) ^{\dagger }\left( 
\mathrm{d}X+\left( K^{\dagger }X+XK-L^{\dagger }XL-\left( XL+L^{\dagger
}X\right) f\right) \mathrm{d}t\right) \left( \mathrm{I}\otimes \delta
_{\varnothing }\right) =0.
\end{equation*}
Obviously this equation coincides with the conditional expectation 
\begin{equation*}
\mathsf{M}\left[ \mathrm{d}B\left( t\right) +\left( \mathrm{K}^{\dagger
}B\left( t\right) +B\left( t\right) \mathrm{K}-\mathrm{L}^{\dagger }B\left(
t\right) \mathrm{L}-\left( B\left( t\right) L+L^{\dagger }B\left( t\right)
\right) f\left( t\right) \right) \mathrm{d}t\right] =0
\end{equation*}
for the stochastic process $B\left( t\right) =V\left( t\right) ^{\dagger }g%
\mathrm{X}V\left( t\right) $ which satisfies the stochastic It\^{o} equation 
\begin{equation*}
\mathrm{d}B+gV^{\dagger }\left( \mathrm{K}^{\dagger }\mathrm{X}+\mathrm{XK}-%
\mathrm{L}^{\dagger }\mathrm{XL}-\left( \mathrm{XL}+\mathrm{L}^{\dagger }%
\mathrm{X}\right) f\right) V\mathrm{d}t=gV^{\dagger }\left( \mathrm{L}%
^{\dagger }\mathrm{X}+\mathrm{XL}+\mathrm{X}\right) V\mathrm{d}w^{t}.
\end{equation*}
This however doesn't give yet the complete solution of the quantum
measurement problem as formulated above because the algebra $\frak{B}_{0}$
generated by $\mathrm{B}\otimes I$ and the Langevin forces $\hat{f}_{k}^{t}$
does not contain the measurement processes $\hat{w}_{k}^{t}$ which do not
commute with $\hat{f}_{k}^{t}$, and the unitary family $W\left( t\right) $
does not form unitary group but only cocycle 
\begin{equation*}
T_{t}W\left( s\right) T_{-t}W\left( t\right) =W\left( s+t\right) ,\quad
\forall s,t>0
\end{equation*}
with respect to the isometric but not unitary right shift semigroup $T_{t}$
in $\mathcal{F}_{0}$.

Let $T_{t}$ be the one parametric continuous unitary shift group on $%
\mathcal{F}^{0]}\otimes \mathcal{F}_{0}$ extending the defining from $%
\mathcal{F}_{0}$. It describes the free evolution by right shifts $\Phi
_{t}\left( \omega \right) =\Phi \left( \omega -t\right) $ in Fock space over
the whole line $\mathbb{R}$ (Here and below we use the notations from the
Sec. 6.3). Then one can easily find the unitary group 
\begin{equation*}
U^{t}=T_{-t}\left( I^{0}\otimes \mathrm{I}\otimes W\left( t\right) \right)
T_{t}
\end{equation*}
on $\mathcal{F}_{0}\otimes \frak{h}\otimes \mathcal{F}^{0]}$ inducing the
quantum stochastic evolution as the interaction representation $U\left(
t\right) =T_{t}U^{t}$ on the Hilbert space $\frak{h}\otimes \mathcal{G}_{0}$%
. In fact this evolution corresponds to an unphysical coordinate
discontinuity problem at the origin $s=0$ which is not invariant under the
reflection of time $t\mapsto -t$. Instead, we shall formulate the unitary
equivalent boundary value problem in the Poisson space $\mathcal{G}=\mathcal{%
G}_{-}\otimes \mathcal{G}_{+}$ for two semi-infinite strings on $\mathbb{R}%
_{+}$, one is the living place for the quantum noise generated by a Poisson
flow of incoming waves of quantum particles of the intensity $\nu >0$, and
the other one is for the outgoing classical particles carrying the
information after a unitary interaction with the measured quantum system at
the origin $r=0$. The probability amplitudes $\Phi \in \mathcal{G}$ are
represented by the $\mathbb{G}^{\otimes }=\frak{g}_{-}^{\otimes }\otimes 
\frak{g}_{+}^{\otimes }$-valued functions $\Phi \left( \upsilon
_{-},\upsilon _{+}\right) $ of two infinite sequences $\upsilon _{\pm
}=\left\{ \pm r_{1},\pm r_{2},\ldots \right\} \subset \mathbb{R}_{+}$ of the
coordinates of the particles in the increasing order $r_{1}<r_{2}<\ldots $
such that 
\begin{equation*}
\left\| \Phi \right\| ^{2}=\iint \left\| \Phi \left( \upsilon _{-},\upsilon
_{+}\right) \right\| ^{2}\mathsf{P}_{\nu }\left( \mathrm{d}\upsilon
_{-}\right) \mathsf{P}_{\nu }\left( \mathrm{d}\upsilon _{+}\right) <\infty
\end{equation*}
with respect to the product of two copies of the Poisson probability measure 
$\mathsf{P}_{\nu }$ defined by the constant intensity $\nu >0$ on $\mathbb{R}%
_{+}$. Here $\frak{g}^{\otimes }$ is the infinite tensor product of $\frak{g}%
=\mathbb{C}^{d}$ obtained by the completion of the linear span of $\chi
_{1}\otimes \chi _{2}\otimes \ldots $ with almost all multipliers\ $\chi
_{n}=\varphi $ given by a unit vector $\varphi \in \mathbb{C}^{d}$ such that
the infinite product $\left\| \Phi \left( \upsilon \right) \right\|
=\prod_{r\in \upsilon }\left\| f\left( r\right) \right\| $ for $\Phi \left(
\upsilon \right) =\otimes _{r\in \upsilon }f\left( r\right) $ with $f\left(
r_{n}\right) =\chi _{n}$ is well defined as it has all but finite number of
multipliers $\left\| \chi _{n}\right\| $ equal $1$. The unitary
transformation $\digamma \mapsto \Phi $ from Fock space $\mathcal{F}\ni
\digamma $ to the corresponding Poisson one $\mathcal{G}$ can be written as 
\begin{equation*}
\Phi =\lim_{t\rightarrow \infty }\mathrm{e}^{\varphi ^{i}A_{i}^{+}\left(
t\right) }\mathrm{e}^{-\varphi _{k}A_{-}^{k}\left( t\right) }\nu ^{-\frac{1}{%
2}A_{i}^{i}\left( t\right) }\digamma \equiv I_{\nu }\left( \varphi \right)
\digamma ,
\end{equation*}
where $\varphi _{k}=\nu \bar{\varphi}^{k}$ for the Poisson intensity $\nu >0$
and the unit vector $\varphi =\left( \varphi ^{i}\right) $ defined by the
initial probability amplitude $\varphi \in \frak{g}$ for the auxiliary
particles to be in a state $k=1,\ldots ,d.$ Here $A_{\iota }^{\kappa }\left(
t\right) $ are the QS integrators defined in Appedix 3, and the limit is
taken on the dense subspace $\cup _{t>0}\mathcal{F}_{0}^{t]}$ of
vacuum-adapted Fock functions $\digamma _{t}\in \mathcal{F}_{0}$ and
extended then onto $\mathcal{F}_{0}$ by easily proved isometry $\left\|
\digamma _{t}\right\| =\left\| \Phi _{t}\right\| $ for $\Phi _{t}=I_{\nu
}\left( \varphi \right) \digamma _{t}$.

The free evolution in $\mathcal{G}$ is the same as in the Section 6.3, 
\begin{equation*}
T_{t}\Phi \left( \upsilon _{-},\upsilon _{+}\right) =\Phi \left( \upsilon
_{-}^{t},\upsilon _{+}^{t}\right) ,
\end{equation*}
where $\upsilon _{\pm }^{t}=\pm \left[ \left( \left[ \left( -\upsilon
_{-}\right) \cup \left( +\upsilon _{+}\right) \right] -t\right) \cap \mathbb{%
R}_{\pm }\right] $. It is given by the second quantization (\ref{5.6}) of
the Dirac Hamiltonian in one dimension on $\mathbb{R}_{+}.$

In order to formulate the boundary value problem in the space $\mathcal{H}=%
\frak{h}\otimes \mathcal{G}$ corresponding to the quantum stochastic
equations of the diffusive type (\ref{6.4}) let us introduce the notation 
\begin{equation*}
\Phi \left( 0^{k}\sqcup \upsilon _{\pm }\right) =\lim_{r\searrow 0}\left(
\langle k|\otimes \mathrm{I}_{1}\otimes \mathrm{I}_{2}\ldots \right) \Phi
\left( \pm r,\pm r_{1},\pm r_{2},\ldots \right) ,
\end{equation*}
where $\langle k|=d^{-1/2}\left( \delta _{1}^{k},\ldots ,\delta
_{l}^{k}\right) $ acts as the unit bra-vector evaluating the $k$-th
projection of the state vector $\Phi \left( \pm r\sqcup \upsilon _{\pm
}\right) $ with $r<r_{1}<r_{2}<\ldots $ corresponding to the nearest to the
boundary $r=0$ particle in one of the strings on $\mathbb{R}_{+}$.

The unitary group evolution $U^{t}$ corresponding to the scattering
interaction at the boundary with the continuously measured system which has
its own free evolution described by the energy operator $\mathrm{E}=\mathrm{E%
}^{\dagger }$ can be obtained by resolving the following generalized
Schr\"{o}dinger equation 
\begin{equation}
\frac{\partial }{\partial t}\Psi ^{t}\left( \upsilon _{-},\upsilon
_{+}\right) -\frac{\mathrm{i}}{\hbar }P\Psi ^{t}\left( t,\upsilon
_{-},\upsilon _{+}\right) =\mathrm{G}_{k}^{-}\Psi ^{t}\left( \upsilon
_{-},0^{k}\sqcup \upsilon _{+}\right) +\mathrm{G}_{+}^{-}\Psi ^{t}\left(
\upsilon _{-},\upsilon _{+}\right)  \label{6.7}
\end{equation}
with the Dirac zero current boundary condition at the origin $r=0$ 
\begin{equation}
\Psi ^{t}\left( 0^{i}\sqcup \upsilon _{-},\upsilon _{+}\right) =\mathrm{G}%
_{+}^{i}\Psi ^{t}\left( \upsilon _{-},\upsilon _{+}\right) +\mathrm{G}%
_{k}^{i}\Psi ^{t}\left( \upsilon _{-},0^{k}\sqcup \upsilon _{+}\right)
,\quad \forall t>0,\upsilon _{\pm }>0.  \label{6.8}
\end{equation}
Here $\mathrm{G}=\left[ \mathrm{G}_{k}^{i}\right] $ is unitary, $\mathrm{G}%
^{-1}=\mathrm{G}^{\dagger }$, like the scattering operator $\mathrm{S}$ in
the simpler quantum jump boundary value problem (\ref{5.8}), and the other
system operators $\mathrm{G}_{\iota }^{\kappa }$, with $\iota =-,i$ and $%
\kappa =k,+$ for any $i,k=1,\ldots ,d$ are chosen as 
\begin{equation}
\mathrm{G}^{-}+\nu \mathrm{G}_{+}^{\dagger }\mathrm{G}=\mathrm{O},\quad 
\mathrm{G}_{+}^{-}+\frac{\nu }{2}\mathrm{G}_{+}^{\dagger }\mathrm{G}_{+}+%
\frac{i}{\hbar }\mathrm{E}=\mathrm{O}.  \label{6.9}
\end{equation}
Note that these conditions can be written as pseudo-unitarity of the
following triangular block-matrix 
\begin{equation*}
\left[ 
\begin{tabular}{lll}
\textrm{I} & \textrm{G}$^{-}$ & \textrm{G}$_{+}^{-}$ \\ 
\textrm{O} & \textrm{G} & \textrm{G}$_{+}$ \\ 
\textrm{O} & \textrm{O} & \textrm{I}
\end{tabular}
\right] ^{-1}=\left[ 
\begin{tabular}{lll}
\textrm{O} & \textrm{O} & \textrm{I} \\ 
\textrm{O} & $\nu $\textrm{I} & \textrm{O} \\ 
\textrm{I} & \textrm{O} & \textrm{O}
\end{tabular}
\right] ^{-1}\left[ 
\begin{tabular}{lll}
\textrm{I} & \textrm{G}$^{-}$ & \textrm{G}$_{+}^{-}$ \\ 
\textrm{O} & \textrm{G} & \textrm{G}$_{+}$ \\ 
\textrm{O} & \textrm{O} & \textrm{I}
\end{tabular}
\right] ^{\dagger }\left[ 
\begin{tabular}{lll}
\textrm{O} & \textrm{O} & \textrm{I} \\ 
\textrm{O} & $\nu $\textrm{I} & $\mathrm{O}$ \\ 
\textrm{1} & \textrm{O} & \textrm{O}
\end{tabular}
\right] .
\end{equation*}
As it was proved in \cite{Be88a, Be92a} this is necessary (and sufficient at
if all operators are bounded) condition for the unitarity $W\left( t\right)
^{-1}=W\left( t\right) ^{\dagger }$ of the cocycle solution resolving the
quantum stochastic differential equation 
\begin{equation*}
\mathrm{d}\Psi _{0}\left( t\right) =\left( \mathrm{G}_{\kappa }^{\iota
}-\delta _{\kappa }^{\iota }\mathrm{I}\right) \Psi _{0}\left( t\right) 
\mathrm{d}A_{\iota }^{\kappa },\quad \Psi _{0}\left( 0\right) =\Psi _{0}
\end{equation*}
in the Hilbert space $\mathcal{H}_{0}=\frak{h}\otimes \mathcal{G}_{0}$ where 
$\mathcal{G}_{0}$ is identified with the space $\mathcal{G}_{+}=\mathbb{G}%
^{\otimes }\otimes L_{\mu }^{2}$ for the Poisson measure $\mu =\mathsf{P}%
_{\nu }$ with the intensity $\nu $ on $\mathbb{R}_{+}$. This is the general
form for the quantum stochastic equation (\ref{6.4}) where $\mathrm{d}%
A_{-}^{+}=\mathrm{d}t$ in the Poisson space see the Appendix 3 for more
detail explanations of these notations. Our recent results partially
published in \cite{Be00, Be01a, BeKo01} prove that this quantum stochastic
evolution extended as the identity $I_{-}$ also on the component $\mathcal{G}%
_{-}$ for the scattered particles, is nothing but the interaction
representation $U^{t}=T_{-t}\left( I_{-}\otimes W\left( t\right) \right) $
for the unitary group $U^{t}$ resolving our boundary value problem in $\frak{%
h}\otimes \mathbb{G}^{\otimes }$ times the Poisson space $L_{\mu }^{2}$%
..Thus the pseudounitarity condition (\ref{6.9}) is necessary (and
sufficient if the operators $\mathrm{G}_{\kappa }^{\iota }$ are bounded) for
the self-adjointness of the Dirac type boundary value problem (\ref{6.7}), (%
\ref{6.8}).

The generators $\mathrm{G}_{\kappa }^{\iota }$ of this boundary value
problem define the generators $\mathrm{S}_{\kappa }^{\iota }$ of the
corresponding quantum stochastic equation in Fock space by the following
transformation 
\begin{eqnarray}
\mathrm{S}_{+}^{i} &=&\nu ^{1/2}\left( \mathrm{G}_{+}^{i}+\mathrm{G}%
_{k}^{i}\varphi ^{k}-\varphi ^{i}\right) ,\;\mathrm{S}_{k}^{-}=\nu
^{-1/2}\left( \mathrm{G}_{k}^{-}+\varphi _{i}\mathrm{G}_{k}^{i}-\varphi
_{k}\right)  \notag \\
\mathrm{S}_{+}^{-} &=&\mathrm{G}_{+}^{-}+\varphi _{i}\mathrm{G}_{+}^{i}+%
\mathrm{G}_{k}^{-}\varphi ^{k}+\varphi _{i}\left( \mathrm{G}_{k}^{i}-\delta
_{k}^{i}\mathrm{I}\right) \varphi ^{k},\quad \quad \mathrm{S}_{k}^{i}=%
\mathrm{G}_{k}^{i},\;
\end{eqnarray}
induced by the canonical transformation $I_{\nu }\left( \varphi \right) $.

The quantum state diffusion equation (\ref{6.6}) for the continuous
measurement of the coordinates $\mathrm{Q}^{k}$ corresponds to the
particular case (\ref{6.2})of the quantum stochastic differential equation
in Fock space, with 
\begin{eqnarray*}
\mathrm{S}_{+}^{i} &=&\nu ^{1/2}\mathrm{G}_{+}^{i},\quad \quad \quad \quad 
\mathrm{S}_{k}^{-}=\nu ^{-1/2}\mathrm{G}_{k}^{-} \\
\mathrm{S}_{+}^{-} &=&\mathrm{G}_{+}^{-}+\varphi _{i}\mathrm{G}_{+}^{i}+%
\mathrm{G}_{k}^{-}\varphi ^{k},\quad \mathrm{S}_{k}^{i}=\delta _{k}^{i}%
\mathrm{I},
\end{eqnarray*}
$\ $and $\mathrm{G}_{+}^{i}=\mathrm{Q}^{i}$, $\mathrm{G}_{k}^{-}=\nu \mathrm{%
Q}^{k}$ such that all coupling constants $\lambda _{k}=\nu ^{1/2}$. are
equal to the square root of the flow intensity $\nu $. The operators $%
\mathrm{G}_{+}^{i}=\varphi ^{i}\mathrm{Q}^{i}$ and $\mathrm{G}_{k}^{-}=%
\mathrm{Q}_{k}\varphi _{k}$ corresponding to the different \ couplings $%
\lambda _{k}$ can also be obtained from the purely jump model in the central
limit $\nu \mapsto \infty $ as it was done in \cite{BeMe96}. In this case 
\begin{equation*}
\mathrm{S}_{+}^{i}=\nu ^{1/2}\left( \mathrm{G}-\mathrm{I}\right)
_{k}^{i}\varphi ^{k}\rightarrow -\mathrm{i}\lambda \varphi ^{i}\mathrm{Q}%
^{i},
\end{equation*}
with $\varphi ^{k}=\mathrm{i}\lambda _{k}/\lambda $.

And finally, we have to find the operator processes $Y_{k}^{s},s\leq 0$ on
the Hilbert space $\mathcal{G}_{-}$ which reproduce the standard Wiener
noises $w_{k}^{t}$ in the state diffusion when our dynamical model is
conditioned (filtered) with respect their nondemolition measurement. As the
candidates let us consider the field coordinate processes 
\begin{equation*}
X_{k}^{-t}=A_{k}^{+}(-t,0]+A_{-}^{k}(-t,0]=T_{-t}\left(
A_{k}^{+}(0,t]+A_{-}^{k}(0,t]\right) T_{t}
\end{equation*}
which are given by the creation and annihilation processes $A^{+}\left(
t\right) $ and $A_{-}\left( t\right) $ shifted from $\mathcal{G}_{+}$. In
our Poisson space model of $\mathcal{G}$ they have not zero expectations 
\begin{equation*}
\Phi ^{\dagger }X_{k}^{-t}\Phi =\Phi ^{\dagger }\left(
A_{k}^{+}(0,t]+A_{-}^{k}(0,t]\right) \Phi =2\nu ^{1/2}t
\end{equation*}
in the ground state $\Phi =I_{\nu }\left( \varphi \right) \delta
_{\varnothing }$ corresponding to the vacuum vector $\delta _{\varnothing }$
in the Fock space. This state is given as the infinite tensor product $\Phi
^{\circ }=\varphi _{-}^{\otimes }\otimes \varphi _{+}^{\otimes }$ of all
equal probability amplitudes $\varphi _{-}=\varphi =\varphi _{+}$ in $\frak{g%
}=\mathbb{C}^{d}$ for each sequence $\upsilon _{-}$ and $\upsilon _{+}$.
Hence the independent increment processes $Y_{k}^{t}=T_{t}Y_{k}^{-t}T_{-t}$
corresponding to the standard Wiener noises $w_{k}^{t}$ represented in Fock
spaces as $\hat{w}_{k}^{t}=A_{k}^{+}\left( t\right) +A_{-}^{k}\left(
t\right) $ are the compensated processes $Y_{k}^{-t}=X_{k}^{-t}-2\nu ^{1/2}t$%
. This unitary equivalence of $Y_{k}^{t}$ and $\hat{w}_{k}^{t}$ under the
Fock-Poisson transformation $I_{\nu }\left( \varphi \right) $, and the
deduction given above of the quantum state diffusion from the quantum
stochastic signal plus noise model (\ref{6.1}) for continuous observation in
Fock space, completes the solution of the quantum measurement model in its
rigorous formulation.

\section{Conclusion: A quantum message from the future}

Although the conventional formulation of quantum mechanics and quantum field
theory is inadequate for the temporal treatment of modern experiments with
the individual quantum system in real time, it has been shown that the
latest developments in quantum probability, stochastics and in quantum
information theory made it possible to reconcile the dynamical and
statistical aspects of its interpretation. All such phenomena as quantum
events, causality, decoherence, quantum jumps, trajectories and state
diffusions which do not exist in usual quantum mechanical formalism but they
do exist in the modern experimental quantum physics, can be interpreted in
the modern mathematical framework of quantum stochastic processes in terms
of the results of the generalized quantum measurements. The problem of
quantum measurement which has been always the greatest problem of
interpretation of the mathematical formalism of quantum mechanics, is
unsolvable in the orthodox formulation of quantum theory. However it has
been recently resolved in a more general framework of the algebraic theory
of quantum systems which admits the superselection rules for the admissible
sets of observables defining the physical systems. The new superselection
rule, which we call quantum causality, or nondemolition principle, can be
formulated in short as a the following resolution of the corpuscular-wave
dualism: \emph{the past is classical (encoded into the trajectories of the
particles), and the future is quantum (encoded into the propensity waves for
these particles). } This principle does not apply, it simply does not exist
in the usual quantum theory with finite degrees of freedom. And there are no
events, jumps and trajectories and other physics in this theory if it is not
supplemented with the additional phenomenological interface rules such as
projection postulate, permanent reduction or a spontaneous localization
theory. This is why it is not applicable for our description of the open
quantum world from inside of this world as we were a part of this world, but
only for the external description of the whole\ of a closed physical system
as we were outside of this world. However the external description doesn't
allow to have a look inside the quantum system as any flow of information
from the quantum world which can be obtained only by performing a
measurement, will require an external measurement apparatus, and it will
inevitable open the system. This is why there is no solution of quantum
measurement and all paradoxes of quantum theory in the conventional,
external description.

As we demonstrated in the Section 5 on the simplest quantum measurement
model for the Schr\"{o}dinger's cat, this new superselection rule of quantum
causality explains the entanglement and decoherence and derives the
projection ``postulate'' in purely dynamical terms of quantum mechanics of
infinitely extended system supplemented with the conventional rules for the
statistical inference (statistical prediction by usual conditioning) from
the classical information theory and statistics. This provides a solution of
the instantaneous quantum measurement problem in its orthodox formulation.
The realistic measurements however are not instantaneous but have a
duration, and physically are performed even in the continuous time.

Recent dynamical models of the phenomenological theories for quantum jumps
and spontaneous localizations, although they all pretend to have a primary
value, extend in fact the instantaneous projection postulate to a certain,
counting class of the continuous in time measurements. The time which
appears in these theories is not the time at which the experimentalist
decides to make a measurement on the system, but the time which the system
does something for the experimenter to be observed. What it actually does
and why, remains unexplained mystery in these theories. As was shown Section
6, there is no need in supplementing the usual quantum mechanics with any of
such mysterious quantum spontaneous localization principles even if they are
formulated in continuous time. They all have been derived from the time
continuous unitary evolution for a generalized Dirac type Schr\"{o}dinger
equation, and `that something' what the system does to be spontaneously
observed, is simply caused by a singular scattering interaction at the
boundary of our Hamiltonian model. The quantum causality principle provides
a time continuous nondemolition counting measurement in the extended system
which enables to obtain `these stupid quantum jumps' simply by time
continuous conditioning called quantum jump filtering.

And even the continuous diffusive trajectories of quantum state diffusion
models have been derived from the usual Hilbert space unitary evolution
corresponding to the Dirac type boundary value problem for a Schr\"{o}dinger
infinite particles equation with a singular scattering interaction. As it is
shown in the Section 7, our causality principle admits to select a
continuous diffusive classical process in the quantum extended world which
satisfies the nondemolition condition with respect to all future of the
measured system. And this allows to obtain the continuous trajectories for
quantum state diffusion by simple filtering of quantum noise exactly as it
was done in the classical statistical nonlinear filtering and prediction
theory. In fact, the quantum state diffusion was first derived over 20 years
ago in the result of solving of a similar quantum prediction problem by
filtering the quantum white noise in a quantum stochastic Langevin model for
the continuous observation and optimal quantum feedback control. Thus the
``primary'' stochastic nonlinear irreversible quantum state diffusion occurs
to be the secondary, as it should be, to the deterministic linear unitary
reversible evolution, but in an extended system containing an infinite
number of auxiliary particles. And this is quantum causality who defines the
arrow of time by selecting what part of the reversible world is related to
the classical past and what is related to the quantum future. And this makes
the unitary group evolution irreversible in terms of the injective semigroup
of the Heisenberg transformations allowing the decoherence and the increase
of entropy in a purely dynamical way without any sort of reservoir averaging.

Our mathematical formulation of the extended quantum mechanics equipped with
the quantum causality to allow events and trajectories in the theory, is
just as continuous as Schr\"{o}dinger could have wished. However it doesn't
exclude the jumps which only appear in the singular interaction picture, are
there as a part of the theory but not only of its interpretation. Although
Schr\"{o}dinger himself didn't believe in quantum jumps, he tried several
times, although unsuccessfully, a possibility to obtain the continuous
reduction from a generalized, relativistic, ``true Schr\"{o}dinger''. He
envisaged that `if one introduces two symmetric systems of waves, which are
traveling in opposite directions; one of them presumably has something to do
with the known (or supposed to be known) state of the system at a later
point in time' \cite{Schr31}, then it would be possible to derive the
`verdammte Quantenspringerei' for the opposite wave as a solution of the
future-past boundary value problem. This desire coincides with the
``transactional'' \ attempt of interpretation of quantum mechanics suggested
in \cite{Crm86} on the basis that the relativistic wave equation yields in
the nonrelativistic limit two Schr\"{o}dinger type equations, one of which
is the time reversed version of the usual equation: `The state vector $\psi $
of the quantum mechanical formalism is a real physical wave with spatial
extension and it is identical with the initial ``offer wave'' of the
transaction. The particle (photon, electron, etc.) and the collapsed state
vector are identical with the completed transaction.' \ There was no proof
of this conjecture, and now we know that it is not even possible to derive
the quantum state diffusions, spontaneous jumps and single reductions from
such models involving only a finite particle state vectors $\psi \left(
t\right) $ satisfying the conventional Schr\"{o}dinger equation.

Our new approach based on the exactly solvable boundary value problems for
infinite particle states described in this paper, resolves this problem
formulated by Schr\"{o}dinger. And thus it resolves the old problem of
interpretation of the quantum theory, together with its famous paradoxes in
a constructive way by giving exact nontrivial models for allowing the
mathematical analysis of quantum observation processes determining the
phenomenological coupling constants and the reality underlying these
paradoxes. Conceptually it is based upon a new idea of quantum causality
called the nondemolition principle \cite{Be94} which divides the world into
the classical past, forming the consistent histories, and the quantum
future, the state of which is predictable for each such history. The
nondemolition principle defines what is actual in the reality and what is
only possible, what are the events and what are just the questions, and
selects from the possible observables the actual ones as the candidates for
Bell's \textit{be}ables. It was unknown to Bell who wrote that ``There is
nothing in the mathematics to tell what is `system' and what is `apparatus', 
$\cdots $'', in \cite{Bell87}, p.174). The mathematics of quantum open
systems and quantum stochastics defines the extended system by the product
of the commutative algebra of the output trajectories, the measured system,
and the noncommutative algebra of the input quantum waves. All output
processes in the apparatus are the beables which ``live'' in the center of
the algebra, and all other observables which are not in the system algebra,
are the input quantum noises of the measurement apparatus whose quantum
states are represented by the offer waves. These are the only possible
conditions when the posterior states exist as the results of inference
(filtering and prediction) of future quantum states upon the measurement
results of the classical past as beables. The act of measurement transforms
quantum propensities into classical realities. As Lawrence Bragg, another
Nobel prize winner, once said, everything in the future is a wave,
everything in the past is a particle.

\section{\textsc{Appendices}\textbf{\ }}

\subsection{On Bell's ``Proof'' that von Neumann's Proof was in Error\textbf{%
.}}

To ``disprove'' the von Neumann's theorem on the nonexistence of hidden
variables in quantum mechanics Bell \cite{Bell66} argued that the
dispersion-free states specified by a hidden parameter $\lambda $ should be
additive only for commuting pairs from the space $\frak{L}$ of all Hermitian
operators on the system Hilbert space $\frak{h}$. One can assume even less,
that the corresponding probability function $E\mapsto \left\langle
E\right\rangle _{\lambda }$ should be additive with respect to only
orthogonal decompositions in the subset $\mathcal{P}\left( \frak{h}\right) $
of all Hermitian projectors $E$, as only orthogonal events are
simultaneously verifiable by measuring an observable $\mathrm{L}\in \frak{L}$%
. In the case of finite-dimensional Hilbert space $\frak{h}$ it is
equivalent to the Bell's assumption, but we shall reformulate his only
counterexample it terms of the propositions, or events $E\in \mathcal{P}%
\left( \frak{h}\right) $ in order to dismiss his argument that this example
`is not dealing with logical propositions, but with measurements involving,
for example, differently oriented magnets' (p.6 in \cite{Bell87}).

Bell's hidden dispersion-free states were designed to reproduce the regular
quantum-mechanical states of two-dimensional space $\frak{h}=\mathbb{C}^{2}$%
. The regular pure quantum states are described by one-dimensional
projectors 
\begin{equation*}
\rho =\frac{1}{2}\left( \mathrm{I}+\sigma \left( \mathbf{r}\right) \right)
\equiv P\left( \mathbf{r}\right) ,\quad \sigma \left( \mathbf{r}\right)
=x\sigma _{x}+y\sigma _{y}+z\sigma _{z}
\end{equation*}
given by the points $\mathbf{r}=x\mathbf{e}_{x}+y\mathbf{e}_{y}+z\mathbf{e}%
_{z}$ on the unit sphere $\mathbf{S}\subset \mathbb{R}^{3}$ and Pauli
matrices $\sigma $. Bell assigned the simultaneously definite values 
\begin{equation*}
s_{\lambda }\left( \mathbf{e}\right) =\pm 1\equiv \left\langle \sigma \left( 
\mathbf{e}\right) \right\rangle _{\lambda },\quad \mathbf{e\in S}_{\lambda
}^{\pm }\left( \mathbf{r}\right)
\end{equation*}
to spin operators $\sigma \left( \mathbf{e}\right) $ describing the spin
projections in the directions $\mathbf{e\in S}$, which is specified by a
split of $\mathbf{S}$ into a positive $\mathbf{S}_{\lambda }^{+}\left( 
\mathbf{r}\right) $ and negative $\mathbf{S}_{\lambda }^{-}\left( \mathbf{r}%
\right) $ parts depending on the polarization $\mathbf{r}$ and a parameter $%
\lambda $. Due to 
\begin{equation*}
\sigma \left( -\mathbf{e}\right) =-\sigma \left( \mathbf{e}\right) ,\quad
\sigma \left( \mathbf{e}\right) ^{2}=\mathrm{I}
\end{equation*}
and $\left\langle \mathrm{I}\right\rangle _{\lambda }=1$, the values $\pm 1$
of $s_{\lambda }\left( \mathbf{e}\right) $ can be taken as dispersion-free
expectations $\left\langle \sigma \left( \mathbf{e}\right) \right\rangle
_{\lambda }$ of the projections $\sigma \left( \mathbf{e}\right) $ if $%
s_{\lambda }\left( -\mathbf{e}\right) =-s_{\lambda }\left( \mathbf{e}\right) 
$. The latter is achieved by a reflection-symmetric partition 
\begin{equation*}
\mathbf{S}_{\lambda }^{-}=-\mathbf{S}_{\lambda }^{+},\quad \mathbf{S}%
_{\lambda }^{-}\cup \mathbf{S}_{\lambda }^{+}=\mathbf{S,\quad S}_{\lambda
}^{-}\cap \mathbf{S}_{\lambda }^{+}=\emptyset
\end{equation*}
of the unit sphere $\mathbf{S}$ for each $\mathbf{r}$ and $\lambda $.
Obviously there are plenty of such partitions, and any will do, but Bell
took a special family 
\begin{equation*}
\mathbf{S}_{\lambda }^{\pm }\left( \mathbf{r}\right) =\left[ \mathbf{S}^{\pm
}\left( \mathbf{r}\right) \backslash \mathbf{S}_{\lambda }\left( \pm \mathbf{%
r}\right) \right] \cup \left[ \mathbf{S}^{\mp }\left( \mathbf{r}\right)
\backslash \mathbf{S}_{-\lambda }\left( \pm \mathbf{r}\right) \right] ,
\end{equation*}
where $\mathbf{S}^{\pm }$ are south and north hemispheres of the standard
reflection-symmetric partition with $\mathbf{r}$ pointing north. Although
this particular choice is not better than any other one, he parametrized 
\begin{equation*}
\mathbf{S}_{\lambda }\left( \mathbf{r}\right) =\left\{ \mathbf{e}\in \mathbf{%
S}:\mathbf{e\cdot r}<2\lambda \right\}
\end{equation*}
by $\lambda \in \left[ -\frac{1}{2},\frac{1}{2}\right] $ in such a way that 
\begin{equation*}
\int_{-1/2}^{1/2}s_{\lambda }\left( \mathbf{e}\right) \mathrm{d}\lambda =%
\mathrm{\Pr }\left\{ \lambda :\mathbf{S}_{\lambda }^{+}\left( \mathbf{r}%
\right) \ni \mathbf{e}\right\} -\mathrm{\Pr }\left\{ \lambda :\mathbf{S}%
_{\lambda }^{-}\left( \mathbf{r}\right) \ni \mathbf{e}\right\} =\mathbf{%
e\cdot r.}
\end{equation*}
Note that in his formula $\mathbf{r=e}_{z}$, but it can be extended also to
the case $\left| \mathbf{r}\right| \leq 1$ of not completely polarized
quantum states $\rho $ defining the quantum-mechanical expectations $%
\left\langle \sigma \left( \mathbf{e}\right) \right\rangle $ and quantum
probabilities $\Pr \left\{ P\left( \mathbf{e}\right) =1\right\} $ of the
propositions $E=P\left( \mathbf{e}\right) $ as the linear and affine forms
in the unit ball of all such $\mathbf{r}$: 
\begin{equation*}
\mathrm{Tr}\sigma \left( \mathbf{e}\right) \rho =\mathbf{e\cdot r,\quad }%
\mathrm{Tr}P\left( \mathbf{e}\right) \rho =\frac{1}{2}\left( 1+\mathbf{%
e\cdot r}\right) .
\end{equation*}
Each $\lambda $ assigns the zero-one probabilities $\left\langle P\left( \pm 
\mathbf{e}\right) \right\rangle _{\lambda }=\chi _{\mathbf{\lambda }}^{\pm
}\left( \mathbf{e}\right) $ given by the characteristic functions $\chi
_{\lambda }^{\pm }$ of $\mathbf{S}_{\lambda }^{\pm }$ simultaneously for all
quantum events $P\left( \pm \mathbf{e}\right) $, the eigen-projectors of $%
\sigma \left( \mathbf{e}\right) $ corresponding to the eigenvalues $\pm 1$: 
\begin{equation*}
P\left( \pm \mathbf{e}\right) =\frac{1}{2}\left( I\pm \sigma \left( \mathbf{e%
}\right) \right) \mapsto \chi _{\lambda }^{\pm }\left( \mathbf{e}\right) =%
\frac{1}{2}\left( 1\pm s_{\lambda }\left( \mathbf{e}\right) \right) \text{.}
\end{equation*}
The additivity of the probability function $E\mapsto \left\langle
E\right\rangle _{\lambda }$ in $\mathcal{P}\left( \frak{h}\right) =\left\{ 
\mathrm{O},P\left( \mathbf{S}\right) ,\mathrm{I}\right\} $ at each $\lambda $
follows from $\left\langle \mathrm{O}\right\rangle _{\lambda }=0$: 
\begin{equation*}
\left\langle \mathrm{O}\right\rangle _{\lambda }+\left\langle \mathrm{I}%
\right\rangle _{\lambda }=1=\left\langle \mathrm{O}+\mathrm{I}\right\rangle
_{\lambda },
\end{equation*}
as $\mathrm{O}+\mathrm{I}=\mathrm{I}$, and from $\chi _{\lambda }^{+}\left( -%
\mathbf{e}\right) =\chi _{\lambda }^{-}\left( \mathbf{e}\right) $: 
\begin{equation*}
\left\langle P\left( \mathbf{e}\right) \right\rangle _{\lambda
}+\left\langle P\left( -\mathbf{e}\right) \right\rangle _{\lambda
}=1=\left\langle P\left( \mathbf{e}\right) +P\left( -\mathbf{e}\right)
\right\rangle _{\lambda },
\end{equation*}
as $P\left( \mathbf{e}\right) +P\left( -\mathbf{e}\right) =\mathrm{I}$.

Thus a classical hidden variable theory reproducing the affine quantum
probabilities $\mathsf{P}\left( \mathbf{e}\right) =\left\langle P\left( 
\mathbf{e}\right) \right\rangle $ as the uniform mean value 
\begin{equation*}
\mathsf{M}\left\langle P\left( \mathbf{e}\right) \right\rangle _{\cdot
}=\int_{-1/2}^{1/2}\frac{1}{2}\left( 1+s_{\lambda }\left( \mathbf{e}\right)
\right) \mathrm{d}\lambda =\frac{1}{2}\left( 1+\mathbf{e\cdot r}\right) =%
\mathrm{Tr}P\left( \mathbf{e}\right) \rho
\end{equation*}
of the classical yes-no observables $\chi _{\cdot }^{+}\left( \mathbf{e}%
\right) =\left\langle P\left( \mathbf{e}\right) \right\rangle _{\cdot }$ was
constructed by Bell.

First let us note that it does not contradict to the von Neumann theorem
even if the latter is strengthened by the restriction of the additivity only
to the orthogonal projectors $E\in \mathcal{P}\left( \frak{h}\right) $. The
constructed dispersion-free expectation function $\mathrm{L}\mapsto
\left\langle \mathrm{L}\right\rangle _{\lambda }$ is not \emph{physically
continuous} on $\frak{L}$ because the value $\left\langle \mathrm{L}%
\right\rangle _{\lambda }=s_{\lambda }\left( \mathbf{l}\right) $ is one of
the eigenvalues $\pm 1$ for each $\lambda $, and it covers both values when
the directional vector $\mathbf{l}$ rotates continuously over the
three-dimensional sphere. A function $\mathbf{l\mapsto }\left\langle \sigma
\left( \mathbf{l}\right) \right\rangle _{\lambda }$ on the continuous
manifold (sphere) with discontinuous values can be continuous only if it is
constant, but this is ruled out by the impossibility to reproduce the
expectations $\left\langle \sigma \left( \mathbf{l}\right) \right\rangle =%
\mathbf{l\cdot r}$, which are linear in $\mathbf{l}$, by averaging the
function $\lambda \mapsto \left\langle \sigma \left( \mathbf{l}\right)
\right\rangle _{\lambda }$, constant in $\mathbf{l}$, over the $\lambda $.
Measurements of the projections of spin on the physically close directions
should be described by close expected values in any physical state specified
by $\lambda $, otherwise it cannot have physical meaning!

Indeed, apart from partial additivity (the sums are defined in $\mathcal{P}%
\left( \frak{h}\right) $ only for the orthogonal pairs from $\mathcal{P}%
\left( \frak{h}\right) $), the von Neumann theorem restricted to $\mathcal{P}%
\left( \frak{h}\right) \subset \frak{L}$ should also inherit the \emph{%
physical continuity}, induced by ultra-strong topology in $\frak{L}$. In the
finite dimensional case it is just ordinary continuity in the projective
topology $\frak{h}$, and in the case $\dim \frak{h}=2$ it is the continuity
on the projective space $\mathbf{S}$ of all one-dimensional projectors $%
P\left( \mathbf{e}\right) $, $\mathbf{e}\in \mathbf{S}$. It is obvious that
the zero-one probability function $E\mapsto \left\langle E\right\rangle
_{\lambda }$ constructed by Bell is not \emph{physically continuous} on the
restricted set: the characteristic function $\chi _{\lambda }^{+}\left( 
\mathbf{e}\right) =\left\langle P\left( \mathbf{e}\right) \right\rangle
_{\lambda }$ of the half-sphere $\mathbf{S}_{\lambda }^{+}\left( \mathbf{r}%
\right) $ is discontinuous in $\mathbf{e}$ on the whole sphere $\mathbf{S}$
for any $\lambda $ and $\mathbf{r}$. Measurements of the spin projections in
the physically close directions $\mathbf{e}_{n}\rightarrow \mathbf{e}$
should be described by close probabilities $\left\langle P\left( \mathbf{e}%
_{n}\right) \right\rangle _{\lambda }\rightarrow \left\langle P\left( 
\mathbf{e}\right) \right\rangle _{\lambda }$ in any physical state specified
by $\lambda $, otherwise the state cannot have physical meaning!

The continuity argument might be considered to be as purely mathematical,
but in fact it is not: even in classical probability theory with a discrete
phase space the pure states defined by Dirac $\delta $-measure, are
uniformly continuous, as any positive probability measure is on the space of
classical observables defined by bounded measurable functions on any
continuous phase space. In quantum theory an expectation defined as a linear
positive functional on $\frak{L}$ is also uniformly continuous, hence the
von Neumann assumption of physical (ultra-weak) continuity is only a
restriction in the infinite-dimensional case. Even if the state is defined
only on $\mathcal{P}\left( \frak{h}\right) \subset \frak{L}$ as a
probability function which is additive only on the orthogonal projectors,
the uniform continuity follows from its positivity in the case of $\dim 
\frak{h}\geq 3$. This follows from the Gleason's theorem \cite{Glea57} with
implication that Bell's dispersion-free states could exist only if the
Hilberet space of our whole universe had the dimensionality not more than
two.

In fact, Gleason obtained more than this: He proved that the case $\dim 
\frak{h}=2$ is the only exceptional one when a probability function on $%
\mathcal{P}\left( \frak{h}\right) $ (which should be countably additive in
the case $\dim \frak{h}=\infty $) may not be induced by a density operator $%
\rho $, and thus cannot be extended to a linear expectation on the operator
space $\frak{L}$. The irregular states cannot be extended by linearity on
the algebra of all (not just Hermitian) operators in $\frak{h}=\mathbb{C}%
^{2} $ even if they are continuous.

To rule out even this exceptional case form the Gleason's theorem we note
that an irregular state $E\mapsto \left\langle E\right\rangle $ on $\mathcal{%
P}\left( \mathbb{C}^{2}\right) $ cannot be composed with any state of an
additional quantum system even if the latter is given by a regular
probability function $\left\langle F\right\rangle =\mathrm{Tr}F\sigma $ on a
set $\mathcal{P}\left( \frak{f}\right) $ of ortho-projectors of another
Hilbert space. There is no additive probability function on the set $%
\mathcal{P}\left( \mathbb{C}^{2}\otimes \frak{f}\right) $of all verifiable
events for the compound quantum system described by a nontrivial Hilbert
space $\frak{f}$ such that 
\begin{equation*}
\left\langle E\right\rangle =\left\langle E\otimes \mathrm{I}\right\rangle
,\quad \left\langle \mathrm{I}\otimes F\right\rangle =\mathrm{Tr}F\sigma ,
\end{equation*}
where $\sigma =P_{\varphi }$ is the density operator of wave function $%
\varphi \in \frak{f}$. Indeed, if it could be possible for some $\frak{f}$
with $\dim \frak{f}>1$, it would be possible for $\frak{f}=\mathbb{C}^{2}$.
By virtue of Gleason's theorem any probability function which is additive
for orthogonal projectors on $\mathbb{C}^{2}\otimes \mathbb{C}^{2}=\mathbb{C}%
^{4}$ is regular on $\mathcal{E}\left( \mathbb{C}^{4}\right) $, given by a
density operator $\hat{\varrho}$. Hence 
\begin{equation*}
\left\langle E\right\rangle =\mathrm{Tr}\left( \mathrm{I}\otimes E\right) 
\hat{\varrho}=\mathrm{Tr}E\rho
\end{equation*}
i.e. the state on $\mathcal{P}\left( \mathbb{C}^{2}\right) $ is also
regular, with the density operator in $\frak{h}=\mathbb{C}^{2}$ given by the
partial trace 
\begin{equation*}
\rho =\mathrm{Tr}\left[ \hat{\varrho}|\frak{h}\right] =\mathrm{Tr}_{\frak{f}}%
\hat{\varrho}.
\end{equation*}

In order to obtain an additive product-state on $\mathcal{P}\left( \mathbb{C}%
^{2}\otimes \frak{f}\right) $ satisfying 
\begin{equation*}
\left\langle E\otimes F\right\rangle =\left\langle E\right\rangle \mathrm{Tr}%
FP_{\varphi },\quad E\in \mathcal{P}\left( \mathbb{C}^{2}\right) ,F\in 
\mathcal{P}\left( \frak{f}\right)
\end{equation*}
for a finite-dimensional $\frak{f}=\mathbb{C}^{n}$ with $n>1$ it is
necessary to define the state as an expectation on the whole unit ball $%
\mathcal{B}\left( \mathbb{C}^{2}\right) $ of the algebra $\frak{B}=\mathcal{B%
}\left( \mathbb{C}^{2}\right) $ of all (not just Hermitian) operators in $%
\mathbb{C}^{2}$. Indeed, any one-dimensional Hermitian projector in $\mathbb{%
C}^{2}\otimes \mathbb{C}^{n}=\mathbb{C}^{2n}$ can be described as an $%
n\times n$-matrix $\mathbf{E}=\left[ \mathrm{A}_{j}\mathrm{A}_{i}^{\dagger }%
\right] $ with $2\times 2$-entries $\mathrm{A}_{j}\in \mathcal{B}\left( 
\mathbb{C}^{2}\right) $, $j=1,\ldots ,n$ satisfying the normalization
condition 
\begin{equation*}
\sum_{j=1}^{n}\mathrm{A}_{j}^{\dagger }\mathrm{A}_{j}=P\left( \mathbf{e}%
\right) =\frac{1}{2}\left( I+\sigma \left( \mathbf{e}\right) \right)
\end{equation*}
for some $\mathbf{e\in S}$. These entries have the form 
\begin{equation*}
\mathrm{A}=\alpha P\left( \mathbf{e}\right) +aQ\left( \mathbf{e}_{\bot
}\right) ,\quad Q\left( \mathbf{e}_{\bot }\right) =\frac{1}{2}\sigma \left( 
\mathbf{e}_{\bot }\right) ,
\end{equation*}
where $\mathbf{e}_{\bot }$ is an orthogonal complex vector such that 
\begin{equation*}
i\mathbf{e}_{\bot }\mathbf{\times e}=\mathbf{e}_{\bot }\quad \mathbf{\bar{e}}%
_{\bot }\cdot \mathbf{e}_{\bot }=2,\quad i\mathbf{\bar{e}}_{\bot }\times 
\mathbf{e}_{\bot }=2\mathbf{e},
\end{equation*}
and $\sum \left( \left| \alpha _{j}^{2}\right| +\left| a_{j}^{2}\right|
\right) =1$ corresponding to $\mathrm{Tr}\mathbf{E}=1$. The matrix elements 
\begin{equation*}
\mathrm{A}_{j}\mathrm{A}_{i}^{\dagger }=\alpha _{j}\bar{\alpha}_{i}P\left( 
\mathbf{e}\right) +a_{j}\bar{a}_{i}P\left( -\mathbf{e}\right) +\alpha _{j}%
\bar{a}_{i}Q\left( \mathbf{\bar{e}}_{\bot }\right) +a_{j}\bar{\alpha}%
_{i}Q\left( \mathbf{e}_{\bot }\right)
\end{equation*}
for these orthoprojectors in $\mathbb{C}^{2n}$ are any matrices from the
unit ball $\mathcal{B}\left( \mathbb{C}^{2}\right) $, not just Hermitian
orthoprojectors. By virtue of Gleason's theorem the product-state of such
events $\mathbf{E}$ must be defined by the additive probability 
\begin{equation*}
\left\langle \mathbf{E}\right\rangle =\sum_{i,j=1}^{n}\varphi ^{j}\varrho
\left( \mathrm{A}_{j}\mathrm{A}_{\iota }^{\dagger }\right) \varphi
^{i}=\varrho \left( \mathrm{B}\right) ,
\end{equation*}
where $\mathrm{B}=\mathrm{A}\left( \varphi \right) \mathrm{A}\left( \varphi
\right) ^{\dagger }=\beta I+\sigma \left( \mathbf{b}\right) $ is given by $%
\alpha \left( \varphi \right) =\varphi ^{j}\alpha _{j}$, $a\left( \varphi
\right) =\varphi ^{j}a_{j}$ for $\varphi \in \mathbb{C}^{n}$ with the
components $\varphi ^{j}=\bar{\varphi}_{j}$, and $\varrho \left( \mathrm{B}%
\right) =\mathrm{TrB}\rho $ is the linear expectation 
\begin{equation*}
\varrho \left( \mathrm{B}\right) =\frac{1}{2}\left( \beta _{+}\left(
1+r_{1}\right) +\beta _{-}\left( 1-r_{1}\right) +b_{\bot }\bar{r}_{\bot }+%
\bar{b}_{\bot }r_{\bot }\right) =\beta +\mathbf{b\cdot r}
\end{equation*}
with $r_{1}=\mathbf{e\cdot r}$, $\mathbf{r}_{\bot }=\mathbf{e}_{\bot }%
\mathbf{\cdot r}$, $\beta _{+}=\left| \alpha \left( \varphi \right) \right|
^{2}$, $\beta _{-}=\left| a\left( \varphi \right) \right| ^{2}$, $b_{\bot
}=\alpha \left( \varphi \right) \overline{a\left( \varphi \right) }$. It
these terms we can formulate the definition of a regular state without
assuming a priori the linearity and even continuity conditions also for the
case $\frak{h}=\mathbb{C}^{2}$.

Thus we proved that in order to formulate the \emph{quantum composition
principle} for a physical system described by a Hilbert space $\frak{h}$ we
need the quantum state to be defined on the unite ball $\mathcal{B}\left( 
\frak{h}\right) $ rather than just on the set $\mathcal{P}\left( \frak{h}%
\right) $ of the orthoprojectors in $\frak{h}$. The following definition
obviously rules out the Bell's hidden variable states even in the case $%
\frak{h}=\mathbb{C}^{2}$ as unphysical.

A complex-valued map $\mathrm{B}\mapsto \varrho \left( \mathrm{B}\right) $
on the unit ball $\mathcal{B}\left( \frak{h}\right) $ normalized as $\varrho
\left( \mathrm{I}\right) =1$ is called\emph{\ state} for a quantum system
described by the Hilbert space $\frak{h}$ (including the case $\dim \frak{h}%
=2$) if it is positive on all Hermitian projective matrices $\mathbf{E}=%
\left[ \mathrm{A}_{j}\mathrm{A}_{k}^{\dagger }\right] $ with entries $%
\mathrm{A}_{j}\in \mathcal{B}\left( \frak{h}\right) $ in the sense 
\begin{equation*}
\sum_{j}\mathrm{A}_{j}^{\dagger }\mathrm{A}_{j}=P\in \mathcal{P}\left( \frak{%
h}\right) \Rightarrow \varrho \left( \mathbf{E}\right) =\left[ \varrho
\left( \mathrm{A}_{j}\mathrm{A}_{k}^{\dagger }\right) \right] \geq 0,
\end{equation*}
of positive-definiteness of the matrices $\varrho \left( \mathbf{E}\right) $
with the complex entries $\left[ \varrho \left( \mathrm{A}_{j}\mathrm{A}%
_{k}^{\dagger }\right) \right] .$ It is called a \emph{regular state} if 
\begin{equation*}
\varrho \left( E\otimes P_{\varphi }\right) =\varrho \left( E\right)
P_{\varphi }
\end{equation*}
for any one-dimensional projector $P_{\varphi }=\left[ \varphi _{j}\varphi
_{i}^{\dagger }\right] $, and\ if it is countably-additive with respect to
the orthogonal decompositions $\mathbf{E}=\sum \mathbf{E}\left( k\right) $: 
\begin{equation*}
\sum_{j}\mathrm{A}_{j}\left( i\right) ^{\dagger }\mathrm{A}_{j}\left(
k\right) =0,\forall i\neq k\Rightarrow \varrho \left( \sum_{k}\mathrm{A}%
_{j}\left( k\right) \mathrm{A}_{j}\left( k\right) ^{\dagger }\right)
=\sum_{k}\varrho \left( \mathrm{A}_{j}\left( k\right) \mathrm{A}_{i}\left(
k\right) ^{\dagger }\right) .
\end{equation*}

It obvious that the state thus defined can be uniquely extended to a regular
product-state on $\mathcal{P}\left( \frak{h}\otimes \mathbb{C}^{n}\right) $
by 
\begin{equation*}
\sum_{j,k}\bar{\varphi}_{j}\varrho \left( \mathrm{A}_{j}\mathrm{A}%
_{i}^{\dagger }\right) \varphi _{i}\geq 0,\quad \forall \varphi _{j}\in 
\mathbb{C},\quad \sum \left| \varphi _{j}\right| =1,
\end{equation*}
which proves that it is continuous and is given by a density operator: $%
\varrho \left( \mathrm{B}\right) =\mathrm{TrB}\rho $. Thus the composition
principle rules out the existence of the hidden variable representation for
the quantum systems.

In conclusion we note that the Bells example in fact doesn't prove the
existence of hidden variables from any reasonable probabilistic point of
view even in this exceptional case $\frak{h}=\mathbb{C}^{2}$. Indeed, his
mean $\mathsf{M}$ over $\lambda $ cannot be considered as the conditional
averaging of a classical partially hidden world with respect to the quantum
observable part. If this were so, not only the mean values of the classical
hidden variables but also their momenta would reproduce the regular quantum
expectations as linear functionals of $\rho $. Bell's model however gives
nonlinear expectations with respect to the states $\rho $ even if it is
restricted to the smallest commutative algebra generated by the
characteristic functions $\left\{ \chi _{\cdot }^{+}\left( \mathbf{e}\right)
:\mathbf{e}\in \mathbf{S}\right\} $ of the subsets $\left\{ \lambda :\mathbf{%
S}_{\lambda }^{+}\left( \mathbf{r}\right) \ni \mathbf{e}\right\} $. One can
see this by the uniform averaging of the commutative products $\chi
_{\lambda }^{+}\left( \mathbf{e}\right) \chi _{\lambda }^{+}\left( \mathbf{f}%
\right) $: such mean values ( i.e. the second order moments) are affine with
respect to $\mathbf{r}$ only for colinear $\mathbf{e}$ and $\mathbf{f}\in 
\mathbf{S}$.

\subsection{Quantum Markov Chains and Stochastic Recurrences}

Let $\frak{h}$, $\frak{g}_{o}$ be Hilbert spaces, $\mathrm{V}$ be an
isometry $\frak{h}\mapsto \frak{g}_{o}\otimes \frak{h}$, $\mathrm{V}%
^{\dagger }\mathrm{V}=\mathrm{I}$, and $\mathcal{A}$ and $\mathcal{B}$ be
von Neumann algebras respectively on $\frak{g}$ and $\frak{h}$ and there
exists a normal representation $\iota $ of the commutant $\mathcal{B}%
^{\prime }=\left\{ \mathrm{B}^{\prime }:\left[ \mathrm{B}^{\prime },\mathrm{B%
}\right] =0,\forall \mathrm{B}\in \mathcal{B}\right\} $ into the commutant
of $\mathcal{A}\otimes \mathcal{B}$ which is intertwined by this isometry$:$%
\begin{equation}
\iota \left( \mathrm{B}^{\prime }\right) \mathrm{V}=\mathrm{VB}^{\prime
}\quad \forall \mathrm{B}^{\prime }\in \mathcal{B}^{\prime }.  \label{A2.0}
\end{equation}
The pair $\left( \mathrm{V},\iota \right) $ defines the \emph{standard
representation} of a $\mathcal{B}$\emph{-transitional probability map} 
\begin{equation}
\pi \left( \mathrm{A},\mathrm{B}\right) =\mathrm{V}^{\dagger }\left( \mathrm{%
A}\otimes \mathrm{B}\right) \text{\textrm{V}}\in \mathcal{B}\quad \forall 
\mathrm{A}\in \mathcal{A}_{o},\mathrm{B}\in \mathcal{B}  \label{A2.1}
\end{equation}
\emph{\ }with an \emph{output algebra} $\mathcal{A}$. The transitional maps
with the output $\mathcal{A}=\mathcal{B}$ on the Hilbert space copy $\frak{g}%
=\frak{h}$ were introduced by Accardi \cite{Acc74} for the case $\mathcal{B}=%
\mathcal{B}\left( \frak{h}\right) $ (in this simple case one can take $%
\mathcal{A}=\mathcal{B}\left( \frak{g}\otimes \frak{h}\right) $ and $\iota
\left( \alpha \mathrm{I}\right) =\alpha \mathrm{I}\otimes \mathrm{I}$ on the
one-dimensional algebra $\mathcal{B}^{\prime }=\mathbb{C}\mathrm{I}$). Note
that the intertwining commutant condition (\ref{A2.0}) which was established
for any conditional state in \cite{BeSt84} is nontrivial if only $\mathcal{B}
$ is smaller than the whole operator algebra $\mathcal{B}\left( \frak{h}%
\right) $. Every normal completely positive unital map $\kappa :\mathcal{A}%
\rightarrow \mathcal{B}$ can be extended to a $\mathcal{B}$-transitional
probability map in the standard representation of $\mathcal{A}$ such that $%
\kappa \left( \mathrm{A}\right) =\mathrm{V}^{\dagger }\left( \mathrm{I}%
\otimes \mathrm{A}\right) \mathrm{V}$ \cite{BeSt84}.

We shall say that the $\mathcal{B}$-transitional map $\pi $ is given in a 
\emph{stochastic representation} if 
\begin{equation}
\pi \left( \mathrm{A},\mathrm{B}\right) =\mathrm{F}^{\dagger }\left( \tau
\left( \mathrm{A}\otimes \mathrm{B}\right) \right) \text{\textrm{F}}\in 
\mathcal{B}\quad \forall \mathrm{A}\in \mathcal{A},\mathrm{B}\in \mathcal{B},
\label{A2.2}
\end{equation}
where $\mathrm{F}$ is an isometry $\frak{h}\rightarrow \frak{f}\otimes \frak{%
h}$ from $\frak{f}\otimes \mathcal{B}$ in the sense $\mathrm{FB}^{\prime
}=\left( \mathrm{I}\otimes \mathrm{B}^{\prime }\right) \mathrm{F}$ for all $%
\mathrm{B}^{\prime }\in \mathcal{B}^{\prime }$, and $\tau $ is a normal, not
necessarily unital representation of $\mathcal{A}$ in the algebra $\mathcal{B%
}\left( \frak{f}\right) \otimes \mathcal{B}$. Every normal completely
positive unital map $\kappa :\mathcal{A}\rightarrow \mathcal{B}$ has a
stochastic representation ( \cite{Dix81}, p. 61) with $\mathrm{F}=\mathrm{T}%
^{\dagger }\mathrm{V}$ and $\tau \left( \mathrm{A}\right) =\mathrm{T}%
^{\dagger }\mathrm{AT}$ given by a partial isometry $\mathrm{T}:\frak{f}%
\otimes \frak{h}\rightarrow \frak{g}\otimes \frak{h}$. such that 
\begin{equation}
\mathrm{T}\left( \mathrm{I}\otimes \mathrm{B}^{\prime }\right) \mathrm{T}%
^{\dagger }=\iota \left( \mathrm{B}^{\prime }\right) ,\quad \forall \mathrm{B%
}^{\prime }\in \mathcal{B}^{\prime }  \label{A2.3}
\end{equation}
The representations (\ref{A2.1}) and (\ref{A2.2}) are called respectively 
\emph{stochastic} and \emph{standard} if $\frak{g}=\frak{h}\otimes \frak{k}=%
\frak{f}$ and 
\begin{equation*}
\left( \mathrm{B}^{\prime }\otimes \mathrm{A}_{o}\otimes \mathrm{B}\right) 
\mathrm{T}=\mathrm{T}\left( \mathrm{B}\otimes \mathrm{A}_{o}\otimes \mathrm{B%
}^{\prime }\right) ,\quad \forall \mathrm{A}_{o}\in \mathcal{A}_{o},\mathrm{B%
}^{\prime }\in \mathcal{B}^{\prime }.
\end{equation*}
such that $\mathcal{A}=\mathrm{I}\otimes \mathcal{A}_{o}$ and $\iota \left( 
\mathrm{B}^{\prime }\right) =\mathrm{B}^{\prime }\otimes \mathrm{I}$ (e.g. $%
\mathrm{T}$ is tensor transposition $\mathrm{T}\left( \varphi \otimes \chi
\otimes \psi \right) =\psi \otimes \chi $ $\otimes \varphi $). Every normal
completely positive unital map $\pi $ has a standard stochastic
representation if it is dominated by a normal faithful state (or weight) on $%
\mathcal{A}_{o}\otimes \mathcal{B}$ in the sense of \cite{BeSt86}. If $\frak{%
g}=L_{\mu }^{2}$ and $\mathcal{B}$ is commutative, the diagonal algebra $%
\mathcal{B}=\mathcal{D}\left( \frak{h}\right) \simeq L_{\mu }^{\infty }$ on
the space $\frak{h}=L_{\mu }^{2}$ of square-integrable function $\psi \left(
y\right) $ say, the conditional probability in the standard stochastic
representation is described by the transposition $\mathrm{V}=\mathrm{T}_{o}%
\hat{v}$ of the operator 
\begin{equation*}
\left( \hat{v}\psi \right) \left( y,x\right) =v\left( y,x\right) \psi \left(
y\right) ,\quad \int \left| \nu \left( y,x\right) \right| ^{2}\mathrm{d}\mu
_{x}=1
\end{equation*}
from $L_{\mu }^{2}$ into $L_{\mu }^{2}\otimes \frak{g}$, the multiplication
by a measurable function $v$ with normalized vector values $v\left( y\right)
\in \frak{g}$, $\left\| v\left( y\right) \right\| =1$ for almost all $y$: 
\begin{equation*}
\left( \mathrm{V}\psi \right) \left( x,y\right) =\left( \mathrm{T}_{o}\hat{v}%
\psi \right) \left( x,y\right) =\left( \hat{v}\psi \right) \left( x,y\right)
=v\left( x,y\right) \psi \left( x\right) ,\quad \forall \psi \in L_{\mu
}^{2}.
\end{equation*}
The conditional probability amplitude $v$ defines the usual conditional
probability with the density $\left| v\left( x,y\right) \right| ^{2}$ for
the transitions $y_{0}=x\mapsto y$ with respect to a given measure $\mu $,
and only such conditional probabilities which are absolutely continuous have
standard stochastic representation.

Every conditional probability map defines a quantum $\mathcal{B}$\emph{%
-Markov chain} 
\begin{equation}
\pi \left( t,\mathrm{A}^{1},\ldots ,\mathrm{A}^{t},\mathrm{B}\right) =%
\mathrm{V}^{\dagger }\left( \mathrm{A}^{1}\otimes \cdots \left( \mathrm{V}%
^{\dagger }\left( \mathrm{A}^{t}\otimes \mathrm{B}\right) \mathrm{V}\right)
\ldots \right) \mathrm{V},\quad t\in \mathbb{N}  \label{A2.4}
\end{equation}
with an output algebra $\mathcal{A}_{o}\subseteq \mathcal{B}\left( \frak{g}%
\right) $ over an operator algebra $\mathcal{B}\subseteq \mathcal{B}\left( 
\frak{h}\right) $. Quantum Markov chains with an output were introduced in 
\cite{Be79} as effectively described by the \emph{quantum recurrency equation%
} 
\begin{equation*}
\pi \left( t,\mathrm{A}_{0}^{t]},\mathrm{B}\right) =\mathrm{V}\left(
t\right) ^{\dagger }\left( \mathrm{A}^{1}\otimes \cdots \otimes \mathrm{A}%
^{t}\otimes \mathrm{B}\right) \mathrm{V}\left( t\right) =\pi \left( t-1,%
\mathrm{A}_{0}^{t-1]},\mathrm{V}^{\dagger }\left( \mathrm{A}^{t}\otimes 
\mathrm{B}\right) \mathrm{V}\right) .
\end{equation*}
(see also an earlier definition in \cite{Acc74} for the simple standard case 
$\mathcal{B}=\mathcal{B}\left( \frak{h}\right) $). Here $\mathrm{A}%
_{0}^{t]}=\left( \mathrm{A}^{1},\ldots ,\mathrm{A}^{t}\right) $ is a
sequence of indexed elements $\mathrm{A}^{t}\in \mathcal{A}_{o}$ and $%
\mathrm{V}\left( t\right) =\mathrm{V}_{t}\mathrm{V}\left( t-1\right) $ with $%
\mathrm{V}\left( 0\right) =\mathrm{I}$ is the isometry $\frak{h}\rightarrow 
\frak{g}_{0}^{t]}\otimes \frak{h}$, where $\frak{g}_{0}^{t]}=\otimes
_{0<r\leq t}\frak{g}_{s}$ is the tensor product of the copies of $\frak{g}%
_{o}$.. This isometry is given by the recurrency equation 
\begin{equation}
\mathrm{V}\left( t\right) =\mathrm{V}_{t}\mathrm{V}\left( t-1\right) ,\quad 
\mathrm{V}\left( 0\right) =\mathrm{I}  \label{A2.9}
\end{equation}
defined by\emph{\ }$\mathrm{V}_{t}=\mathrm{I}_{0}^{t}\otimes \mathrm{V}$,
where $\mathrm{I}_{0}^{t}=\otimes _{0<r<t}\mathrm{I}_{r}$ with all $\mathrm{I%
}_{s}=\mathrm{I}_{o}$ is the identity in $\frak{g}_{0}^{t}=\otimes _{0<r<t}%
\frak{g}_{r}$. In particular, if there is no output, $\mathcal{A}_{o}=%
\mathbb{C}$ corresponding to $\frak{g}=\mathbb{C}$, the operator $\mathrm{V}$
is an isometry $\frak{h}\rightarrow \frak{h}$ which is often unitary, $%
\mathrm{V}^{\dagger }=\mathrm{V}^{-1}$, as it is always in the case of
finite dimensional $\frak{h}$. The usual, Heisenberg dynamics $\pi \left( t,%
\mathrm{B}\right) =\mathrm{V}^{-t}\mathrm{BV}^{t}$ induced by the unitary $%
\mathrm{V}$ in $\frak{h}$ over the algebra $\mathcal{B}=\mathcal{B}\left( 
\frak{h}\right) $ doesn't admit a nontrivial output $\mathcal{A}_{o}\neq 
\mathbb{C}$. Note that Markov chains over the maximal commutative algebras $%
\mathcal{A}=\mathcal{C}=\mathcal{B}$ on $\frak{g}=L_{\mu }^{2}=\frak{h}$,
given as the standard stochastic representation by $\mathrm{V}=\hat{v}$, are
equivalent to the classical Markov chains described by the conditional
probability densities $p\left( y|x\right) =\left| v\left( y|x\right) \right|
^{2}$ with the standard classical output $\mathcal{A}_{o}=\mathcal{C}\simeq
L_{\mu }^{\infty }$.

Now we shall prove that every Markov chain with an output $\mathcal{A}$ and
arbitrary $\mathcal{B}$ can be induced by a Heisenberg dynamics on the
infinite tensor product algebra $\frak{B}=\frak{A}^{0]}\otimes \mathcal{B}%
\otimes \frak{A}_{0}$ embedding all tensor powers $\mathcal{A}_{o}^{\otimes
t}$ into the output algebra $\frak{A}^{0]}=\otimes _{s\leq 0}\mathcal{A}_{s}$
as $I^{s}\otimes \mathcal{A}_{0}^{\otimes t}$ by $\mathcal{A}_{s}=\mathbb{C}%
\oplus \mathcal{A}_{o}$, $s\leq 0$. This induction can be achieved by a
unitary semigroup $U^{t}$ for all $t\in \mathbb{N}$ as 
\begin{equation}
\pi \left( t,\mathrm{A}_{0}^{t]},\mathrm{B}\right) =\left( \Phi ^{0]}\otimes 
\mathrm{I}\otimes \Phi _{0}^{\circ }\right) ^{\dagger }U^{-t}\left(
I^{t}\otimes \mathrm{A}^{1}\otimes \ldots \otimes \mathrm{A}^{t}\otimes 
\mathrm{B}\otimes I_{0}\right) U^{t}\left( \Phi ^{0]}\otimes \mathrm{I}%
\otimes \Phi _{0}^{\circ }\right) ,  \label{A2.5}
\end{equation}
where $\Phi _{0}^{\circ }=\otimes _{s>0}\chi _{s}^{\circ }$ is the input
vacuum-vector given by the infinite tensor product of the copies of a unit
vector $\chi ^{\circ }\in \frak{f}$, and $\Phi ^{0]}$ is arbitrary
(normalized) in the output Hilbert space $\mathcal{G}^{0]}=\otimes _{s\leq 0}%
\frak{g}_{s}$ with all $\frak{g}_{s}=\frak{g}$, . In the case of commutative 
$\mathcal{A}=\mathcal{D}\left( \frak{g}\right) $ this construction will
provide the dynamical solution to the quantum measurement problem as stated
in the Section 5. We shall give an explicit, canonical construction of this
dilation. Other constructions are also possible and known in the simple case 
$\mathcal{B=B}\left( \frak{h}\right) $ at least for finite-dimensional $%
\frak{h}$ \cite{Kum85, KuMa87}. In general we need an extension of the
spaces $\frak{g}$ and $\frak{f}$, and this is why we denote them here as $%
\frak{g}_{o}$ and $\frak{f}_{o}$. For simplicity we shall also assume that
the stochastic and standard representations are intertwined by a unitary
operator $\mathrm{T}$ called transposition and denoted also as $\mathrm{T}%
_{o}$.

First, let us consider the canonical unitary dilation of the stochastic
isometry $\mathrm{V}$ intertwining the algebras $\iota \left( \mathcal{B}%
^{\prime }\right) $ and $\mathcal{B}^{\prime }$. It is uniquely defined on $%
\frak{h}\oplus \left( \frak{h}\otimes \frak{f}_{o}\right) $ up to a unitary
transformation $\mathrm{e}^{-i\mathrm{E}/\hbar }\in \mathcal{B}$ by 
\begin{equation}
\mathrm{U}=\left[ 
\begin{tabular}{ll}
$\mathrm{O}$ & $\mathrm{V}^{\ddagger }$ \\ 
$\mathrm{V}$ & $\mathrm{G}$%
\end{tabular}
\right] ,\quad \mathrm{G}=\left( \mathrm{I}_{o}\otimes \mathrm{e}^{-i\mathrm{%
E}/\hbar }\right) \mathrm{T}_{o}-\mathrm{Ve}^{i\mathrm{E}/\hbar }\mathrm{V}%
^{\ddagger },  \label{A2.6}
\end{equation}
where $\mathrm{V}^{\ddagger }=\mathrm{V}^{\ast \dagger }$ and $\mathrm{V}%
^{\ast }=\mathrm{T}_{o}^{\dagger }\left( \mathrm{I}\otimes \mathrm{e}^{i%
\mathrm{E}/\hbar }\right) \mathrm{Ve}^{i\mathrm{E}/\hbar }$. Here for the
notational convenience $\mathrm{U}$ is represented as an operator on $\frak{h%
}\otimes \frak{f}$ into $\frak{g}\otimes \frak{h}$ with $\frak{f}=\mathbb{%
C\oplus }\frak{f}_{o}$ and $\frak{g}=\mathbb{C\oplus }\frak{g}_{o}$ which is
done by the ``transposition'' $\mathrm{T}_{o}:\frak{h}\otimes \frak{f}%
_{o}\rightarrow \frak{g}_{o}\otimes \frak{h}$ intertwining the algebra and $%
\mathcal{B}\otimes \mathrm{I}_{o}$ and \textrm{I}$_{o}\otimes \mathcal{B}$
as in the standard stochastic case, where $\frak{f}_{o}=\frak{k}\otimes 
\frak{h}$ and $\frak{g}_{o}=\frak{h}\otimes \frak{k}$:. The unitarity $%
\mathrm{U}^{-1}=\mathrm{U}^{\dagger }$ of the operator $\mathrm{U}$ follows
from $\mathrm{T}^{\dagger }\mathrm{T}=\mathrm{I}\oplus \left( \mathrm{I}%
\otimes \mathrm{I}_{o}\right) =\mathrm{I}\otimes \mathrm{I}$, 
\begin{equation*}
\mathrm{U}^{\dagger }\mathrm{U}=\left[ 
\begin{tabular}{ll}
$\mathrm{O}$ & $\mathrm{V}^{\dagger }$ \\ 
$\mathrm{V}^{\ast }$ & $\mathrm{G}^{\dagger }$%
\end{tabular}
\right] \left[ 
\begin{tabular}{ll}
$\mathrm{O}$ & $\mathrm{V}^{\ddagger }$ \\ 
$\mathrm{V}$ & $\mathrm{G}$%
\end{tabular}
\right] =\left[ 
\begin{tabular}{ll}
$\mathrm{V}^{\dagger }\mathrm{V}$ & $\mathrm{V}^{\dagger }\mathrm{G}$ \\ 
$\mathrm{G}^{\dagger }\mathrm{V}$ & $\mathrm{V}^{\ast }$\textrm{V}$%
^{\ddagger }+\mathrm{G}^{\dagger }\mathrm{G}$%
\end{tabular}
\right] =\left[ 
\begin{tabular}{ll}
$\mathrm{I}$ & $\mathrm{O}$ \\ 
$\mathrm{O}$ & $\mathrm{I}\otimes \mathrm{I}_{o}$%
\end{tabular}
\right]
\end{equation*}
and the selfadjointness of $\mathrm{S}=\mathrm{e}^{i\mathrm{E}/\hbar }%
\mathrm{T}^{\dagger }\mathrm{U}$. Moreover, as it follows straightforward
from the construction of $\mathrm{U}$, 
\begin{equation*}
\left( \mathrm{B}^{\prime }\oplus \iota ^{-1}\left( \mathrm{B}^{\prime
}\right) \right) \mathrm{U}=\mathrm{U}\left( \mathrm{B}^{\prime }\oplus
\left( \mathrm{B}^{\prime }\otimes \mathrm{I}_{o}\right) \right) =\mathrm{U}%
\left( \mathrm{B}^{\prime }\otimes \mathrm{I}\right) \quad \forall \mathrm{B}%
^{\prime }\in \mathcal{B}^{\prime }
\end{equation*}
on $\frak{h}\otimes \frak{f}$, where $\mathrm{I}=1\oplus \mathrm{I}_{o}$ is
the identity on $\frak{f}$. This implies $\mathrm{U}^{\dagger }\left( 
\mathrm{A}\otimes \mathrm{B}\right) \mathrm{U}\in \left( \mathcal{B}^{\prime
}\otimes \mathrm{I}\right) ^{\prime }$ i.e. $\mathrm{U}^{\dagger }\left( 
\mathcal{A}_{0}\otimes \mathcal{B}\right) \mathrm{U}\subseteq \mathcal{B}%
\otimes \mathcal{A}_{1}$ in the sense \ 
\begin{equation*}
\mathrm{U}^{\dagger }\left( \mathrm{A}\otimes \mathrm{B}\right) \mathrm{U}%
\in \mathcal{B}\otimes \mathcal{A}_{1},\quad \forall \mathrm{A}\in \mathcal{A%
}_{0},\mathrm{B}\in \mathcal{B},
\end{equation*}
where $\mathcal{A}_{0}=\mathbb{C\oplus }\mathcal{A}_{o}$ and $\mathcal{A}%
_{1}=\mathcal{B}\left( \frak{f}\right) $. One can always obtain such $%
\mathrm{U}$ as $\mathrm{U}=\mathrm{TW}$ from its \emph{stochastic
representation} $\mathrm{W}\in \mathcal{B}\otimes \mathcal{B}\left( \frak{f}%
\right) $ as commuting with all $\mathrm{B}^{\prime }\otimes \mathrm{I}$ by
the ``transposition'' $\mathrm{T}=\mathrm{I}\oplus \mathrm{T}_{o}$ of $\frak{%
h}\otimes \frak{f}$ onto $\frak{g}\otimes \frak{h}$.

Second, we construct the local auxiliary spaces $\frak{g}_{r}^{t]}$ as the
Hilbert tensor products $\otimes _{r<z\leq t}\frak{g}_{z}$ of the copies $%
\frak{g}_{z}=\frak{f}$ for $z>0$ and $\frak{g}_{z}=\frak{g}$ for $z\leq 0$
indexed by $z\in \mathbb{Z}$, and define the Hilbert spaces of the past
output $\mathcal{G}^{t]}$, and its future $\mathcal{G}_{t}$ as the
state-vector spaces of semifinite discrete strings generated by the infinite
tensor products $\Phi ^{t]}=\otimes _{z\leq t}\chi _{z}$, $\Phi _{t}=\otimes
_{z>t}\chi _{z}$ with all but finite number of $\chi _{z}\in \frak{g}_{z}$
equal to the initial states $\chi _{z}^{\circ }$, the copies of $\chi
^{\circ }=1\oplus 0$ in $\frak{g}$ or in $\frak{f}$. Denoting by $\mathcal{A}%
_{z}$ the copies of $\mathcal{A}_{1}=\mathcal{B}\left( \frak{f}\right) $ if $%
z>0$, and the copies of $\mathcal{A}_{0}=\mathbb{C}\oplus \mathcal{A}_{o}$
if $z\leq 0$ on the corresponding spaces $\frak{g}_{z}$, we construct the
local algebras $\mathcal{A}_{r}^{t]}=\otimes _{r<z\leq t}\mathcal{A}_{z}$ on 
$\frak{g}_{r}^{t]}$, and the past (output) $\frak{A}^{t]}$ and the future
(input) $\frak{A}_{t}$ algebras generated respectively on $\mathcal{G}^{t]}$
and on $\mathcal{G}_{t}$ by the operators $I^{t}\otimes \mathrm{A}\otimes 
\mathrm{I}_{s}^{t]}$ for all $s<t$, and by $\mathrm{I}_{t}^{s}\otimes 
\mathrm{A}\otimes I_{s}$ for all $s>t$, where $\mathrm{A}\in \mathcal{A}_{s}$%
, $\mathrm{I}_{r}^{t]}=\otimes _{r<s\leq t}\mathrm{I}_{s}$, $I^{t}=\mathrm{I}%
_{-\infty }^{t-1]}$, and $I_{t}=\otimes _{s>t}\mathrm{I}_{s}$ is the
identity operator in $\mathcal{G}_{t}$. The Hilbert space $\mathcal{G}=%
\mathcal{G}^{0]}\otimes \mathcal{G}_{0}$, the von Neumann algebra $\frak{A}=%
\frak{A}^{0]}\otimes \frak{A}_{0}$, and the product vector $\Phi =\otimes
\chi _{s}^{\circ }\equiv \Phi ^{\circ }$, which is an eigen-vector for any
operator $A_{s}=I^{s}\otimes \mathrm{A}\otimes I_{s}$ with $s<0$ in $%
\mathcal{G}$, stand as candidates for the triple $\left( \mathcal{G},\frak{A}%
,\Phi \right) $.

Third, we define the unitary evolution on the product space $\mathcal{H}=%
\mathcal{G}^{0]}\otimes \frak{h}\otimes \mathcal{G}_{0}$ by 
\begin{equation}
U:\cdots \chi _{-1}\otimes \chi _{0}\otimes \psi \otimes \chi _{1}\otimes
\cdots \mapsto \cdots \chi _{0}\otimes \mathrm{U}\left( \psi \otimes \chi
_{1}\right) \otimes \chi _{2}\cdots ,  \label{A2.7}
\end{equation}
incorporating the shift in the right and left strings. They are connected by
the conservative boundary condition $\mathrm{U}:\frak{h}\otimes \frak{g}%
_{1}\rightarrow \frak{g}_{0}\otimes \frak{h}$ given by the unitary dilation (%
\ref{4.4}). We have obviously 
\begin{equation*}
\left( \langle \ldots ,y_{-1},y|\otimes \mathrm{I}\otimes \langle
x_{1},\ldots |\right) U\left( \Phi ^{0}\otimes \psi \otimes |0,0,\ldots
\rangle \right) =\langle \ldots ,y_{-1}|\Phi ^{0}V\left( y\right) \delta
_{0}^{x_{1}}\cdots
\end{equation*}
in any orthonormal basis $\left\{ |x\rangle \right\} $ of $\frak{f}$ and $%
\left\{ |y\rangle \right\} $ of $\frak{g}$ which contain the vector $%
|0\rangle =\chi ^{\circ }$. Thus the extended unitary operator $U$ still
reproduces the isometry $\mathrm{V}=\sum_{y\neq 0}|y\rangle V\left( y\right) 
$, and 
\begin{equation*}
\pi \left( \mathrm{A},\mathrm{B}\right) =\left( \Phi ^{0]}\otimes \mathrm{I}%
\otimes \Phi _{0}^{\circ }\right) ^{\dagger }U\left( I^{0}\otimes \mathrm{A}%
\otimes \mathrm{B}\otimes I_{0}\right) U^{t}\left( \Phi ^{0]}\otimes \mathrm{%
I}\otimes \Phi _{0}^{\circ }\right) =\mathrm{V}^{\dagger }\left( \mathrm{A}%
\otimes \mathrm{B}\right) \mathrm{V}
\end{equation*}
for all $\mathrm{A}\in \mathcal{A}_{o}$ and $\mathrm{B}\in \mathcal{B}$,
where $\Phi _{0}^{\circ }=|0,\ldots \rangle $ is the vacuum vector (the
infinite tensor power of vector $|0\rangle \in \frak{f}$ ) and $\Phi ^{0]}$
is arbitrary state-vector in $\mathcal{G}^{0]}$.

Fourth, we prove the dynamical invariance $U^{\dagger }\frak{B}U\subseteq 
\frak{B}$ of the algebra $\frak{B}=\frak{A}^{0]}\otimes \mathcal{B}\otimes 
\frak{A}_{0}$. This follows straightforward from the definition of $U$ as 
\begin{equation*}
U^{\dagger }\left( \cdots \mathrm{A}_{0}\otimes \mathrm{B}\otimes \mathrm{A}%
_{1}\otimes \mathrm{A}_{2}\cdots \right) U=\cdots \mathrm{A}_{-1}\otimes 
\mathrm{U}^{\dagger }\left( \mathrm{A}_{0}\otimes \mathrm{B}\right) \mathrm{U%
}\otimes \mathrm{A}_{1}\cdots \in \frak{B}
\end{equation*}
due to $\mathrm{U}^{\dagger }\left( \mathrm{A}\otimes \mathrm{B}\right) 
\mathrm{U}\in \mathcal{B}_{1}$ for all $\mathrm{A}\in \mathcal{A}_{0}$, $%
\mathrm{B}\in \mathcal{B}$ where $\mathcal{B}_{1}=\mathcal{B\otimes A}_{1}$.
However this algebra is not invariant under the inverse transformation as $%
\mathcal{A}_{0}\subset \mathcal{A}$, $\mathcal{B}_{0}=\mathcal{A}_{0}\otimes 
\mathcal{B}\subset \mathcal{A}_{1}\otimes \mathcal{B}$, and if $\mathrm{U}%
\left( \mathrm{B}\otimes \mathrm{A}\right) \mathrm{U}^{\dagger }\notin 
\mathcal{B}_{0}$ for some $\mathrm{A}\in \mathcal{B}\left( \frak{g}\right) $
and $\mathrm{B}\in \mathcal{B}$, then 
\begin{equation*}
U\left( \cdots \mathrm{A}_{-1}\otimes \mathrm{A}_{0}\otimes \mathrm{B}%
\otimes \mathrm{A}_{1}\cdots \right) U^{\dagger }=\cdots \mathrm{A}%
_{0}\otimes \mathrm{U}\left( \mathrm{B}\otimes \mathrm{A}_{1}\right) \mathrm{%
U}^{\dagger }\otimes \mathrm{A}_{2}\cdots \notin \frak{B}.
\end{equation*}

And fifth, we prove that the power semigroup $U^{t}$, $t\in \mathbb{N}$ of
the constructed unitary operator induces the whole quantum Markov chain. We
shall do this using the \emph{Schr\"{o}dinger recurrency equation} $\Psi
^{t}=U\Psi ^{t-1}$ with the initial condition $\Psi ^{0}=\Phi ^{0]}\otimes
\psi \otimes \Phi _{0}^{\circ }$ for $\Psi ^{t}=U^{t}\Psi ^{0}$ representing
it as $T_{-t}\Psi \left( t\right) $, where $T_{-t}=T_{-1}^{t}$ is the right
shift defined by the unitary operator 
\begin{equation*}
T_{-1}:\cdots \chi _{0}\otimes \psi \otimes \chi _{1}\otimes \chi _{2}\cdots
\mapsto \cdots \chi _{-1}\otimes \mathrm{T}^{\dagger }\left( \chi
_{0}\otimes \psi \right) \otimes \chi _{1}\cdots .
\end{equation*}
The interaction representation evolution $\Psi \left( t\right) =U\left(
t\right) \Psi ^{0}$ is obviously adapted in the sense $U\left( t\right)
=I^{0]}\otimes \mathrm{W}_{0}^{t]}\otimes I_{t}$, where $\mathrm{W}_{0}^{t]}$
is a unitary operator in $\frak{h}\otimes \mathcal{G}_{0}^{t]}$. It can be
found from the recurrency equation 
\begin{equation*}
\mathrm{W}_{0}^{t]}=\mathrm{W}_{t}\left( \mathrm{W}_{0}^{t-1]}\otimes 
\mathrm{I}\right) ,\quad \mathrm{W}_{0}^{0]}=\mathrm{I}
\end{equation*}
as the chronological product $\mathrm{W}_{0}^{t]}=\mathrm{W}_{t}\left( 
\mathrm{W}_{t-1}\otimes \mathrm{I}\right) \cdots \left( \mathrm{W}%
_{1}\otimes \mathrm{I}_{1}^{t]}\right) $ generated by 
\begin{equation*}
\mathrm{W}_{t}=\mathrm{T}_{0}^{t]\dagger }\left( \mathrm{I}_{0}^{t}\otimes 
\mathrm{U}\right) \left( \mathrm{T}_{0}^{t-1]}\otimes \mathrm{I}\right) =%
\mathrm{T}_{0}^{t]\dagger }\left( I_{0}^{t}\otimes \mathrm{W}_{0}\right) 
\mathrm{T}_{0}^{t]},
\end{equation*}
Here $\mathrm{T}_{0}^{t]}$ is the transposition of $\frak{h}$ and $\frak{g}%
_{t}$ which can also be obtained by the recurrency 
\begin{equation*}
\mathrm{T}_{0}^{t]}=\left( \mathrm{I}_{0}^{t-1]}\otimes \mathrm{T}\right)
\left( \mathrm{T}_{0}^{t-1]}\otimes \mathrm{I}\right) ,\quad \mathrm{T}%
_{0}^{0]}=\mathrm{I}
\end{equation*}
and $\mathrm{W}_{0}=\mathrm{TWT}^{\dagger }=\mathrm{UT}^{\dagger }$ is the
initial scattering operator obtained by transposing $\mathrm{W}=\mathrm{W}%
_{1}$ onto $\frak{g}\otimes \frak{h}$ such that $\mathrm{W}_{t}$ is applied
only to the system at $s=0$ and point of the string at $s=t$ by transposing
the system from the boundary $s=0$ to $t$.

Thus for the given initial condition $\Psi \left( 0\right) =\Psi ^{0}$ in
the interaction picture we have $\Psi \left( t\right) =\Phi ^{0]}\otimes
\Psi _{0}\left( t\right) $, where $\Psi _{0}\left( t\right) =\psi \left(
t\right) \otimes \Phi _{t}^{\circ }$ and $\psi \left( t\right) =\mathrm{W}%
_{0}^{t]}\left( \psi \otimes \Phi _{0}^{t]}\right) $. This $\Psi _{0}\left(
t\right) =W\left( t\right) \Psi _{0}$ with $W\left( t\right) =\mathrm{W}%
_{0}^{t]}\otimes I_{t}$ is the solution to the quantum stochastic recurrency
equation 
\begin{equation}
\Psi _{0}\left( t\right) =W_{t}\Psi _{0}\left( t-1\right) ,\quad \Psi
_{0}\left( 0\right) =\psi \otimes \Phi _{0}^{\circ }  \label{A2.10}
\end{equation}
given by the quantum stochastic unitary generator $W_{t}=\mathrm{W}%
_{t}\otimes I_{t}$ in $\mathcal{H}_{0}=\frak{h}\otimes \mathcal{G}_{0}$. Now
we can derive the \emph{quantum filtering recurrency equation} 
\begin{equation*}
\psi \left( t,y_{0}^{t]}\right) =V\left( y_{t}\right) \psi \left(
t-1,y_{0}^{t-1]}\right) ,\quad \psi \left( 0\right) =\psi
\end{equation*}
for $\psi \left( t,y_{0}^{t]}\right) =\left( \mathrm{I}\otimes \langle
y_{0}^{t]}|\right) \psi \left( t\right) $ from the equation 
\begin{equation*}
\psi \left( t\right) =\mathrm{W}_{t}\left( \psi \left( t-1\right) \otimes
\chi ^{\circ }\right) ,\quad \psi \left( 0\right) =\psi
\end{equation*}
for $\psi \left( t\right) =\mathrm{W}_{0}^{t]}\psi _{0}^{t]}$, where $\psi
_{0}^{t]}=\psi \otimes \chi _{1}^{\circ }\otimes \ldots \otimes \chi
_{t}^{\circ }$, and $\left\{ |y\rangle \right\} $ is an orthonormal basis of 
$\frak{g}$ with $|0\rangle =\chi ^{\circ }=1\oplus 0$. Indeed, 
\begin{equation*}
\psi \left( t,y_{0}^{t]}\right) =\left( \mathrm{I}\otimes \langle
y_{0}^{t]}|\right) \left( \mathrm{W}_{t}\otimes \mathrm{I}\right) \psi
\left( t-1\right) =\left( \mathrm{I}\otimes \langle y_{t}|\right) \mathrm{V}%
\psi \left( t-1,y_{0}^{t-1]}\right) ,
\end{equation*}
where we took into account that $\psi \left( t-1\right) =\mathrm{W}%
_{0}^{t-1]}\Psi _{0}^{t-1]}\otimes |0\rangle \otimes \Phi _{t}^{\circ }$ and
denoted 
\begin{equation*}
V\left( y\right) \psi =\left( \mathrm{I}\otimes \langle y|\right) \mathrm{W}%
_{t}\left( \psi \otimes |0\rangle \right) =\left( \mathrm{I}\otimes \langle
y|\right) \mathrm{V}\psi .
\end{equation*}
This proves the filtering equation in the abstract form $\psi \left(
t\right) =\mathrm{V}\psi \left( t-1\right) $ which has the solution $\psi
\left( t,y_{0}^{t]}\right) =V\left( t,y_{0}^{t]}\right) \psi $ corresponding
to $\psi \left( 0\right) =\psi $, with 
\begin{equation*}
V\left( t,y_{0}^{t]}\right) =V\left( y^{t}\right) \cdots V\left(
y^{1}\right) =\left( \langle y_{0}^{t]}|\otimes \mathrm{I}\right) \mathrm{V}%
\left( t\right) .
\end{equation*}
Moreover, if $\mathrm{A}_{0}^{t]}=\left( \mathrm{A}^{1},\ldots ,\mathrm{A}%
^{t}\right) $ is any sequence of the operators \textrm{A}$^{r}\in \mathcal{A}%
_{o}$, then obviously 
\begin{eqnarray*}
\pi \left( t,\mathrm{A}_{0}^{t]},\mathrm{B}\right) &=&\left( \mathrm{I}%
\otimes \langle 0,\ldots ,0|\right) \mathrm{W}_{0}^{t]\dagger }\left( 
\mathrm{B}\otimes \mathrm{A}^{1}\otimes \cdots \otimes \mathrm{A}^{t}\right) 
\mathrm{W}^{t]}\left( \mathrm{I}\otimes |0,\ldots ,0\rangle \right) \\
&=&U^{-t}\left( I^{-t]}\otimes \mathrm{A}^{1}\otimes \cdots \otimes \mathrm{A%
}^{t}\otimes \mathrm{B}\otimes I_{0}\right) U^{t}\in \mathcal{B},\quad
\forall \mathrm{B}\in \mathcal{B}\text{.}
\end{eqnarray*}
This gives a dynamical endomorphic realization of any quantum Markov chain $%
\pi \left( t\right) $ induced by a unitary semigroup $U^{t}$.

Note that we also proved that any stochastic $\mathcal{B}$-Markov chain has
a \emph{quantum stochastic} realization $W\left( t\right) =\mathrm{W}%
_{0}^{t]}\otimes I_{t}$ given by a unitary generator $\mathrm{W}$ in $\frak{h%
}\otimes \frak{g}$ which belongs to $\mathcal{B}\otimes \mathcal{B}\left( 
\frak{g}\right) $ as commuting with $\mathcal{B}^{\prime }\otimes \mathrm{I}$%
. It is simply the interaction evolution induced on $\mathcal{H}_{0}=\frak{h}%
\otimes \mathcal{G}_{0}$ by $U\left( t\right) =I^{0}\otimes W\left( t\right) 
$ for the positive $t\in \mathbb{N}$. It defines an adapted \emph{quantum
stochastic cocycle} of authomorphisms $\omega \left( t,B_{0}\right) =W\left(
t\right) ^{\dagger }B_{0}W\left( t\right) $ on the algebra $\frak{B}_{0}=%
\mathcal{B}\otimes \frak{A}_{0}$ as a family of ``system $\mathcal{B}$ plus
noise $\frak{A}_{0}$'' adapted representations 
\begin{equation*}
\omega \left( t,\mathrm{B}\otimes \mathrm{A}_{0}^{t]}\otimes A_{t}\right) =%
\mathrm{W}_{0}^{t]\dagger }\left( \mathrm{B}\otimes \mathrm{A}%
_{0}^{t]}\right) \mathrm{W}_{0}^{t]}\otimes A_{t}\quad \forall A_{t}\in 
\mathcal{B}\left( \mathcal{G}_{t}\right)
\end{equation*}
of this algebra, where $\mathrm{A}_{0}^{t]}\in \mathcal{B}\left( \frak{g}%
_{0}^{t]}\right) $. Their compositions $\vartheta ^{t}=\omega \left(
t\right) \circ \tau _{-t}$ with the right shifts $\tau _{-t}\left(
B_{-t}\right) =T_{t}\left( I^{t}\otimes B_{[-t}\right) T_{-t}$ of $B_{-t}\in 
\frak{B}_{-t}$ into $\frak{B}_{0}$ obviously extends to the semigroup of
injective endomorphisms 
\begin{equation*}
U^{-t}\left( A^{-t]}\otimes \mathrm{A}_{-t}^{0]}\otimes \mathrm{B}\otimes
A_{0}\right) U^{t}=A^{0]}\otimes \mathrm{U}_{0}^{t]\dagger }\left( \mathrm{A}%
_{-t}^{0]}\otimes \mathrm{B}\right) \mathrm{U}_{0}^{t]}\otimes A_{t}
\end{equation*}
on the algebra $\frak{B}$, where $\mathrm{U}_{0}^{t]}=\mathrm{T}_{0}^{t]}%
\mathrm{W}_{0}^{t]}$ are given by the same recurrency as $\mathrm{T}%
_{0}^{t]} $ but with the generator $\mathrm{U}$ instead of $\mathrm{T}$.
This semigroup in general is not invertible on $\mathcal{B}$ if $\mathcal{A}%
_{0}$ is smaller then $\mathcal{A}_{1}$ (the shifts $\tau _{t}$ are not
invertible in this case) despite the invertibility of the interaction
evolution described by the authomorphisms $\omega \left( t\right) $.

\subsection{Symbolic Quantum Calculus and Stochastic Differential Equations.}

In order to formulate the differential nondemolition causality condition and
to derive a filtering equation for the posterior states in the
time-continuous case we need quantum stochastic calculus.

The classical differential calculus for the infinitesimal increments 
\begin{equation*}
\mathrm{d}x=x\left( t+\mathrm{d}t\right) -x\left( t\right)
\end{equation*}
became generally accepted only after Newton gave a simple algebraic rule $%
\left( \mathrm{d}t\right) ^{2}=0$ for the formal computations of the
differentials $\mathrm{d}x$ for smooth trajectories $t\mapsto x\left(
t\right) $. In the complex plane $\mathbb{C}$ of phase space it can be
represented by a one-dimensional algebra $\frak{\alpha }=\mathbb{C}\mathrm{d}%
_{t}$ of the elements $a=\alpha \mathrm{d}_{t}$ with involution $a^{\star }=%
\bar{\alpha}\mathrm{d}_{t}$. Here 
\begin{equation*}
\text{$\mathrm{d}_{t}$}=\left[ 
\begin{array}{ll}
0 & 1 \\ 
0 & 0
\end{array}
\right] =\frac{1}{2}\left( \sigma _{x}+i\sigma _{y}\right)
\end{equation*}
for $\mathrm{d}t$ is the nilpotent matrix, which can be regarded as
Hermitian $\mathrm{d}_{t}^{\star }=\mathrm{d}_{t}$ with respect to the
Minkowski metrics $\left( \mathbf{z}|\mathbf{z}\right) =2\func{Re}z_{-}\bar{z%
}_{+}$ in $\mathbb{C}^{2}$.

This formal rule was generalized to non-smooth paths early in the last
century in order to include the calculus of forward differentials $\mathrm{d}%
w\simeq\left( \mathrm{d}t\right) ^{1/2}$ for continuous diffusions $w_{t}$
which have no derivative at any $t$, and the forward differentials $\mathrm{d%
}n\in\left\{ 0,1\right\} $ for left continuous counting trajectories $n_{t}$
which have zero derivative for almost all $t$ (except the points of
discontinuity where $\mathrm{d}n=1$). The first is usually done by adding
the rules 
\begin{equation*}
\left( \mathrm{d}w\right) ^{2}=\mathrm{d}t,\quad\mathrm{d}w\mathrm{d}t=0=%
\mathrm{d}t\mathrm{d}w
\end{equation*}
in formal computations of continuous trajectories having the first order
forward differentials $\mathrm{d}x=\alpha\mathrm{d}t+\beta\mathrm{d}w$ with
the diffusive part given by the increments of standard Brownian paths $w $.
The second can be done by adding the rules 
\begin{equation*}
\left( \mathrm{d}n\right) ^{2}=\mathrm{d}n,\quad\mathrm{d}n\mathrm{d}t=0=%
\mathrm{d}t\mathrm{d}n
\end{equation*}
in formal computations of left continuous and smooth for almost all $t$
trajectories having the forward differentials $\mathrm{d}x=\alpha \mathrm{d}%
t+\gamma\mathrm{d}m$ with jumping part given by the increments of standard
compensated Poisson paths $m_{t}=n_{t}-t$. These rules were developed by
It\^{o} \cite{Ito51} into the form of a stochastic calculus.

The linear span of $\mathrm{d}t$ and $\mathrm{d}w$ forms the Wiener-It\^{o}
algebra $\frak{b}=\mathbb{C}\mathrm{d}_{t}+\mathbb{C}\mathrm{d}_{w}$, while
the linear span of $\mathrm{d}t$ and $\mathrm{d}n$ forms the Poisson-It\^{o}
algebra $\frak{c}=\mathbb{C}\mathrm{d}_{t}+\mathbb{C}\mathrm{d}_{m}$, with
the second order nilpotent $\mathrm{d}_{w}=\mathrm{d}_{w}^{\star }$ and the
idempotent $\mathrm{d}_{m}=\mathrm{d}_{m}^{\star }$. They are represented
together with $\mathrm{d}_{t}$ by the triangular Hermitian matrices 
\begin{equation*}
\text{$\mathrm{d}_{t}$}=\left[ 
\begin{array}{lll}
0 & 0 & 1 \\ 
0 & 0 & 0 \\ 
0 & 0 & 0
\end{array}
\right] ,\quad \mathrm{d}_{w}=\left[ 
\begin{array}{lll}
0 & 1 & 0 \\ 
0 & 0 & 1 \\ 
0 & 0 & 0
\end{array}
\right] ,\emph{\quad }\mathrm{d}_{m}\mathbf{=}\left[ 
\begin{array}{lll}
0 & 1 & 0 \\ 
0 & 1 & 1 \\ 
0 & 0 & 0
\end{array}
\right] ,
\end{equation*}
on the Minkowski space $\mathbb{C}^{3}$ with respect to the inner Minkowski
product $\left( \mathbf{z}|\mathbf{z}\right) =z_{-}z^{-}+z_{\circ }z^{\circ
}+z_{+}z^{+}$, where $z^{\mu }=\bar{z}_{-\mu }$, $-\left( -,\circ ,+\right)
=\left( +,\circ ,-\right) $.

Although both algebras $\frak{b}$ and $\frak{c}$ are commutative, the matrix
algebra $\frak{a}$ generated by $\frak{b}$ and $\frak{c}$ on $\mathbb{C}^{3}$
is not: 
\begin{equation*}
\mathrm{d}_{w}\mathrm{d}_{m}=\left[ 
\begin{array}{lll}
0 & 1 & 1 \\ 
0 & 0 & 0 \\ 
0 & 0 & 0
\end{array}
\right] \neq \left[ 
\begin{array}{lll}
0 & 0 & 1 \\ 
0 & 0 & 1 \\ 
0 & 0 & 0
\end{array}
\right] =\mathrm{d}_{m}\mathrm{d}_{w}.
\end{equation*}
The four-dimensional $\star $-algebra $\frak{a}=\mathbb{C}\mathrm{d}_{t}+%
\mathbb{C}\mathrm{d}_{-}+\mathbb{C}\mathrm{d}^{+}+\mathbb{C}\mathrm{d}$ of
triangular matrices with the canonical basis 
\begin{equation*}
\mathrm{d}_{-}=\left[ 
\begin{array}{lll}
0 & 1 & 0 \\ 
0 & 0 & 0 \\ 
0 & 0 & 0
\end{array}
\right] ,\,\mathrm{d}^{+}\mathbf{=}\left[ 
\begin{array}{lll}
0 & 0 & 0 \\ 
0 & 0 & 1 \\ 
0 & 0 & 0
\end{array}
\right] ,\,\mathrm{d}=\left[ 
\begin{array}{lll}
0 & 0 & 0 \\ 
0 & 1 & 0 \\ 
0 & 0 & 0
\end{array}
\right] ,
\end{equation*}
given by the algebraic combinations 
\begin{equation*}
\mathrm{d}_{-}=\mathrm{d}_{w}\mathrm{d}_{m}-\text{$\mathrm{d}_{t}$},\;%
\mathrm{d}^{+}=\mathrm{d}_{m}\mathrm{d}_{w}-\text{$\mathrm{d}_{t}$},\;\text{%
\textrm{d}}=\text{\textrm{d}}_{m}-\text{\textrm{d}}_{w}
\end{equation*}
is the canonical representation of the differential $\star $-algebra for
one-dimensional vacuum noise in the unified quantum stochastic calculus \cite
{Be88a, Be92a}. It realizes the HP (Hudson-Parthasarathy) table \cite{HuPa84}
\begin{equation*}
\mathrm{d}A_{-}\mathrm{d}A^{+}=\mathrm{d}t,\quad \mathrm{d}A_{-}\mathrm{d}A=%
\mathrm{d}A_{-},\quad \mathrm{d}A\mathrm{d}A^{+}=\text{\textrm{d}}%
A^{+},\quad \left( \mathrm{d}A\right) ^{2}=\mathrm{d}A,
\end{equation*}
with zero products for all other pairs, for the multiplication of the
canonical counting $\mathrm{d}A=\lambda \left( \mathrm{d}\right) $, creation 
$\mathrm{d}A^{+}=\lambda \left( \mathrm{d}^{+}\right) $, annihilation $%
\mathrm{d}A_{-}=\lambda \left( \mathrm{d}_{-}\right) $, and preservation $%
\mathrm{d}t=\lambda \left( \text{$\mathrm{d}_{t}$}\right) $ quantum
stochastic integrators in Fock space over $L^{2}\left( \mathbb{R}_{+}\right) 
$. As was proved recently in \cite{Be98}, any generalized It\^{o} algebra
describing a quantum noise can be represented in the canonical way as a $%
\star $-subalgebra of a quantum vacuum algebra 
\begin{equation*}
\mathrm{d}A_{\mu }^{\kappa }\mathrm{d}A_{\iota }^{\nu }=\delta _{\iota
}^{\kappa }\mathrm{d}A_{\mu }^{\nu },\quad \iota ,\mu \in \left\{ -,1,\ldots
,d\right\} ;\;\kappa ,\nu \in \left\{ 1,\ldots ,d,+\right\} ,
\end{equation*}
in the Fock space with several degrees of freedom $d$, where $\mathrm{d}%
A_{-}^{+}=\mathrm{d}t$ and $d$ is restricted by the doubled dimensionality
of quantum noise (could be infinite), similar to the representation of every
semi-classical system with a given state as a subsystem of quantum system
with a pure state. Note that in this quantum It\^{o} product formula $\delta
_{\kappa }^{\iota }=0$ if $\iota =+$ or $\kappa =-$ as $\delta _{\kappa
}^{\iota }\neq 0$ only when $\iota =\kappa $.

The quantum It\^{o} product gives an explicit form 
\begin{equation*}
\mathrm{d}\psi \psi ^{\dagger }+\psi \mathrm{d}\psi ^{\dagger }+\mathrm{d}%
\psi \mathrm{d}\psi ^{\dagger }=\left( \alpha _{\kappa }^{\iota }\psi
^{\dagger }+\psi \alpha _{\kappa }^{\star \iota }+\alpha _{j}^{\iota
}a_{\kappa }^{\star j}\right) _{\kappa }^{\iota }\mathrm{d}A_{\iota
}^{\kappa }
\end{equation*}
of the term $\mathrm{d}\psi \mathrm{d}\psi ^{\dagger }$ for the adjoint
quantum stochastic differentials 
\begin{equation*}
\mathrm{d}\psi =\alpha _{\kappa }^{\iota }\mathrm{d}A_{\iota }^{\kappa
},\quad \mathrm{d}\psi ^{\dagger }=\alpha _{\kappa }^{\star \iota }\mathrm{d}%
A_{\iota }^{\kappa },
\end{equation*}
for evaluation of the product differential 
\begin{equation*}
\mathrm{d}\left( \psi \psi ^{\dagger }\right) =\left( \psi +\mathrm{d}\psi
\right) \left( \psi +\mathrm{d}\psi \right) ^{\dagger }-\psi \psi ^{\dagger
}.
\end{equation*}
Here $\alpha _{-\kappa }^{\star \iota }=\alpha _{-\iota }^{\kappa \dagger }$
is the quantum It\^{o} involution with respect to the switch $-\left(
-,+\right) =\left( +,-\right) $, $-\left( 1,\ldots ,d\right) =\left(
1,\ldots ,d\right) $, introduced in \cite{Be88a}, and the Einstein summation
is always understood over $\kappa =1,\ldots ,d,+$; $\iota =-,1,\ldots ,d$
and $k=1,\ldots ,d$. This is the universal It\^{o} product formula which
lies in the heart of the general quantum stochastic calculus \cite{Be88a,
Be92a} unifying the It\^{o} classical stochastic calculi with respect to the
Wiener and Poisson noises and the quantum differential calculi \cite{HuPa84,
GaCo85} based on the particular types of quantum It\^{o} algebras for the
vacuum or finite temperature noises. It was also extended to the form of
quantum functional It\^{o} formula and even for the quantum nonadapted case
in \cite{Be91, Be93}.

Every stationary classical (real or complex) process $x^{t}$, $t>0$ with $%
x^{0}=0$ and independent increments $x^{t+\Delta }-x^{t}$ has mean values $%
\mathsf{M}\left[ x^{t}\right] =\lambda t$. The compensated process $%
y^{t}=x^{t}-\lambda t$, which is called noise, has an operator
representation $\hat{x}^{t}$ in Fock space $\mathcal{F}_{0}$ the Hilbert
space $L^{2}\left( \mathbb{R}_{+}\right) $ in the form of the integral with
respect to basic processes $A_{j}^{+},A_{-}^{j},A_{k}^{i}$\ such that $%
\digamma =f\left( \hat{x}\right) \delta _{\varnothing }\simeq f\left(
x\right) $\ in terms of the $L_{\mu }^{2}$ -- Fock isomorphism $%
f\longleftrightarrow \digamma $ of the chaos expansions 
\begin{equation*}
f\left( x\right) =\sum_{n=0}^{\infty }\idotsint_{0<r_{1}<\ldots
<r_{n}}\digamma \left( r_{1},\ldots r_{n}\right) \mathrm{d}y^{r_{1}}\cdots 
\mathrm{d}y^{r_{n}}\equiv \int \digamma \left( \upsilon \right) \mathrm{d}%
y^{\upsilon }
\end{equation*}
of the stochastic functionals $f\in L_{\mu }^{2}$ having the finite second
moments $\mathsf{M}\left[ \left| f\right| ^{2}\right] =\left\| \digamma
\right\| ^{2}$ and the Fock vectors $\digamma \in \mathcal{F}_{0}$. The
expectations of the Fock operators $f\left( \hat{x}\right) $ given by the
iterated stochastic integrals $f$ coincides on the vacuum state-vector $%
\delta _{\varnothing }\in \mathcal{F}_{0}$ with their expectation given by
the probability measure $\mu $: 
\begin{equation*}
\mathsf{M}\left[ f\left( x\right) \right] =\langle \delta _{\varnothing
}|f\left( \hat{x}\right) \delta _{\varnothing }\rangle =\digamma \left(
\varnothing \right) .
\end{equation*}
If its differential increments $\mathrm{d}x^{t}$ form a two dimensional
It\^{o} algebra, $\hat{x}^{t}$ can be represented in the form of a
commutative combination of the three basic quantum stochastic increments $%
A=A_{0}^{0},A_{-}=A_{-}^{0},A^{+}=A_{0}^{+}$. The It\^{o} formula for the
process $x^{t}\;$given by the quantum stochastic differential 
\begin{equation*}
\mathrm{d}\hat{x}^{t}=\alpha \mathrm{d}A+\alpha ^{-}\mathrm{d}A_{-}+\alpha
_{+}\mathrm{d}A^{+}\mathrm{d}\psi +\alpha _{+}^{-}\mathrm{d}t
\end{equation*}
can be obtained from the HP product \cite{HuPa84} 
\begin{equation*}
\mathrm{d}\hat{x}^{t}\mathrm{d}\hat{x}^{t\dagger }=\alpha \alpha ^{\dagger }%
\mathrm{d}A+\alpha ^{-}\alpha ^{\dagger }\mathrm{d}A_{-}+\alpha \alpha
^{-\dagger }\mathrm{d}A^{+}+\alpha ^{-}\alpha ^{-\dagger }\mathrm{d}t.
\end{equation*}

The noises $y_{k}^{t}=x_{k}^{t}-\lambda _{k}t$ with stationary independent
increments are called standard if they have the standard variance $\mathsf{M}%
\left[ \left( x^{t}\right) ^{2}\right] =t$. In this case 
\begin{equation*}
\hat{y}_{k}^{t}=\left( A_{k}^{+}+A_{-}^{k}+\varepsilon _{k}A_{k}^{k}\right)
\left( t\right) =\varepsilon _{k}m_{k}^{t}+\left( 1-\varepsilon _{k}\right)
w_{k}^{t},
\end{equation*}
where $\varepsilon _{k}\geq 0$ is defined by the equation $\left( \mathrm{d}%
x_{k}^{t}\right) ^{2}-\mathrm{d}t=\varepsilon \mathrm{d}x_{k}^{t}$. Such,
and indeed higher dimensional, quantum noises for continual measurements in
quantum optics were considered in \cite{GPZ92, DPZG92}.

The general form of a quantum stochastic decoherence equation, based on the
canonical representation of the arbitrary It\^{o} algebra for a quantum
noise in the vacuum of $d$ degrees of freedom, can be written as 
\begin{equation*}
\mathrm{d}\hat{\psi}\left( t\right) =\left( \mathrm{S}_{\kappa }^{\iota
}-\delta _{\kappa }^{\iota }\mathrm{I}\right) \mathrm{d}A_{\iota }^{\kappa }%
\hat{\psi}\left( t\right) ,\quad \hat{\psi}\left( 0\right) =\psi \otimes
\delta _{\varnothing },\;\psi \in \frak{h}.
\end{equation*}
Here $\mathrm{L}_{\kappa }^{\iota }$ are the operators in the system Hilbert
space $\frak{h}$ $\ni \psi $ with $\mathrm{S}_{\kappa }^{\star -}\mathrm{S}%
_{+}^{\kappa }=0$ for the mean square normalization 
\begin{equation*}
\hat{\psi}\left( t\right) ^{\dagger }\hat{\psi}\left( t\right) =\mathsf{M}%
\left[ \psi \left( t,\cdot \right) ^{\dagger }\psi \left( t,\cdot \right) %
\right] =\psi ^{\dagger }\psi
\end{equation*}
with respect to the vacuum of Fock space of the quantum noise, where the
Einstein summation is understood over all $\kappa =-,1,\ldots ,d,+$ with the
agreement 
\begin{equation*}
\mathrm{S}_{-}^{-}=\mathrm{I}=\mathrm{S}_{+}^{+},\quad \,\mathrm{S}_{-}^{j}=%
\mathrm{O}=\mathrm{S}_{j}^{+},\quad j=1,\ldots ,d
\end{equation*}
and $\delta _{\kappa }^{\iota }=1$ for all coinciding $\iota ,\kappa \in
\left\{ -,1,\ldots ,d,+\right\} $ such that $\mathrm{L}_{\kappa }^{\iota }=%
\mathrm{S}_{\kappa }^{\iota }-\delta _{\kappa }^{\iota }\mathrm{I}=0$
whenever $\iota =+$ or $\kappa =-$. In the notations $\mathrm{S}_{+}^{j}=%
\mathrm{L}^{j}$, $\mathrm{S}_{+}^{-}=-\mathrm{K}$, $\mathrm{S}_{j}^{-}=-%
\mathrm{K}_{j}$, $j=1,\ldots ,d$ the decoherence wave equation takes the
standard form \cite{Be95, Be97} 
\begin{equation*}
\mathrm{d}\hat{\psi}\left( t\right) +\left( \mathrm{Kd}t+\mathrm{K}_{j}%
\mathrm{d}A_{-}^{j}\right) \hat{\psi}\left( t\right) =\left( \mathrm{L}^{j}%
\mathrm{d}A_{j}^{+}+\left( \mathrm{S}_{k}^{i}-\delta _{k}^{i}\mathrm{I}%
\right) \mathrm{d}A_{i}^{k}\right) \hat{\psi}\left( t\right) ,
\end{equation*}
where $A_{j}^{+}\left( t\right) ,A_{-}^{j}\left( t\right) ,A_{i}^{k}\left(
t\right) $ are the canonical creation, annihilation and exchange processes
respectively in Fock space, and the normalization condition is written as $%
\mathrm{L}_{k}\mathrm{L}^{k}=\mathrm{K}+\mathrm{K}^{\dagger }$ with $\mathrm{%
L}_{k}^{\dagger }=\mathrm{L}^{k}$ (the Einstein summation is over $%
i,j,k=1,\ldots ,d$).

Using the quantum It\^{o} formula one can obtain the corresponding equation
for the quantum stochastic density operator $\hat{\varrho}=\psi \psi
^{\dagger }$ which is the particular case $\kappa =-,1,\ldots ,d,+$ of the
general quantum stochastic Master equation 
\begin{equation*}
\mathrm{d}\hat{\varrho}\left( t\right) =\left( \mathrm{S}_{\gamma }^{\iota }%
\hat{\varrho}\left( t\right) \mathrm{S}_{\kappa }^{\star \gamma }-\hat{%
\varrho}\left( t\right) \delta _{\kappa }^{\iota }\right) \mathrm{d}A_{\iota
}^{\kappa },\quad \hat{\varrho}\left( 0\right) =\rho ,
\end{equation*}
where the summation over $\kappa =-,k,+$ is extended to infinite number of $%
k=1,2,\ldots $. This general form of the decoherence equation with $\mathrm{L%
}_{\kappa }^{\star -}\mathrm{L}_{+}^{\kappa }=\mathrm{O}$ corresponding to
the normalization condition $\left\langle \hat{\varrho}\left( t\right)
\right\rangle =\mathrm{Tr}\rho $ in the vacuum mean, was recently derived in
terms of quantum stochastic completely positive maps in \cite{Be95, Be97}.
Denoting $\mathrm{L}_{\kappa }^{-}=-\mathrm{K}_{\kappa }$, $\mathrm{L}%
_{+}^{\star \iota }=-\mathrm{K}^{\iota }$ such that $\mathrm{K}_{\iota
}^{\dagger }=\mathrm{K}^{\iota }$, this can be written as 
\begin{equation*}
\mathrm{d}\hat{\varrho}\left( t\right) +\mathrm{K}_{\kappa }\hat{\varrho}%
\left( t\right) \mathrm{d}A_{-}^{\kappa }+\hat{\varrho}\left( t\right) 
\mathrm{K}^{\iota }\mathrm{d}A_{\iota }^{+}=\left( \mathrm{L}_{\kappa }^{j}%
\hat{\varrho}\left( t\right) \mathrm{L}_{j}^{\star \iota }-\hat{\varrho}%
\left( t\right) \delta _{\kappa }^{\iota }\right) \mathrm{d}A_{\iota
}^{\kappa },
\end{equation*}
or in the notation above, $\mathrm{K}_{+}=\mathrm{K},\mathrm{K}^{-}=\mathrm{K%
}^{\dagger }$, $\mathrm{L}_{+}^{k}=\mathrm{L}^{k}$, $\mathrm{L}_{k}^{\star
-}=\mathrm{L}_{k}$, $\mathrm{L}_{k}^{\star i}=\mathrm{L}_{i}^{k\dagger }$ as 
\begin{equation*}
\mathrm{d}\hat{\varrho}\left( t\right) +\left( \mathrm{K}\hat{\varrho}\left(
t\right) +\hat{\varrho}\left( t\right) \mathrm{K}^{\dagger }-\mathrm{L}^{j}%
\hat{\varrho}\left( t\right) \mathrm{L}_{j}\right) \mathrm{d}t=\left( 
\mathrm{S}_{k}^{j}\hat{\varrho}\left( t\right) \mathrm{S}_{j}^{\dagger i}-%
\hat{\varrho}\left( t\right) \delta _{k}^{i}\right) \mathrm{d}A_{i}^{k}
\end{equation*}
\begin{equation*}
+\left( \mathrm{S}_{k}^{j}\hat{\varrho}\left( t\right) \mathrm{L}_{j}-%
\mathrm{K}_{k}\hat{\varrho}\left( t\right) \right) \mathrm{d}%
A_{-}^{k}+\left( \mathrm{L}^{j}\hat{\varrho}\left( t\right) \mathrm{S}%
_{j}^{\dagger i}-\hat{\varrho}\left( t\right) \mathrm{K}^{i}\right) \mathrm{d%
}A_{i}^{+},
\end{equation*}
with $\mathrm{K}+\mathrm{K}^{\dagger }=\mathrm{L}_{j}\mathrm{L}^{j}$, $%
\mathrm{L}^{j}=\mathrm{L}_{j}^{\dagger }$, $\mathrm{L}_{k}^{\dagger i}=%
\mathrm{L}_{i}^{k\dagger }$ for any number of $j$'s, and arbitrary $\mathrm{K%
}^{j}=\mathrm{K}_{j}^{\dagger }$, $\mathrm{L}_{k}^{i}$, $i,j,k=1,\ldots ,d$.
This is the quantum stochastic generalization of the general form \cite
{Lin76} for the non-stochastic (Lindblad) Master equation corresponding to
the case $d=0$. In the case $d>0$ with pseudo-unitary block-matrix $\mathrm{S%
}\mathbf{=}\left[ \mathrm{S}_{\kappa }^{\iota }\right] _{\nu =-,\circ
,+}^{\iota =-,\circ ,+}$ in the sense $\mathbf{S}^{\star }=\mathbf{S}^{-1}$,
it gives the general form of quantum stochastic Langevin equation
corresponding to the HP unitary evolution for $\psi \left( t\right) $ \cite
{HuPa84}.

The nonlinear form of this decoherence equation for the exactly normalized
density operator $\hat{\rho}\left( t\right) =\hat{\varrho}\left( t\right) /%
\mathrm{Tr}_{\frak{h}}\hat{\varrho}\left( t\right) $ was obtained for
different commutative It\^{o} algebras in \cite{Be90c, BaBe, Be92a}.

\medskip

\textbf{Acknowledgment:}

I would like to acknowledge the help of Robin Hudson and some of my students
attending the lecture course on Modern Quantum Theory who were the first who
read and commented on these notes containing the answers on some of their
questions. The best source on history and drama of quantum theory is in the
biographies of the great inventors, Schr\"{o}dinger, Bohr and Heisenberg 
\cite{Schr, Bohr, Heis}, and on the conceptual development of this theory
before the rise of quantum probability --\ in \cite{Jamm}. An excellent
essay ``The quantum age begins'', as well as short biographies with posters
and famous quotations of all mathematicians and physicists mentioned here
can be found on the mathematics website at St Andrews University --
http://www-history.mcs.st-and.ac.uk/history/, the use of which is
acknowledged.


\medskip

\begin{thebibliography}{999}
\bibitem{Bell87}  J. S. Bell, \textit{Speakable and Unspeakable in Quantum
Mechanics}. Camb. Univ. Press, 1987.

\bibitem{Be94}  V.P. Belavkin, \textit{Nondemolition Principle of Quantum
Measurement Theory.} Foundations of Physics, \textbf{24}, No. 5, 685--714
(1994).

\bibitem{Be79}  V. P. Belavkin, \textit{Optimal Measurement and Control in
Quantum Dynamical Systems}. Preprint No. 411, Inst. of Phys.., Nicolaus
Copernicus University, Torun', February 1979. Published in Rep. Math. Phys., 
\textbf{43}, No. 3, 405--425, (1999).

\bibitem{Be80}  V. P. Belavkin, \textit{Quantum Filtering of Markov Signals
with Wight Quantum Noise. }Radiotechnika and Electronika, \textbf{25},
1445--1453 (1980). Full English translation in: \textit{Quantum
Communications and Measurement.} V. P. Belavkin et al, eds., 381--392
(Plenum Press, 1994).

\bibitem{Be85}  V. P. Belavkin, In: \textit{Information Complexity and
Control in Quantum Physics}, ed. A. Blaqui\`{e}re, 331--336
(Springer-Verlag, Udine 1985).

\bibitem{BaLu85}  A. Barchielli \& G. Lupierri. J Math Phys., \textbf{26},
2222--30 (1985).

\bibitem{Be88}  V. P. Belavkin, In: \textit{Modelling and Control of Systems}%
, ed. A. Blaqui\`{e}re, Lecture Notes in Control and Information Sciences, 
\textbf{121}, 245--265, Springer 1988.

\bibitem{Dio88}  L. Diosi, Phys. Rev. A, \textbf{40}, 1165--74 (1988).

\bibitem{Be89a}  V. P. Belavkin,\textit{\ A Continuous Counting Observation
and Posterior Quantum Dynamics}. J. Phys. A: Math. Gen. \textbf{22},
L1109--L1114 (1989).

\bibitem{Gis89}  N. Gisin. Helv. Phys. Acta, \textbf{62}, 363 (1989).

\bibitem{Be90b}  V. P. Belavkin, In: \textit{Stochastic Methods in
Experimental Sciences}, W. Kasprzak and A. Weron eds., 26--42 (World
Scientific, 1990); \textit{A Posterior Schr\"{o}dinger Equation for
Continuous Nondemolition Measurement}. J. Math. Phys. \textbf{31},
2930--2934 (1990).

\bibitem{GPR90}  G. C. Ghirardi, P. Pearl \& A. Rimini, \textit{Markov
Processes in Hilbert Space and Continuous Spontaneous Localization of
Systems of Identical Particles.} Phys. Rev. A \textbf{42,} 78--89 (1990).

\bibitem{Be90c}  V. P. Belavkin, \textit{Stochastic Posterior Equations for
Quantum Nonlinear Filtering}. In: Prob. Theory and Math. Stat., ed. B
Grigelionis et al., \textbf{1}, 91--109, VSP/Mokslas, Vilnius 1990.

\bibitem{Wig63}  E. Wigner. Am. J. Phys. \textbf{31}, 6 (1963).

\bibitem{Dav76}  E. B. Davies, \textit{Quantum Theory of Open Systems}.
Academic, 1976.

\bibitem{Be78}  {\ V. P. Belavkin,} \textit{Operational theory of quantum
stochastic processes}. In: Proc. of VII-th Conference on Coding Theory and
Information transmission, \textbf{1}, 23--28 (Moscow--Vilnius, 1978).

\bibitem{BLP82}  A.~Barchielli, L.~Lanz \& G.M. Prosperi. Nuovo Cimento, 
\textbf{72}B, 79 (1982).

\bibitem{Hol82}  A. S. Holevo. Soviet Mathematics \textbf{26}, 1--20 (1982);
Izvestia Vuzov, Matematica, \textbf{26}, 3--19 (1992).

\bibitem{Be83}  V. P. Belavkin, \textit{Towards Control Theory of Quantum
Observable Systems}. Automatica and Remote Control, \textbf{44},178--188
(1983).

\bibitem{Bar83}  A. Barchielli. Nuovo Cimento, \textbf{74}B, 113--138 (1983).

\bibitem{Zeh70}  H. D. Zeh. Foundation of \ Physics. \textbf{1}, 69 (1970).

\bibitem{UnZu89}  W. G. Unruh \& W. G. Zurek. Physical Review D, \textbf{40}%
, 1071 (1989).

\bibitem{Per76}  P. Pearle, \textit{Reduction of the State Vector by a
Nonlinear Schr\"{o}dinger Equation.} Phys. Rev. D, \textbf{13}, No. 4,
857--868 (1976).

\bibitem{Gis83}  N. Gisin. J. Math. Phys., \textbf{24}, 1779--82 (1983).

\bibitem{GeHa90}  M. Gell-Mann \& J. B. Hartle. In: \textit{Complexity,
Entropy and Physics of Information}, W. H. Zurek, ed. (Addison-Wesley, 1990).

\bibitem{Haa95}  R. Haag, \textit{An evolutionary picture for quantum
physics.} Com. Math. Phys., \textbf{180}, 733--743 (1995).

\bibitem{GRW86}  G. C. Ghirardi, A. Rimini \& T. Weber. Phys. Rev. D \textbf{%
34}, No 2, 470--491 (1986).

\bibitem{BlJa95}  Ph. Blanchard \& A. Jadczyk, \textit{%
Event-Enhanced-Quantum Theory and Piecewise deterministic Dynamics}. Annalen
der Physik, \textbf{4}, 583--599 (1995).

\bibitem{Men93}  M. B. Mensky, \textit{Continuous Quantum Measurements and
Paths Integrals,} IOP, Bristol 1993.

\bibitem{AKS97}  S. Albeverio, V. N. Kolokol'tsov \& O. G. Smolyanov, 
\textit{Continuous Quantum Measurement: Local and Global Approaches.} Rew.
Math. Phys., \textbf{9}, 907--920 (1997).

\bibitem{MiWa84}  G. J. Milburn \& D. E. Walls. Phys. Rev., \textbf{30}(A),
56--60 (1984).

\bibitem{WCM85}  D. F. Walls, Collet M.J., and Milburn G.J. Phys. Rev., 
\textbf{32}(D),3208--15 (1985).

\bibitem{Car86}  H. Carmichael, \textit{Open Systems in Quantum Optics}.
Lecture Notes in Physics, m18, (Springer-Verlag, 1986).

\bibitem{ZMW87}  P. Zoller, M. A. Marte \& D. F. Walls. Phys. Rev., \textbf{%
35}(A), 198--207 (1987).

\bibitem{Bar87}  A. Barchielli. J. Phys. A: Math. Gen., \textbf{20,}
6341--55 (1987).

\bibitem{HMW89}  C. A. Holmes, G. J. Milburn \& D. F. Walls. Phys. Rev., 
\textbf{39}(A), 2493--501 (1989).

\bibitem{Ued90}  M. Ueda. Phys. Rev., \textbf{41}(A), 3875--90 (1990).

\bibitem{MiGa92}  G. {Milburn} \& Gagen. Phys Rev. A, \textbf{46,}1578
(1992).

\bibitem{Gis84}  N. Gisin, \textit{Quantum Measurement and Stochastic
Processes}. Phys. Rev. Lett., \textbf{52}, No. 19, 1657--60 (1984).

\bibitem{GiPe92}  N. Gisin \& I. C. Percival, \textit{The Quantum State
Diffusion Model Applied to Open systems.} J. Phys. A: Math. Gen. \textbf{25}%
, 5677--91 (1992).

\bibitem{GiPe93}  N. Gisin \& I. C. Percival, \textit{The Quantum State
Diffusion Picture of Physical Processes.} J. Phys. A: Math. Gen. \textbf{26}%
, 2245--60 (1993).

\bibitem{Car93}  H. J. Carmichael, \textit{An Open System Approach to
Quantum Optics,} Lecture Notes in Physics, m\textbf{18} (Springer, Berlin,
1993)$.$

\bibitem{WiMi93}  H. M. Wiseman \& G. J. Milburn. Phys. Rev. A \textbf{47},
642 (1993).

\bibitem{GoGr93}  P. Goetsch \& R. Graham, \textit{Quantum Trajectories for
Nonlinear Optical Processes}. Ann. Physik \textbf{2}, 708--719 (1993).

\bibitem{WiMi94}  H. M. Wiseman \& G. J. Milburn. Phys. Rev. A \textbf{49},
1350 (1994).

\bibitem{Car94}  H. J. Carmichael. In: \textit{Quantum Optics VI,} ed J. D.
Harvey and D. F. Walls (Springer, Berlin, 1994).

\bibitem{GoGr94}  P. Goetsch \& R. Graham, \textit{Linear Stochastic Wave
Equation for Continuously Measurement Quantum Systems}. Phys. Rev. A \textbf{%
50}, 5242--55 (1994).

\bibitem{GGH95}  P. Goetsch, R. Graham, \& F. Haake, \textit{Schr\"{o}dinger
Cat and Single Runs for the Damped Harmonic Oscillator}. Phys. Rev. A, 
\textbf{51}, No. 1, 136--142 (1995).

\bibitem{Plnkab}  M. Planck, \textit{Scientific Autobiography, and Other
Papers.} Williams \& Norgate LTD. London 1949.

\bibitem{Be01b}  V. P. Belavkin. \textit{Quantum Noise, Bits and Jumps.}
Progress in quantum Electronics, \textbf{25}, No. 1, 1--53 (2001).

\bibitem{Heis25}  W. Heisenberg. Z. Phys. \textbf{33}, 879--93 (1925).

\bibitem{BHJ26}  M. Born, W. Heisenberg \& P. Z. Jordan, Phys. \textbf{36},
557--615 (1926).

\bibitem{Schr26}  E. Schr\"{o}dinger, \textit{Quantization as an Eigenvalue
Problem}. Ann. Phys. \textbf{79}, 361--76 (1926).

\bibitem{Schr26c}  E. Schr\"{o}dinger, \textit{Abhandlundgen zur
Wellenmechanik}. Leipzig: J.A. Barth (1926).

\bibitem{Schr}  W. Moore, \textit{Schr\"{o}dinger life and thought}.
Cambridge University Press (1989).

\bibitem{Boh52}  D. Bohm. Phys. Rev. \textbf{85}, 166, 180 (1952).

\bibitem{Eve57}  H. Everett, Rev. Mod. Phys. \textbf{29}, 454 (1957).

\bibitem{Heis27}  W. Heisenberg, \textit{On the Perceptual Content of
Quantum Theoretical Kinematics and Mechanics}. Z. Phys. \textbf{43},
172--198 (1925). English translation in: J. A. Wheeler and Wojciech Zurek,
eds. \textit{Quantum Theory and Measurement.} (Princeton University Press,
1983), pp. 62--84.

\bibitem{AhBo61}  Y. Aharonov \& D. Bohm, Phys. Rev. \textbf{122}, 1649
(1961).

\bibitem{Be76}  V. P. Belavkin, \textit{Generalized Uncertaity Relations and
Efficient Measurements in Quantum Systems.} Theoretical and Mathematical
Physics, \textbf{26}, No 3, 316--329 (1976), \textit{The Nondemolition
Measurement of Quantum Time.} International Journal of Theoretical Physics, 
\textbf{37}, No1, 219--226 (1998).

\bibitem{Hol80}  A. S. Holevo, \textit{Probabilistic Aspects of Quantum
Theory}, Kluwer Publisher, 1980.

\bibitem{BePer98}  V. P. Belavkin \& M. G. Perkins, \textit{Nondemolition
Measurement of Quantum Time.} Int. J. of Theor. Phys. \textbf{37}, No 1
219--226 (1998).

\bibitem{Neum32}  J. von Neumann, \textit{Mathematische Grundlagen der
Quantummechanik.} Springer, Berlin, 1932. (English translation : Prinston
University Press, 1955).

\bibitem{Bell66}  J. S. Bell, \textit{On the Problem of Hidden Variables in
Quantum Theory}. Rev. Mod. Phys., \textbf{38}, 447--452 (1966).

\bibitem{Glea57}  A. M. Gleason. J. Math. \& Mech., \textbf{6}, 885 (1957).

\bibitem{BiNe36}  G. Birkhoff \& J. von Neumann, \textit{The Logic of
Quantum Mechanics.} Annals of Mathematics \textbf{37}, 823--843 (1936).

\bibitem{WWW52}  G. C. Wick, A. S. Wightman \& E. P. Wigner, \textit{The
Intrinsic Parity of Elementary Particles}. Phys. Rev. \textbf{88}, 101--105
(1952).

\bibitem{JaPi63}  J. P. Jauch \& G. Piron, \textit{Can Hidden Variables be
Excluded in Quantum mechanics? Helv. Phys. Acta \textbf{36}}, 837\textit{\
(1963).}

\bibitem{Russ}  B. Russel, \textit{Mysticism and Logic}, p.75. Penguin,
London (1953).

\bibitem{EPR}  A. Einstein, B. Podolski \& N. Rosen, \textit{Can
Quantum-Mechanical Description of Physical Reality be Considered Complete?}
Phys.Rev. \textbf{47}, 777--800 (1935).

\bibitem{Bohr35}  N. Bohr, Phys. Rev. \textbf{48}, 696--702 (1935).

\bibitem{Schr35}  E. Schr\"{o}dinger, Naturwis. \textbf{23}, 807--12,
823--8, 844--9 (1935).

\bibitem{Lud51}  G. L\"{u}ders, Ann. Phys. (Leipzig) \textbf{8}, 322 (1951).

\bibitem{BuLa91}  P. Bush, P. J. Lahti, P. Mittelst\"{a}dt, \textit{The
Quantum Measurement Theory.} Lecture Notes in Physics, Springer--Verlag,
Berlin, 1991.

\bibitem{StBe96}  R. L. Stratonovich \& V. P. Belavkin, \textit{Dynamical
Interpretation for the Quantum Measurement Projection Postulate.} Int. J. of
Theor. Phys., \textbf{35}, No. 11, 2215--2228 (1996).

\bibitem{Gar91}  C. W. Gardiner, \textit{Quantum Noise}. Springer-Verlag,
Berlin Heidelberg 1991.

\bibitem{HuPa84}  R. L. Hudson \& K. R. Parthasarathy, \textit{Quantum
It\^{o}'s Formula and Stochastic Evolution.} Comm.\thinspace Math.\thinspace
Phys., \textbf{93},\thinspace 301--323 (1984).

\bibitem{GaCo85}  C. W. Gardiner \& M. J. Collett. Phys. Rev. A, \textbf{31}%
, 3761 (1985).

\bibitem{Be88a}  V. P. Belavkin, \textit{A New Form and a }$\star $\textit{%
-Algebraic Structure of Quantum Stochastic Integrals in Fock Space}.
Rediconti del Sem. Mat. e Fis. di Milano, LVIII, 177--193 (1988).

\bibitem{Lud68}  G. Ludwig, Math. Phys., \textbf{4}, 331 (1967), \textbf{9},
1 (1968).

\bibitem{DaLe70}  E. B. Davies \& J. Lewis, Comm. Math. Phys., \textbf{17},
239--260 (1970).

\bibitem{Oz84}  E. B. Ozawa, J. Math. Phys., \textbf{25}, 79--87 (1984).

\bibitem{Be92a}  V. P. Belavkin, \textit{Quantum Stochastic Calculus and
Quantum Nonlinear Filtering.} Journal of Multivariate Analysis, \textbf{42},
No. 2, 171--201 (1992).

\bibitem{BeSt91}  V. P. Belavkin \& P. Staszewski. Rep. Math. Phys., \textbf{%
29,} 213 (1991).

\bibitem{BeMe96}  V. P. Belavkin \& O. Melsheimer, \textit{A Stochastic
Hamiltonian Approach for Quantum Jumps, Spontaneous Localizations, and
Continuous Trajectories.} Quantum and Semiclassical Optics \textbf{8,}
167--187 (1996).

\bibitem{Be92b}  V. P. Belavkin, \textit{Quantum Continual Measurements and
a Posteriori Collapse on CCR.} Com. Math. Phys., \textbf{146}, 611--635
(1992).

\bibitem{AFL82}  L. Accardi, A. Frigerio \& J. Lewis, Publ. RIMS Kyoto
Univ., \textbf{18}, 97 (1982).

\bibitem{Str66}  R. L. Stratonovich, \textit{Conditional Markov Processes
and Their Applications to Optimal Control.} Moscow State University, Moscow
1966.

\bibitem{Be89b}  V. P. Belavkin, \textit{A New Wave Equation for a
Continuous Nondemolition Measurement}. Phys. Lett. A, \textbf{140}, 355--358
(1989).

\bibitem{Be90a}  V. P. Belavkin, \textit{A Stochastic Posterior
Schr\"{o}dinger Equation for Counting Nondemolition Measurement}. Letters in
Math. Phys. \textbf{20,} 85--89 (1990).

\bibitem{BaBe}  A. Barchielli \& V. P. Belavkin, \textit{Measurements
Continuous in Time and a posteriori States in Quantum Mechanics}. J. Phys.
A: Math. Gen. \textbf{24}, 1495--1514 (1991).

\bibitem{Perc99}  I. Persival, \textit{Quantum State Diffusion}. Cambridge
University Press, 1999.

\bibitem{BeSt89}  V. P. Belavkin \& P. Staszewski, \textit{A Quantum
Particle Undergoing Continuous Observation.} Phys. Let. \textbf{140},
359--362 (1989).

\bibitem{BeSt92}  V. P. Belavkin \& P. Staszewski, \textit{Nondemolition
Observation of a Free Quantum Particle.} Phys. Rev. \textbf{45}, No. 3,
1347--1356 (1992).

\bibitem{ChSt92}  D. Chruscinski \& P. Staszewski, \textit{On the Asymptotic
Solutions of the Belavkin's Stochastic Wave Equation}. Physica Scripta. 
\textbf{45}, 193--199 (1992).

\bibitem{Kol95}  V. N. Kolokoltsov, \textit{Scattering Theory for the
Belavkin Equation Describing a Quantum Particle with Continuously Observed
Coordinate.} J. Math. Phys.\textit{\ }\textbf{36} (6), 2741--2760 (1995).

\bibitem{Be95}  V. P. Belavkin, \textit{A Dynamical Theory of Quantum
Continuous Measurements and Spontaneous Localizations.} Russ Journ of Math
Phys \textbf{3} (1), 3--24 (1995).

\bibitem{BeSt00}  V. P. Belavkin \& P. Staszewski, \textit{Quantum
Stochastic Differential Equation for Unstable Systems}. J. Math. Phys.%
\textit{\ }\textbf{41} (11), 7220--7233 (2000).

\bibitem{Be00}  V. P. Belavkin, In: \textit{New Development of
Infinite-Dimensional Analysis and Quantum Probability}, RIMS Kokyuroku 1139,
54--73, April, 2000; \textit{Quantum Stochastics, Dirac Boundary Value
Problem, and the Inductive Stochastic Limit}. Rep. Math. Phys., \textbf{46},
No.3, 2000.

\bibitem{Be01a}  V. P. Belavkin, In: \textit{Evolution Equations and Their
Applications in Physical and Life Sciences}, G Lumer and L. Weis eds., Lect.
Notes in Pure and Appl. Math. \textbf{215}, 311--327 (Marcel Dekker, Inc.,
2001).

\bibitem{BeKo01}  V. P. Belavkin \& V. N. Kolokoltsov, \textit{Stochastic
Evolutions as Boundary Value Problems.} In: Infinite Dimensional Analysis
and Quantum Probability, \ Research Institute for Mathematical Studies,
Kokyuroku \textbf{1227}, 83--95 (2001).

\bibitem{Schr31}  E. Schr\"{o}dinger, Sitzberg Press Akad. Wiss.
Phys.--Math. Kl. 144-53 (1931).

\bibitem{Crm86}  J. G. Cramer, Rev. Mod. Phys., \textbf{58}, 647--87 (1986).

\bibitem{Dix81}  J. Dixmier, \textit{Von Neumann Algebras}.North Holand,
Amsterdam, (1981)

\bibitem{BeSt84}  V. P. Belavkin, P. Staszewski, \textit{Relative Entropy in
C}$^{\ast }$-algebraic Statistical Mechanics. Reports in Mathematical
Physics, \textbf{20}, 373--384 (1984).

\bibitem{Acc74}  L. Accardi, \textit{Noncommutative Markov Chains,} Proc. of
Int. School of Mathematical Physics, Camerino,\textit{\ 266--295 (1974).}

\bibitem{Kum85}  B. K\"{u}mmerer, \textit{On the Structure of Markov
Dilations on W*-algebras.} In: Quantum Probability and Applications II, eds.
L. Accardi and W. von Waldenfels, 228--244 (Springer, 1988).

\bibitem{KuMa87}  B. K\"{u}mmerer \& H. Maassen, \textit{The Essentially
Commutative Dilations of Dynamical Semigroups on M}$_{n},$ Commun. Math.
Phys. \textbf{109}, 1--22 (1987).

\bibitem{BeSt86}  V. P. Belavkin, P. Staszewski, \textit{A Radon-Nikodym
Theorem for Completely Positive Maps}. Reports in Mathematical Physics, 
\textbf{24}, 49--55 (1986).

\bibitem{Ito51}  K. It\^{o}, \textit{On a Formula Concerning Stochastic
Differentials.} Nagoya Math . J., \textbf{3}, 55-65, (1951).

\bibitem{Be98}  V. P. Belavkin, \textit{On Quantum It\^{o} Algebras and
Their Decompositions.} Lett. in Math. Phys. \textbf{45}, 131--145 (1998).

\bibitem{Be91}  V. P. Belavkin, \textit{A Quantum Nonadapted It\^{o} Formula
and Stochastic Analysis in Fock Scale. J. Func. Anal., }\textbf{102}, No. 2,
414--447 (1991).

\bibitem{Be93}  V. P. Belavkin, \textit{The Unified It\^{o} Formula Has the
Pseudo-Poisson Structure.} J. Math. Phys. \textbf{34}, No. 4, 1508--18
(1993).

\bibitem{GPZ92}  C. W. Gardiner, A. S. Parkins, \& P. Zoller. Phys. Rev. A 
\textbf{46}, 4363 (1992).

\bibitem{DPZG92}  R. Dunn, A. S. Parkins, P. Zoller, \& C. W. Gardiner.
Phys. Rev. A \textbf{46}, 4382 (1992).

\bibitem{Be95a}  V. P. Belavkin, \textit{On Stochastic Generators of
Completely Positive Cocycles.} Rus. J. Math. Phys., \textbf{3}, 523--528
(1995).

\bibitem{Be97}  V. P. Belavkin, \textit{Quantum Stochastic Positive
Evolutions: Characterization, Construction, Dilation.} Comm. Math. Phys., 
\textbf{184}, 533--566 (1997).

\bibitem{Lin76}  G. Lindblad, \textit{On The Generators of Quantum
Stochastic Semigroups}. Commun. Math. Phys., \textbf{48}, pp. 119--130
(1976).

\bibitem{Bohr}  A. Pais, \textit{Niels Bohr's Times}, Clarendon Press -
Oxford 1991.

\bibitem{Heis}  D. C. Cassidy, \textit{Uncertainty. Werner Heisenberg}. W.
H. Freeman, New-York, 1992.

\bibitem{Jamm}  M. Jammer, \textit{The Conceptual Development of Quantum
Mechanics.} McGraw-Hill, 1966.
\end{thebibliography}
\end{document}